\documentclass[aps,prb,onecolumn,nofootinbib,citeautoscript,10pt]{revtex4-2}  
\usepackage{amsmath,amssymb} 
\usepackage{graphicx,comment}

\usepackage[tight]{subfigure} 
\usepackage{ragged2e}

\usepackage[dvipsnames]{xcolor} 
\usepackage[papersize={8.5in,11in}]{geometry}
\usepackage[colorlinks=true]{hyperref}
\hypersetup{
    bookmarks=true,         
    unicode=false,          
    pdftoolbar=true,        
    pdfmenubar=true,        
    pdffitwindow=false,     
    pdfstartview={FitH},    
    pdfkeywords={keyword1} {key2} {key3}, 
    pdfnewwindow=true,      
    colorlinks=true,       
    linkcolor=magenta, 
    citecolor=blue,        
    filecolor=magenta,      
    urlcolor=blue           
} 

\usepackage{graphicx}
\usepackage{dcolumn}
\usepackage{color}
\usepackage{tabularx, makecell}
\usepackage{epstopdf}
\usepackage{latexsym}
\usepackage{colortbl}
\usepackage{psfrag}
\usepackage{bbm,bm,array,physics}
\usepackage{dsfont}
\usepackage{float}

\def \nn{\nonumber \\}

\def\*#1{\mathbf{#1}} 

\begin{document}

\title{Anisotropic conductivity for the type-I and type-II phases of Weyl/multi-Weyl semimetals in planar Hall set-ups}

\author{Ipsita Mandal}
\email{ipsita.mandal@snu.edu.in}

\affiliation{Department of Physics, Shiv Nadar Institution of Eminence (SNIoE), Gautam Buddha Nagar, Uttar Pradesh 201314, India}

\begin{abstract} 
We compute the non-Drude part of the conductivity tensor in planar Hall set-ups, for tilted Weyl and multi-Weyl semimetals, considering both the type-I and type-II phases. We do so in three distinct set-ups, taking into account the possible relative orientations of the plane spanned by the electric and magnetic fields ($\mathbf E $ and $\mathbf B $) and the direction of the tilt-axis. We derive the analytical expressions for the response tensor, including the effects of the Berry curvature (BC) and the orbital magnetic moment (OMM), both of which arise due to a nontrivial topology of the three-dimensional manifold defined by the Brillouin zone. We exhibit the interplay of the BC-only and the OMM-dependent parts in the nonzero components of the magnetoelectric conductivity, and outline whether the contributions from the former or the latter dominate the overall response.
Our results also show that, depending on the configuration of the planar Hall set-up, one may or may not get terms which have a linear-in-$ B$ dependence.
\end{abstract}

\maketitle

\tableofcontents

\section{Introduction}

Over the last decade, there have been continuous intensive efforts to understand the transport properties of semimetals, which represent materials demonstrating nodal points in their bandstructure, implying that two or more bands cross at these points where the density of states vanishes. Among the three-dimensional (3d) semimetals with twofold band-crossings, the well-known examples include the Weyl semimetals (WSMs) \cite{burkov11_weyl,yan17_topological} and the multi-Weyl semimetals (mWSMs) \cite{bernevig,bernevig2,dantas18_magnetotransport}, whose Brillouin zone (BZ) forms a manifold exhibiting nontrivial topology, due to the Berry phase. A node of an mWSM is a straightforward generalization of that of a WSM \cite{bernevig,bernevig2,dantas18_magnetotransport}, with the dispersion of the former being linear along one direction (which we choose to be the $z$-direction, without any loss of generality) and quadratic/cubic in the plane perpendicular to it (which we label as the $xy$-plane).
The band-crossing points for both the WSMs and the mWSMs are protected by the point-group symmetries of the crystal lattice \cite{bernevig2}. We use the notion of the Berry-scurvature (BC) flux, with each nodal point acting as a source or sink in the momentum space, thus mimicking the elusive magnetic monopole. The value of the monopole charge is equal to the Chern number arising from the Berry connection. Obeying the Nielsen-Ninomiya theorem \cite{nielsen}, such nodal points appear in pairs, with each pair carrying Chern numbers $\pm J$. Thus, whatever BC flux emanates from one partner of the pair, disappears into the singular point represented by the other partner. The sign of the monopole charge (which equals the Chern number) is labelled as the chirality $\chi $ of the corresponding node. For Weyl (e.g., TaAs \cite{huang15_observation, lv_Weyl, yang_Weyl} and HgTe-class materials~\cite{ruan_Weyl}), double-Weyl (e.g., $\mathrm{HgCr_2Se_4}$~\cite{Gang2011} and $\mathrm{SrSi_2}$~\cite{hasan_mweyl16, singh18_tunable}), and triple-Weyl nodes (e.g., transition-metal monochalcogenides~\cite{liu2017predicted}), $J$ takes the values of one, two, and three, respectively.

In an experimental set-up, where a WSM/mWSM is subjected to externally-applied uniform electric ($ \mathbf E \equiv  E  \,\hat{\mathbf r}_E $) and  magnetic ($ \mathbf B \equiv  B \, \hat{\mathbf r}_B $) fields, oriented perpendicular to each other, a potential difference (known as the Hall voltage) is generated along the axis perpendicular to both $\mathbf E $ and $\mathbf B $. This phenomenon is the well-known Hall effect. Generalizing the alignment directions, if we apply $ \mathbf B $ making an angle $ \theta $ with $ \mathbf E $, where $  \theta  \neq \pi/2 \text{ or } 3\pi/2$, the conventional Lorentz-force-induced Hall voltage is zero along the $\hat{\mathbf r}_E $-$ \hat{\mathbf r}_B$ plane. However, due to the nontrivial topology in the BZ, an in-plane voltage difference $V_{PH}$ appears along the axis perpendicular to $ \hat{\mathbf r}_E$, which is known as the planar Hall effect (PHE). This is a consequence of the so-called chiral anomaly~\cite{son13_chiral, burkov17_giant, li_nmr17, nandy_2017_chiral, nandy18_Berry, Nag_2020, ips-mwsm-floquet}, which refers to the charge pumping from one node to its partner with opposite chirality, when $\mathbf E \cdot \mathbf B \neq 0 $. In other words, the planar Hall current originates from a local non-conservation of electric charge in the vicinity of an individual node. The rate of change of the number density of chiral quasiparticles is proportional to $ J \left( \mathbf E \cdot \mathbf  B \right) $, analogous to the Adler-Bell-Jackiw anomaly of the relativistic Weyl fermions \cite{chiral_ABJ, chiral_ano_mWSM}. 
The associated in-plane components of the conductivity tensor are referred to as the longitudinal magnetoconductivity (LMC) and the planar Hall conductivity (PHC), which of course are functions of the mutual angle $ \theta $. The literature currently comprises an extensive number of theoretical works investigating various aspects of such transport coefficients \cite{
ips-mwsm-floquet, amit-magneto, ips-rahul-ph, onofre, ips-rahul-tilt, ips-medel, das20_thermal, ips-rsw-ph, ips-shreya}. Extending the definition of the net magnetic field to include artificial gauge fields, the effects of pseudomagnetic fields (induced by elastic deformations), on the type-I phases of nodal-point semimetals have been studied in Refs.~\cite{ips-rahul-ph, onofre, ips-medel, ips-rsw-ph}.

\begin{figure}[t]
\centering 
\includegraphics[width= 0.5 \textwidth]{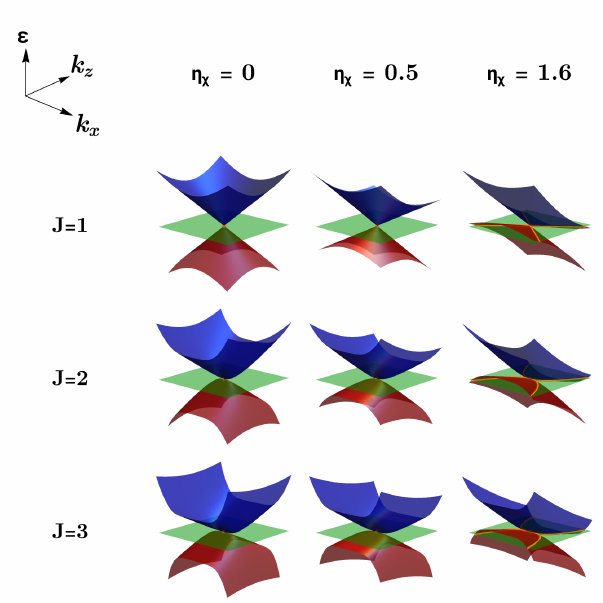}
\caption{\label{fig_bands}
Schematic dispersion ($\varepsilon $) of a single node [cf. Eq.~\eqref{eqham}] plotted against the $k_z k_x $-plane (highlighted with a green colour), where $\eta_\chi $ represents the tilt parameter. The tilting is taken with respect to the $k_z$-direction, along which the dispersion is linear-in-momentum. While the values $\eta_\chi =0 $ (untilted) and $\eta_\chi = 0.5 $ represent the type-I phase, $\eta_\chi = 1.6 $ corresponds to the type-II phase. The yellow points and the yellow lines demarcate the Fermi points and the projections of the open Fermi pockets, respectively, when the chemical potential cuts the band-crossing points.
}
\end{figure}

While the WSM exhibits isotropic dispersion, the mWSMs are inherently anisotropic. However, the bandstructures generically show tilted nodes \cite{emil_tilted, trescher17_tilted, herring} when the system does not possess certain discrete symmetries (e.g., particle-hole and crystal's point-group symmetries). Tilting causes an anisotropy even in the WSMs, making the response dependent on the tilt direction. The response for the untilted mWSMs are already anisotropic, because of the presence of the hybrid of linear dispersion (along the $z$-axis) and quadratic/cubic dispersion (in the $xy$-plane) --- the tilting introduces another source of anisotropy, providing more possibilities of direction-dependence. Here, we would like to point out that, since the tilt parameter enters the Hamiltonian via an identity matrix [cf. Eq.~\eqref{eqham}], the eigenspinors and, hence, the topological quantities (e.g., BC and OMM) for a node remain unchanged. In this paper, we consider a tilt with respect to the $z$-axis, along which both the WSMs and the mWSMs have linear-in-momentum dispersion. If the tilt is small enough, the chemical potential $\mu $ cutting the nodal point (taken to be the zero of the energy) gives a Fermi point rather than a Fermi surface, and the resulting system is said to be in the type-I phase. However, if the tilt is increased to a point such that $\mu = 0$ gives rise to electron-like and hole-like pockets, we call it a type-II phase \cite{soluyanov2015type}. This is also known as the  overtilted situation, which is characterized by the presence of open (i.e., unbounded) Fermi pockets. It is important to realize that, in reality, the open Fermi pockes are unphysical, as they arise as artifacts of considering effective continuum models. Since such models are valid only in the low-energy regimes, in the vicinity of a nodal point, we need to introduce momentum (or energy) cutoffs while performing the momentum integrals appearing in the expressions for the response tensors. The nature of the dispersion in the tiltless, type-I, and type-II phases is illustrated schematically in Fig.~\ref{fig_bands}.

\begin{figure}[t]
\centering 
\subfigure[]{\includegraphics[width=0.25 \textwidth]{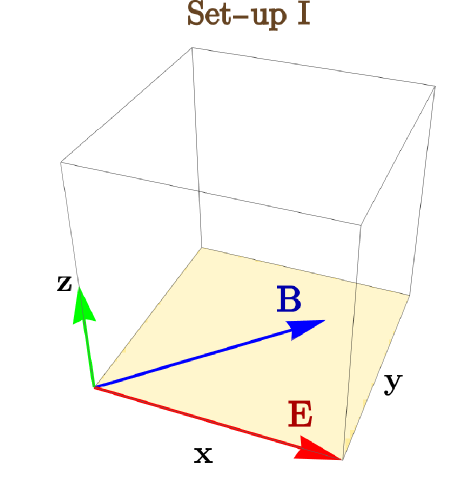}} \hspace{1 cm}
\subfigure[]{\includegraphics[width=0.25\textwidth]{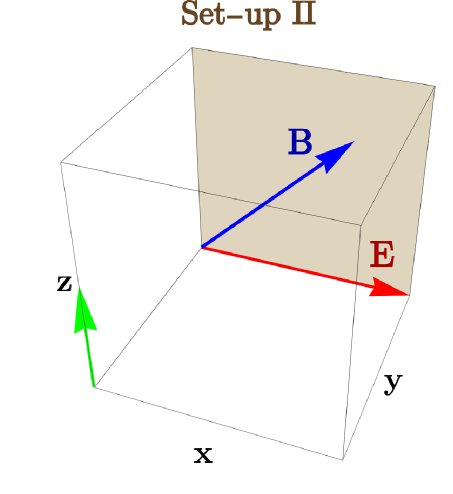}} \hspace{1 cm}
\subfigure[]{\includegraphics[width=0.25 \textwidth]{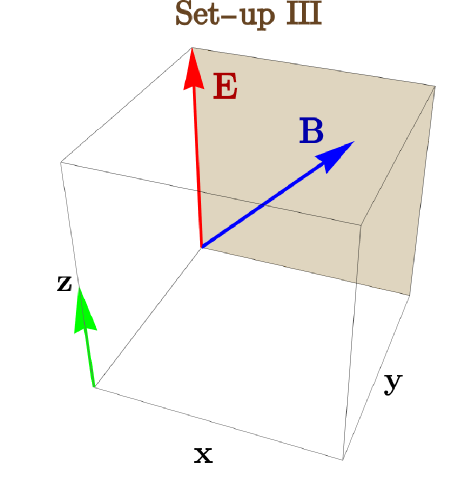}} 
\caption{Schematics of the three set-ups that we have used for investigating the planar Hall effect in WSMs/mWSMs, showing the relative orientations of the external homogeneous electric (red arrow) and magnetic blue arrow) fields, which we label as (a) set-up I, (b) set-up II, and (c) set-up III, respectively. The plane containing the $\mathbf E $ and $\mathbf B $  vectors (making an angle $\theta$ with each other) in each set-up has been highlighted by a background colour-shading. The green arrow denotes the tilting axis, which is fixed throughout the paper. Each type of semimetal has a direction along which the dispersion is linear-in-momentum, chosen here to be the $z$-direction, which is also the axis with respect to which the dispersion has a tilt [cf. Fig.~\ref{fig_bands} and Eq.~\eqref{eqham}].  
\label{figsetup}}
\end{figure}

It has been found earlier that \cite{ma19_planar, kundu20, konye21_microscopic, shao22_plane, amit-magneto, ips-rahul-tilt} tilting can lead to the emergence of linear-in-$ B $ terms for the LMC and PHE, depending on the orientation of the $\hat{\mathbf r}_E $-$ \hat{\mathbf r}_B$ plane with respect to the tilt-axis. With the $z$-axis being chosen as the tilt-axis, we consider three distinct configurations for orienting $ \hat{\mathbf r}_E $ and $\hat{\mathbf r}_B $, with $\mathbf E \cdot  \mathbf B \neq 0$, which are depicted schematically in Fig.~\ref{figsetup}. In the first two set-ups, which we label as I and II, $\hat{\mathbf r}_E$ is oriented perpendicular to the $z$-axis. In set-up I (II), we align $\hat{\mathbf r}_B$ to lie along the $xy$-plane ($zx$-plane). In set-up III, $ \mathbf E$ is applied along the tilt-axis, with $ \hat{\mathbf r}_B $ lying along the $zx$-plane. In order to compute the linear response, we use the semiclassical Boltzmann formalism, which applies in the regime of low-magnitude magnetic fields, leading to a small cyclotron frequency $\omega_c = e\,B / (m^* c) $, where $m^* $ is the effective mass $\sim 0.11 \, m_e$ \cite{params2} and $m_e$ denotes the electron mass. More specifically, we must have $\hbar \, \omega_c \ll \mu$, so that we need not take into account the energy levels being modified into quantized Landau levels.

The information contained in the behaviour of the conductivity tensors includes the signatures of the nontrivial BC, as we will see explicitly from our expressions of the net currents. Additionally, the orbital magnetic moment (OMM) \cite{xiao_review, sundaram99_wavepacket} is another physical property arising from the nontrivial topology of the BZ, which also contributes to the response tensors \cite{timm, onofre, ips-medel, ips-rsw-ph, ips-shreya}. In an earlier work \cite{ips-rahul-tilt}, we computed the in-plane components of the response tensors considering the three distinct set-ups explained above, but neglecting the OMM and restricting to the type-I phases. In this paper, we will derive all the relevant components of the magnetoelectric conductivity (including the out-of-plane components) systematically, which constitute a complete description incorporating the effects of both the BC and the OMM. Furthermore, we will show the final answers both for the type-I and type-II phases. In this context, we would like to point out that complementary signatures of nontrivial topology of the BZ appear as intrinsic anomalous-Hall effect~\cite{haldane, pallab_axionic, burkov_intrinsic_hall}, magneto-optical conductivity when quantized Landau levels determine the conductivity~\cite{gusynin06_magneto, marcus-emil, ips_optical_cond}, Magnus Hall effect~\cite{papaj_magnus, amit-magnus, ips-magnus}, circular dichroism \cite{ips-cd1, ips-cd}, circular photogalvanic effect \cite{moore18_optical, guo23_light, kozii, ips_cpge}, and transmission of quasiparticles across potential barriers/wells \cite{ips-aritra, ips-sandip, ips-sandip-sajid, ips-jns}. 

The paper is organized as follows: In Sec.~\ref{secmodel}, we describe the low-energy effective continuum moded for the WSMs and mWSMs. In Sec.~\ref{secsigma}, we show the generic expressions for the components of the magnetoelectric conductivity, applicable for arbitrary orientations of the $\mathbf E $ and $ \mathbf B $ vectors. The contents of Secs.~\ref{secset1}, \ref{secset2}, and \ref{secset3} are devoted to describing the behaviour for set-ups I, II, and III, respectively. The subsections therein contain the answers obtained for the individual components of the response tensors. In what follows, we will use the natural units, which implies that the reduced Planck's constant ($\hbar $), the speed of light ($c$), the Boltzmann constant ($k_B $), and the magnitude of electron's charge are each set to unity. The appendices contain the explicit derivations and final expressions for the conductivity tensor.

\section{Model}
\label{secmodel}

In the vicinity of a nodal point with chirality $\chi$ and Berry-monopole charge of magnitude $J$, the low-energy effective continuum Hamiltonian is given by \cite{liu2017predicted,bernevig,bernevig2}
\begin{align} 
\label{eqham}
\mathcal{H}_\chi ( \mathbf k) & = 
\mathbf d_\chi( \mathbf k) \cdot \boldsymbol{\sigma}
 +  \eta_\chi \,v_z\, k_z \, \sigma_0 \,,
\quad k_\perp=\sqrt{k_x^2 + k_y^2}\,, \quad
\phi_k = \arctan({\frac{k_y}{k_x}})\,,
\quad \alpha_J=\frac{v_\perp}{k_0^{J-1}} \,, \nn
\mathbf d_\chi( \mathbf k) &=
\left \lbrace
\alpha_J \, k_\perp^J \cos(J\phi_k), \,
\alpha_J \, k_\perp^J \sin(J\phi_k), \,
\chi \, v_z \, k_z \right \rbrace.
\end{align}
Here, $ \boldsymbol{\sigma} = \lbrace \sigma_x, \, \sigma_y, \, \sigma_z \rbrace $ is the vector-operator consisting of the three Pauli matrices, $\sigma_0$ is the $2 \times 2$ identity matrix, $\chi \in \lbrace 1, -1 \rbrace $ denotes the chirality of the node, and $v_z$ ($v_\perp$) is the Fermi velocity along the $z$-direction ($xy$-plane). The parameter $k_0$ has the dimension of momentum, whose value depends on the microscopic details of the material under consideration. Lastly, $\eta_\chi $ is the tilt parameter, with the tilt-axis chosen to be along the $z$-direction.

The eigenvalues of the Hamiltonian are given by
\begin{align} 
\label{eigenvalues_kperp_kz_phik}
\varepsilon_{\chi, s} ({ \mathbf k})= 
\eta_\chi \, v_z \, k_z + (-1)^{s+1} \, \epsilon_{\mathbf k} \,, \quad
s \in \lbrace 1,2 \rbrace ,
\quad 
\epsilon_{\mathbf k}
= \sqrt{\alpha_J^2 \, k_\perp^{2J} + v_z^2 \, k_z^2}\,,
\end{align}
where the value $1$ ($2$) for $s$ represents the conduction (valence) band. 
We note that we recover the linear and isotropic nature of a WSM by setting $J=1$ and $\alpha_1= v_z$.

The band-velocity of the chiral quasiparticles is given by
\begin{align}
{\boldsymbol v}^{(0,s)}_\chi ( \mathbf k) 
\equiv \nabla_{\mathbf k} \varepsilon_{\chi, s}   (\mathbf k)
=  \frac{(-1)^{s+1} }
{  \epsilon_{\mathbf k} }  \left\lbrace J\, 
 \alpha_J^2 \,  k_\perp^{2J-2} \,  k_{x} , \, J \,  \alpha_J^2 \,  k_\perp^{2J-2} 
 \,  k_{y}  , \, v_z^2\,  k_z \right \rbrace
+ \left \lbrace 0, 0, v_z \,\eta_\chi \right \rbrace .
\end{align}
The Berry curvature (BC) and the orbital magnetic moment (OMM), associated with the $s^{\rm{th}}$ band, are expressed by  \cite{xiao_review,xiao07_valley,konye21_microscopic}
\begin{align} 
\label{eqomm}
& {\mathbf \Omega}_{\chi, s}( \mathbf k)  = 
    i \, \langle  \nabla_{ \mathbf k}  \psi_s^\chi ({ \mathbf k})| \, 
    \cross  \, | \nabla_{ \mathbf k}  \psi_s^\chi ({ \mathbf k})\rangle
\Rightarrow
\Omega^i_{\chi, s}( \mathbf k)  \overset{\text{two}-} {\underset{\text{band}} {=}}
 \frac{  (-1)^{s} \,  
\epsilon^i_{\,\,\,jl}}
 {4\,| \mathbf d_\chi (\mathbf k) |^3} \, 
 \mathbf d_\chi (\mathbf k) \cdot
 \left[   \partial_{k_j} \mathbf d_\chi (\mathbf k) \cross  
 \partial_{k_l } \mathbf d_\chi (\mathbf k) \right ],  \text{ and }    
\nn & 
{\boldsymbol{m}}_{\chi,s} ( \mathbf k) 
\overset{\text{two}-} {\underset{\text{band}} {=}}
 \,   \frac{ -\, i \, e} {2 } \,
\langle  \mathbf \nabla_{ \mathbf k} \psi_s({ \mathbf k})| \cross
\left [\,
\left \lbrace \mathcal{H}({ \mathbf k}) -\mathcal{E}_{\chi, s}
({ \mathbf k}) 
\right \rbrace
| \mathbf \nabla_{ \mathbf k} \psi_s({ \mathbf k})\rangle \right ]
\Rightarrow
m^i_{\chi,s} ( \mathbf k) =
  \frac{ -\, e \, \epsilon^i_{\,\,\,jl}
  } 
{4 \, | \mathbf d_\chi (\mathbf k) |^2} \,  
\mathbf d_\chi (\mathbf k) \cdot
 \left[   \partial_{k_j} \mathbf d_\chi (\mathbf k) \cross  \partial_{k_l} \mathbf d_\chi (\mathbf k) \right ],
\end{align}
respectively. The indices $i$, $j$, and $l$ $ \in \lbrace x, y, z \rbrace $, and are used to denote the Cartesian components of the 3d vectors and tensors. The symbol $ |  \psi_s^\chi ({ \mathbf k}) \rangle $ denotes the normalized eigenvector corresponding to the band labelled by $s$, with $ \lbrace |  \psi_1^\chi \rangle, \,  \lbrace |  \psi_2^\chi \rangle \rbrace $ forming an orthonornomal set for each node.

On evaluating the expressions in Eq.~\eqref{eqomm}, using Eq.~\eqref{eqham}, we get
\begin{align}
\mathbf \Omega_{\chi, s }({ \mathbf k})= 
 \frac{ \chi \,(-1)^{s}
J \,v_z \, \alpha_J^2 \, k_\perp^{2J-2} }
{2 \,\epsilon^3_{\mathbf k}
} 
\left
\lbrace k_x, \, k_y, \, J\, k_z \right \rbrace , \quad 
   {\boldsymbol{m}}_{\chi, s }({ \mathbf k}) 
=    \frac{ -\,\chi \, e\, J \,v_z \, \alpha_J^2 \, k_\perp^{2\,J-2} }
{2 \, \epsilon^2_{\mathbf k}} 
 \left \lbrace k_x, \, k_y, \, J \, k_z \right \rbrace .
\end{align}
From these expressions, we immediately observe the identity
\begin{align}
{\boldsymbol{m}}_{\chi, s }({ \mathbf k}) 
=   (-1)^{s+1} \,  e \, \epsilon_{\mathbf k} \, \mathbf \Omega_{\chi, s }({ \mathbf k}) \, . 
\end{align}
While the BC changes sign with $s$, the OMM does not. Hence, we will remove the subscript ``$s$'' from ${\boldsymbol{m}}_{\chi, s }({ \mathbf k})$.

\begin{figure}[t]
\centering 
\includegraphics[width=0.8 \textwidth]{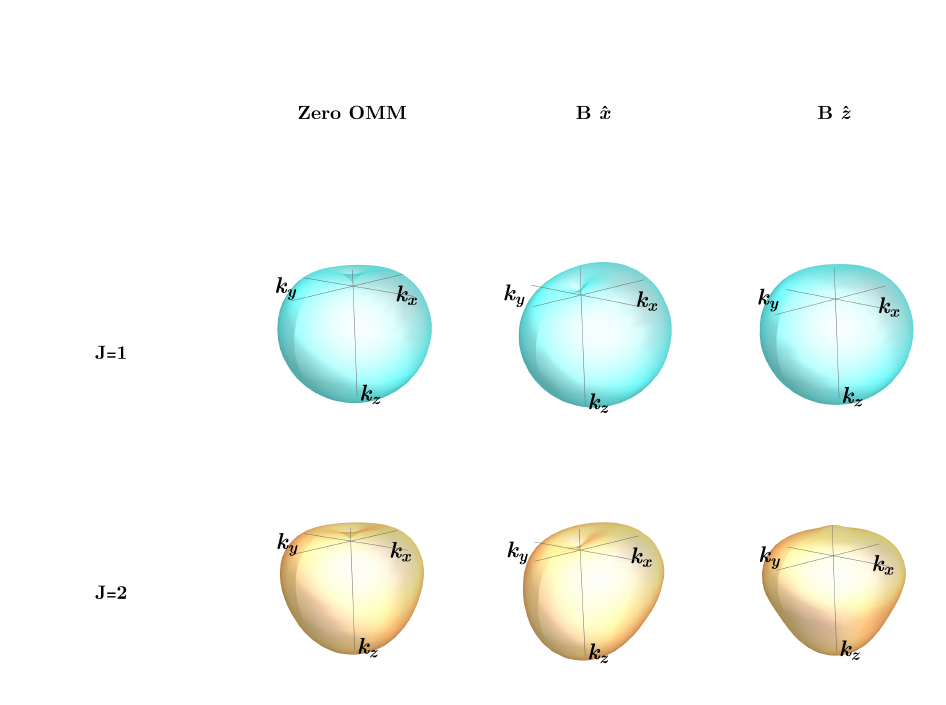} \hspace{ 2.5 cm }
\caption{Schematics of the Fermi surfaces for the tilted node of a WSM and a double-Weyl semimetal (in the type-I phase), without and with the OMM-corrections for the effective energy dispersion [cf. Eq.~\eqref{eqmodi}]. Here, the effective magnetic field is directed purely along the $x$-axis ($z$-axis) for the resulting Fermi surfaces shown in the second (third) column.
\label{figfs}}
\end{figure}

\section{Magnetoelectric conductivity}
\label{secsigma}

In order to include the effects from the OMM and the BC, we first define the quantitites
\begin{align}
\label{eqmodi}
& \mathcal{E}_{\chi, s} (\mathbf k) 
= \varepsilon_{\chi, s}  (\mathbf k) + \varepsilon_\chi^{ (m) }  (\mathbf k) \, ,
\quad 
\varepsilon_\chi^{(m)}   (\mathbf k) 
= - \,{\mathbf B} \cdot \boldsymbol{m}_\chi  (\mathbf k) \,, \quad
{\boldsymbol  v}_{\chi, s} ({\mathbf k} ) \equiv 
 \nabla_{{\mathbf k}}   \mathcal{E}_{\chi, s} ({\mathbf k})
 = {\boldsymbol  v}^{(0, s)}_\chi ({\mathbf k} ) + {\boldsymbol  v}^{(m)}_\chi ({\mathbf k} ) \,,\nn
& {\boldsymbol  v}^{(m)}_\chi ({\mathbf k} )
= \nabla_{{\mathbf k}} \varepsilon_\chi^{(m)}   (\mathbf k) \,,
\quad  D_{\chi, s} = \left [1 
+ e \,  \left \lbrace 
{\mathbf B} \cdot \mathbf{\Omega }_{\chi, s}  (\mathbf k)
\right \rbrace  \right ]^{-1} ,
\end{align}
where $ \varepsilon_\chi^{ (m) } ({\mathbf k})$ is the Zeeman-like correction to the energy due to the OMM, $ {\boldsymbol  v}_{\chi, s}({\mathbf k} )  $ is the modified band-velocity of the Bloch electrons after including $ \varepsilon_\chi^{ (m) } ({\mathbf k})$, and $ D_{\chi, s} $ is the modification factor of a phase-space volume-element, caused by a nonzero BC. The emergence of a deformed effective Fermi surface, on including the OMM-correction, is given by $\varepsilon_\chi^{ (m) }  (\mathbf k)$, as depicted schematically in Fig.~\ref{figfs}.

The weak-magnetic-field limit implies that 
\begin{align}
 e \, |{\mathbf B} \cdot \mathbf{\Omega }_{\chi, s}  | \ll 1 . 
 \label{Condition}
\end{align}
In our calculations, we will retain terms upto $\order{ |{\mathbf B}|^2}$ and, thus, use
\begin{align}
D_{\chi, s} &=
 1 - e \,  \left( {\mathbf B}  \cdot \mathbf{\Omega }_{\chi, s}  \right) 
+   e^2  \,  \left( {\mathbf B} \cdot \mathbf{\Omega }_{\chi, s}  \right)^2  
+  \order{ |{\mathbf B}|^3} \,. 
 \label{Exp_D}
\end{align}
Also, the condition in Eq.~(\ref{Condition}) implies that $ |\varepsilon_\chi^{ (m) } (\mathbf k) | $ is small compared to $|\varepsilon_{\chi, s} (\mathbf k) |$, i.e.,
$ \vert {\mathbf B} \cdot  \boldsymbol{ m }_\chi \vert \equiv 
e\, \epsilon_{\mathbf k} \, 
 \left| {\mathbf B} \cdot  \mathbf{\Omega}_{\chi, s} \right |
 \ll  |\varepsilon_{\chi, s} |  \,.$
This means that the Fermi-Dirac distribution ($f_0$) and its first-order derivative should be power-expanded up to quadratic order in the magnetic field as follows:
\begin{align}
& f_{0} ( \mathcal{E}_{\chi, s} ) = f_{0} ( \varepsilon_{\chi, s} ) 
+  \varepsilon_\chi^{ (m) } \,  
f'_{0} (\varepsilon_{ \chi, s} ) + \frac{1}{2} \left(  \varepsilon_\chi^{(m)} \right)^2  \,
  f^{\prime \prime}_{0} ( \varepsilon_{\chi, s} )  +  \order{ |\mathbf B|^3} \,,
\nn & f_{\rm prime} ( \mathcal{E}_{\chi, s} ) \equiv 
\frac{ \partial f_0 ( \mathcal{E}_{\chi, s} )} {\partial {\mathcal{E}_{\chi, s} }} 
= f_0^\prime ( \varepsilon_{ \chi, s}  ) 
+  \varepsilon^{ (m) }_\chi \,  
f^{\prime \prime}_0 (\varepsilon_{ \chi, s} ) 
+ \frac{1}{2} \left(  \varepsilon^{(m)}_\chi \right)^2  \,
  f^{\prime \prime \prime}_0 ( \varepsilon_s  )  +  \order{ |\mathbf B|^3} \,,
\label{Exp_f}
\end{align}
where a ``prime'' superscript indicates derivative with respect to the energy argument of $f_0$. The first expansion is needed only for the anomalous-Hall term, which we do not consider in this paper.

Using the semiclassical Boltzmann formalism, the general expression for the magnetoelectric conductivity tensor for an isolated node of chirality $\chi$, contributed by the band with index $s$, is given by \cite{mermin, ips-kush-review}
\begin{align}
\sigma^{\chi, s}_{i j} &= - \, e^2 \,  \tau   
\int \frac{  d^3 \mathbf k } { (2\, \pi )^3  } \,  
D_{\chi, s}  \,  \left[ \left( v_{\chi,s } \right)_i + e  \, 
 (  {\boldsymbol  v}_{\chi,s} \cdot \mathbf{\Omega }_{\chi,s} ) \,  B_{i} \right] 
 \left[ \left( v_{\chi,s } \right)_j + 
 e  \,  (  {\boldsymbol  v}_{\chi, s} \cdot \mathbf{\Omega }_{\chi, s} ) \,  B_{j} \right] 
\frac{\partial f_{0} (\mathcal{E}_{\chi, s}) }
{ \partial \mathcal{E}_{\chi, s}  }  .
\label{eq_elec}
\end{align}
The above expression is valid in the relation-time approximation for the collision integral, ignoring internode scatterings. We ignore the internode scattering on the grounds that such scattering involves momentum-transfers of large magnitude (of the order of the separation of the Weyl nodes in the BZ), which makes it a weaker process compared to the intranode scattering.
The case of internode collisions will be considered in future works, using the same formalism as reported in Ref.~\cite{ips-internode}. Hence, $\tau$ denotes a momentum-independent relaxation time, which is assumed to be determined phenomenologically. Furthermore, we do not include here the parts coming from the so-called ``intrinsic anomalous-Hall'' effect and the Lorentz-force operator~\cite{ips-rsw-ph, ips-spin1-ph}. The detailed steps for obtaining $\sigma^{\chi, s}_{i j} $ can be found in Appendix A of Ref.~\cite{ips-rahul-ph} --- hence, we do not repeat those steps for the sake of brevity.

We will work in the $T \rightarrow 0 $ limit, such that $f_0^\prime (\mathcal E) \rightarrow -\, \delta (\mathcal E - \mu )$. We note that the results for $T>0$ can be easily obtained by using the relation given by \cite{mermin}
\begin{align}
\sigma^\chi_{ij} (T) =  -\int_{-\infty}^\infty d\varepsilon \,\sigma^\chi_{ij} (T=0)\, 
\frac{ \partial f_0 ( \varepsilon , \mu,T   )}
{\partial {\varepsilon }}  \,.
\end{align}
The so-called Drude term (say, $\sigma^{\chi, s}_{{\rm Drude}, i i} $) is the longiudinal part part existing even in absence of an external magnetic field. We will not discuss it any further because it does not change while varying the external magnetic field. For the $B$-dependent terms, we note that a linear-in-$B$ term can emerge only if $\left( \mathbf E \cdot \mathbf B \right) \eta_\chi \, {\mathbf{\hat z}} $, $ \left( \mathbf B \cdot  \eta_\chi \, {\mathbf{\hat z}} \right ) \mathbf E $, or $\left( \mathbf E \cdot \eta_\chi \, {\mathbf{\hat z}} \right)  \mathbf B $ is nonzero (cf. Ref.~\cite{amit-magneto}). More explicitly, if we have $\mathbf E = E \, {\mathbf{\hat r}_E} $ and $\mathbf B = B \, {\mathbf{\hat r}_B} $, where ${\mathbf{\hat r}_E}$ and  ${\mathbf{\hat r}_B}$ $\in \lbrace  {\mathbf{\hat x}}, \,  {\mathbf{\hat y}},\,  {\mathbf{\hat z}} \rbrace$, we will have nonzero currents $\propto B $ if $  \chi\, B \, ({\mathbf{\hat r}_E} \cdot {\mathbf{\hat r}_B}) \, {\mathbf{\hat z}} $, $ \chi\, B_z \, {\mathbf{\hat r}_E} $, or $\chi \,E_z \, B \, {\mathbf{\hat r}_B}$ is nonzero. The reason is obviously the fact that the integrals will give a nonzero answer for a linear-in-$B$ term only when at least one of the above conditions is satisfied.

In the following, we will assume that a positive chemical potential $\mu$ is applied (i.e., $\mu>0$). Hence, we will employ the following coordinate transformation to perform the integrations:
\begin{align}
k_x = k_\perp \cos \phi \,, \quad k_y = k_\perp \sin \phi\,, \quad
k_z = \frac{\epsilon \cos \gamma} {v_z} \,,
\quad k_\perp = \left( \frac{ \epsilon  \sin \gamma  }
{\alpha_J } \right)^{1/J}\,,
\end{align}
where $\phi \in [0, 2 \pi )$, $\epsilon \in [0, \infty )$, and $\gamma \in [0, \pi )$.
The Jacobian for the coordinate-change is given by
$
\mathcal J_0  = \frac {\alpha_J^{-\frac{2} {J}} \epsilon^{\frac{2} {J}}
\sin^{\frac {2} {J} -1} \gamma} {J\, v_z} \,.$
The integrals containing the Dirac-delta functions can be simplified as:
\begin{align}
\label{eqint}
 & \int_0^{\pi} d \gamma \int_0^\infty  d\epsilon \int_{0}^{2\pi} 
 d\phi \; \mathcal J_0 \,
 \delta(\eta_\chi \,\epsilon \cos \gamma + (-1)^{s+1} \,\epsilon -\mu)
\rightarrow  
\int_0^{\pi} d \gamma \int_0^\infty  d\epsilon \int_{0}^{2\pi}  d\phi\;
\mathcal J \, \delta \Big (\epsilon - \frac{\mu}
 { \eta_\chi \cos \gamma + (-1)^{s+1} }  \Big ) 
\nn &  \rightarrow  
\int_{-1}^1 d u  \int_0^\infty  d\epsilon \int_{0}^{2\pi}  d\phi \; 
\frac{   \mathcal J
\, \delta \Big (\epsilon - \frac{\mu} { \eta_\chi \,u \,+ \, (-1)^{s+1} }  \Big ) }
 {\sqrt{1-u^2}}\, , \text{ where }
\mathcal J = \frac{\mathcal J_0 }  { | \eta_\chi \cos \gamma + (-1)^{s+1} | } 
= \frac {\alpha_J^{-\frac{2} {J}} \epsilon^{\frac{2} {J}}
\sin^{\frac {2} {J} - 1} \gamma} 
{J\, v_z \, | \eta_\chi \cos \gamma + (-1)^{s+1} | }\,.
\end{align}
We perform the $\phi$-integral as the first step. Thereafter, we get rid of the $\epsilon$-integral. Observing that the root of the Dirac-delta function imposes the restriction that $u \equiv \cos \gamma =\frac{\mu - (-1)^s\,\epsilon } {\epsilon \, \eta_\chi}$. Therefore, $\epsilon \rightarrow \infty $ implies $ u \rightarrow - (-1)^s / \eta_\chi $, necessitating the need for imposing a cutoff to regularize the integrals for $\eta_\chi >1$. We implement this by using the parameter $\Lambda $, such that $ \Lambda / \mu > 1 $ and, additionally,
\begin{enumerate}
\item for $s=1$, the range of the $u$-integration needs to be restricted to
$ -\left( 1 -\frac{\mu}{\Lambda}\right)/\eta_\chi \leq u \leq 1 $, with $\left( 1 -\frac{\mu}{\Lambda}\right) < \eta_\chi$;

\item for $s=2$, the range of the $u$-integration needs to be restricted to
$ \left( 1 +\frac{\mu}{\Lambda}\right)/\eta_\chi \leq u \leq 1 $, with $ \left( 1 +\frac{\mu}{\Lambda}\right) < \eta_\chi$.
\end{enumerate}
Within the above restricted ranges, we immediately find that, for $s=2 $, $ | \eta_\chi \cos \gamma - (-1)^s | = | \eta_\chi \cos \gamma -1 | =1- \eta_\chi \cos \gamma $.
In order to disentangle the contributions purely from the BC (i.e., when OMM is neglected) from the ones which arise when OMM is included, we define the BC-only part as $\sigma_{ij}^{ (\chi, bc) }$, and the rest as $\sigma_{ij}^{ (\chi, m) }$:
\begin{align}
& \sigma_{ij}^{ (\chi) } =
\sigma_{ij}^{ (\chi, bc) } + \sigma_{ij}^{ (\chi, m) }\,.
\end{align}

The nature of the components for the type-I and type-II phases is summarized in Table~\ref{table_sets}, which provides a glimpse at the final results before delving into the explicit expressions in the sections that follow:
\begin{table}[h!]
\subfigure[\textcolor{blue}{Set-up I}: Comprises $\mathbf E = E \, {\mathbf{\hat x}}$, $\mathbf B = B_x \, {\mathbf{\hat x}} + B_y \, {\mathbf{\hat y}}  $, presented in Sec.~\ref{secset1}. A nonzero linear-in-$B$ part appears only in $\sigma_{zx} $, in the presence of a nonzero tilt, which arises from the current proportional to $ (\mathbf E\cdot \mathbf B )
\,\eta_\chi \, {\mathbf{\hat z}} $.]
{\centering
\begin{tabular}{ |c|c|c|c|}
\hline
    & $\sigma_{xx} $ --- longitudinal
    &  $\sigma_{yx} $ --- in-plane transverse
    & $\sigma_{zx} $ --- out-of-pane \\ \hline
type-I & 
\makecell{terms proportional to $B_x^2$ and $B_y^2$} 
& \makecell{terms proportional to $B_x B_y$}
& 
terms proportional to $\chi B_x $\\ \hline
type-II &
\makecell{terms proportional to $B_x^2$ and $B_y^2$;\\
non-divergent}
&\makecell{terms proportional to $B_x B_y$;\\
non-divergent}
&\makecell{terms proportional to $\chi B_x $;\\
diverges as $\ln \Lambda $}  \\ \hline
\end{tabular}}
\subfigure[\label{xyz}\textcolor{blue}{Set-up II}: Comprises $ \mathbf E = E \, {\mathbf{\hat x}} $, $\mathbf B = B_x \, {\mathbf{\hat x}} + B_z \, {\mathbf{\hat z}} $, presented in Sec.~\ref{secset2}. Nonzero linear-in-$B$ parts appear in $\sigma_{xx} $ and $\sigma_{zx} $ for a nonzero tilt, caused by electric currents which are proportional to $ ( \mathbf B \cdot  \eta_\chi \, {\mathbf{\hat z}} ) \, \mathbf E $
and $(\mathbf E \cdot \mathbf B ) \, \eta_\chi \, {\mathbf{\hat z}} $, respectively.]
{\centering
\begin{tabular}{ |c|c|c|c|}
\hline
    & $\sigma_{xx} $ --- longitudinal 
  & $\sigma_{zx} $ --- in-plane transverse
  & $\sigma_{yx} $ --- out-of-pane \\ \hline
type-I & 
\makecell{terms proportional to $\chi B_z$, $B_x^2$, and $B_z^2$}&
\makecell{terms proportional to $\chi B_x $  and $B_x B_z $}& 
vanishes  \\ \hline
type-II &
\makecell{terms proportional to $\chi B_z$, $B_x^2$, and $B_z^2$;\\
$\propto \chi B_z$-terms diverge as 
$
\begin{cases}
\ln \Lambda &\text{ for } J=1 \\
\,\Lambda &\text{ for } J=2 \\
\Lambda^{ \frac{4} {3} } &\text{ for } J=3
\end{cases} ; $ \\
$\propto B_z^2 $-terms diverge for $J\geq 2$
\\
$ \Rightarrow \begin{cases}
\ln \Lambda &\text{ for } J=2 \\
\Lambda^{ \frac{2} {3} } &\text{ for } J=3
\end{cases} $} &
\makecell{
terms proportional to $\chi  B_x $ and $B_x B_z $;\\
\\
$\propto \chi B_x $-terms diverge as $\ln \Lambda $;\\
\\
$\propto B_x B_z  $-terms diverge as $
\begin{cases}
\ln \Lambda &\text{ for } J=1 \\
\,\Lambda &\text{ for } J=2 \\
\Lambda^{ \frac{4} {3} } &\text{ for } J=3
\end{cases} $} & 
\makecell{vanishes}  \\ \hline
\end{tabular} }
\subfigure[\textcolor{blue}{Set-up III}: Comprises $\mathbf E = E \, {\mathbf{\hat z}}$, $\mathbf B = 
B_x \, {\mathbf{\hat x}} + B_z \, {\mathbf{\hat z}}  $, discussed in Sec.~\ref{secset3}. While linear-in-$B$ parts in $\sigma_{zz} $ are caused by currents $ \propto ( \mathbf E \cdot \mathbf B ) \, \eta_\chi \, {\mathbf{\hat z}} $ and/or $\propto ( \mathbf B \cdot  \eta_\chi \, {\mathbf{\hat z}} )\, \mathbf E $, the linear-in-$B$ part in $\sigma_{xz} $ originates from a current $ \propto (\mathbf E \cdot  \eta_\chi \, {\mathbf{\hat z}})\, B_x \, {\mathbf{\hat x}}$.]
{\centering
\begin{tabular}{ |c|c|c|c|}
\hline
 & $\sigma_{zz} $ --- longitudinal 
 &  $\sigma_{xz} $ --- in-plane transverse 
 & $\sigma_{yz} $ --- out-of-pane \\ \hline
type-I & 
\makecell{terms proportional to $\chi B_z$, $B_x$, and $B_z^2$}
& \makecell{terms proportional to $\chi B_x $ and $B_x B_z $} & 
vanishes  \\ \hline
type-II &
\makecell{terms proportional to $ \chi B_z $, $B_x$, and $B_z^2$;\\
\\$ \propto  \chi B_z $-terms diverge as $\ln \Lambda$} &
\makecell{
terms proportional to $\chi B_x $ and $B_x B_z $;\\
$\propto \chi B_x$-terms diverge as $\ln \Lambda $;\\
$\propto B_x B_z  $-terms diverge as $
\begin{cases}
\ln \Lambda &\text{ for } J=1 \\
\,\Lambda &\text{ for } J=2 \\
\Lambda^{ \frac{4} {3} } &\text{ for } J=3
\end{cases} $} &
\makecell{vanishes}  \\ \hline
\end{tabular} }
\caption{\label{table_sets}Summary of the key characteristics of the response for the three distinct set-ups.}
\end{table}

\section{Set-up I: $\mathbf E = E \, {\mathbf{\hat x}}$, $\mathbf B = B_x \, {\mathbf{\hat x}} + B_y \, {\mathbf{\hat y}}  $}
\label{secset1}

In set-up I, as shown in Fig.~\ref{figsetup}(a), the tilt-axis is perpendicular to the plane spanned by $\mathbf E $ and $\mathbf B $. Due to the rotational symmetry of the dispersion of each semimetallic node within the $xy$-plane, the exact directions of $\mathbf E $ and $\mathbf B $ do not matter --- the only physically relevant parameter is the angle between $ \hat{\mathbf r}_E$ and $ \hat{\mathbf r}_B$. Hence, without any loss of generality, we choose $ \hat{\mathbf r}_E = {\mathbf{\hat x}}  $ and $\hat{\mathbf r}_B = \cos \theta \, {\mathbf{\hat x}} + \sin \theta \, {\mathbf{\hat y}}  $, such that $\mathbf E = E \, {\mathbf{\hat x}}$ and $\mathbf B = B_x \, {\mathbf{\hat x}} + B_y \, {\mathbf{\hat y}} \equiv B  \,\hat{\mathbf r}_B  $.
The details of the generic forms of the integrals are shown in Appendix~\ref{appset1}.
Therein, Appendices~\ref{appset1long}, \ref{appset1_inplane_trs}, and \ref{appset1_outplane} deal with the longitudinal, in-plane transverse, and out-of-plane transverse components, respectively.

\subsection{Set-up I: Longitudinal components}
\label{set1long}

The $J$-dependent expressions for the starting integrals are shown in Eq.~\eqref{eqappset1long} of Appendix~\ref{appset1long}. 

\subsubsection{\underline{Results for the type-I phase for $\mu>0$}
\label{set1_longtype1}}

\begin{figure*}[t!]
\centering 
\subfigure[]{\includegraphics[width=0.32 \textwidth]{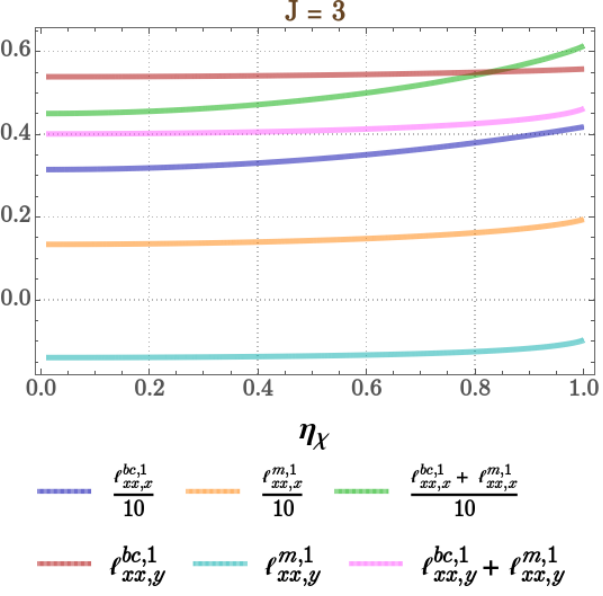}} \hspace{0.1 cm}
\subfigure[]{\includegraphics[width=0.315 \textwidth]{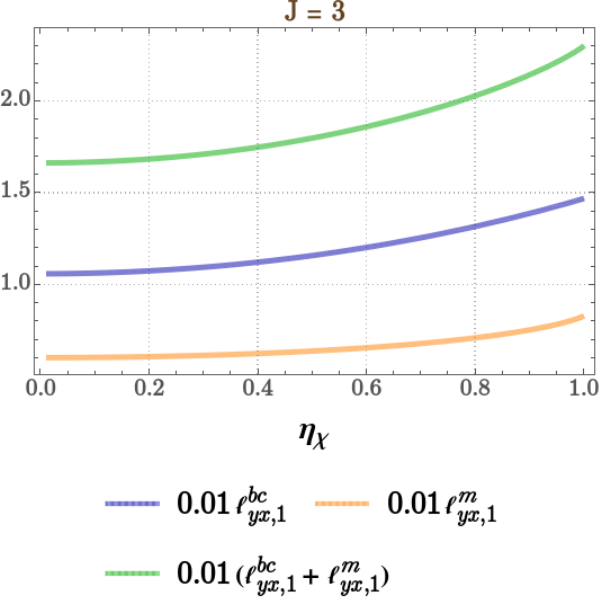}} \hspace{0.1 cm}
\subfigure[]{\includegraphics[width=0.32 \textwidth]{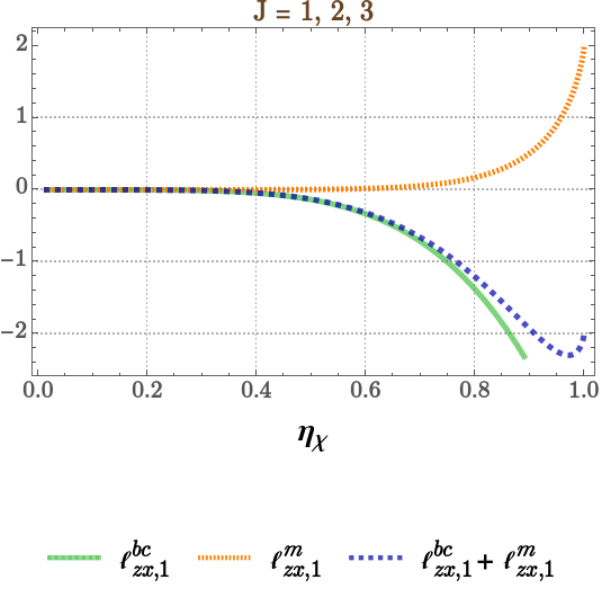}}
\caption{
Response in the type-I phase of set-up I: Subfigures (a) and (b) show the comparision-curves for $J=3 $ --- (a) $ \ell^{bc, 1}_{xx,x} $, $ \ell^{m, 1}_{xx,x} $, $ \ell^{bc, 1}_{xx,y} $, and $ \ell^{m, 1}_{xx,y} $ [cf. Eq.~\eqref{eqsetup1xxtype1}]; (b) $ \ell^{bc}_{yx,1} $ and $ \ell^{m}_{yx,1} $ [cf. Eq.~\eqref{eqsetup1yxtype1}]. Subfigure (c) shows the $J$-independent values of $ \ell^{bc}_{zx,1} $ and $ \ell^{m}_{zx,1}$ [cf. Eq.~\eqref{eqsetup1zxtype1}].
\label{figell1}}
\end{figure*}

For $\mu>0$, only the conduction band contributes in the type-I phase.
The contributions are divided up into BC-only and OMM parts as
\begin{align}
\label{eqsetup1xxtype1}
& \sigma_{xx}^{ (\chi, bc) } =
 \frac {e^4 \,  \tau  \, J^3\, v_z  } 
 {128 \, \pi^{ \frac{3} {2}} } 
\left (\frac {\alpha_J} {\mu} \right)^{\frac{2} {J}} 
 \left(  B_x^2 \,\ell^{bc, 1}_{xx,x} 
 + B_y^2 \, \ell^{bc, 1}_{xx,y} \right ) \text{ and }
\sigma_{xx}^{ (\chi, m) } = 
\frac {e^4 \,  \tau  \,J^3\,  v_z  } 
 {128 \, \pi^{ \frac{3} {2}} } 
\left (\frac {\alpha_J} {\mu} \right)^{\frac{2} {J}} 
 \left( 
B_x^2 \, \ell^{m, 1}_{xx,x} + B_y^2\, \ell^{m, 1}_{xx,y}
\right ) .
\end{align}
Here, $ \ell^{bc, 1}_{xx,x} $ and $\ell^{m, 1}_{xx,x}$ ($ \ell^{bc, 1}_{xx,y} $ and $\ell^{m, 1}_{xx,y}$)
represent the parts proportional to $B_x^2 $ ($ B_y^2 $). From the final expressions shown in Appendix~\ref{appset1_longtype1}, the following characteristics are observed:
\begin{center}
\begin{tabular}{ |c||c|c|c||c|c|c|}
\hline
 & $ \ell^{bc, 1}_{xx,x} $ &  $\ell^{m, 1}_{xx,x}$ & $ \ell^{bc, 1}_{xx,x} + \ell^{m, 1}_{xx,x}$
 & $ \ell^{bc, 1}_{xx,y} $ &  $\ell^{m, 1}_{xx,y}$ & $ \ell^{bc, 1}_{xx,y} + \ell^{m, 1}_{xx,y}$ \\ \hline
 $J=1$ &  $ \frac{16 \, \left( 8 + 13 \,\eta_\chi^2 \right)}{15 \,\sqrt{\pi }} $
 & $ \frac { - \,64 \,\left ( 1 + 2 \, \eta_\chi^2  \right)} {15 \, \sqrt {\pi}}$
 & $ \frac {16 \,\left ( 4 + 5 \,\eta_\chi^2  \right)} {15 \, \sqrt {\pi}}$
 & $  \frac{16} {15\, \sqrt{\pi }} $
 & $\frac{ -\, 16 }{5\, \sqrt{\pi }}$ & $  \frac{-\, 32}{15\, \sqrt{\pi }} $ \\ \hline
 $J=2$ & $  \frac{\sqrt{\pi } \,\left( 31 + 16 \,\eta_\chi^2\right)} {16}$ 
& $ \frac{\sqrt{\pi }}{4} $ & $ \frac {\sqrt {\pi} \,\left ( 35 + 16 \,\eta_\chi^2  \right)} {16} $
& $ \frac {5 \,\sqrt {\pi}} {16}$ & $\frac {- \,\sqrt {\pi}} {4} $ & $ \frac {\sqrt {\pi}} {16}$ \\ \hline
 $J=3 $ & $>0$  & $>0 $ & $> 0$ & $>0$  & $ < 0 $ & $> 0$ \\ \hline
\end{tabular}
\end{center}
For $J=3$, since the expressions involve hypergeometric functions, the individual and overall behaviour is illustrated in Fig.~\ref{figell1}(a). From there, we find that $ \ell^{bc, 1}_{xx,x} > 0$, $\ell^{m, 1}_{xx,x} > 0$, $ \ell^{bc, 1}_{xx,y}>0 $, and $ \ell^{m, 1}_{xx,y}< 0 $, with the net response remaining positive in the range $ 0 \leq \eta_\chi \leq 1 $.

\subsubsection{\underline{Results for the type-II phase for $\mu>0$}}
\label{set1_longtype2}

\begin{figure*}[t!]
\centering 
\subfigure{\includegraphics[width=0.335 \textwidth]{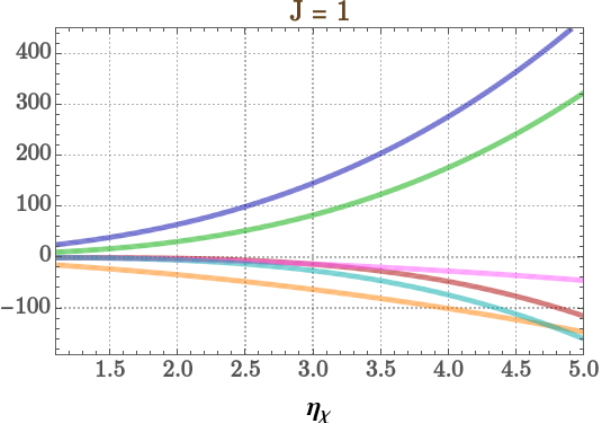} \hspace{0.1 cm}
\includegraphics[width=0.32 \textwidth]{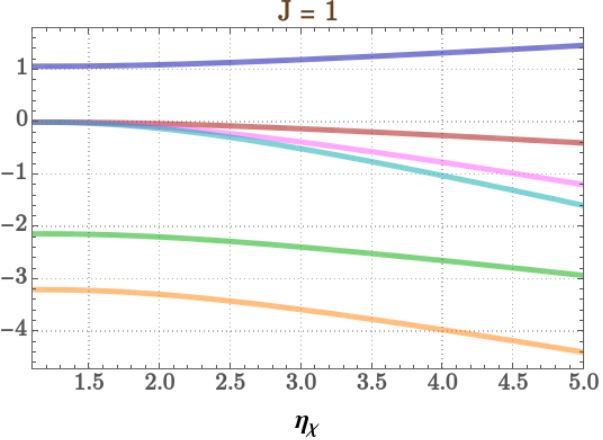} \hspace{0.1 cm}
\includegraphics[width=0.325 \textwidth]{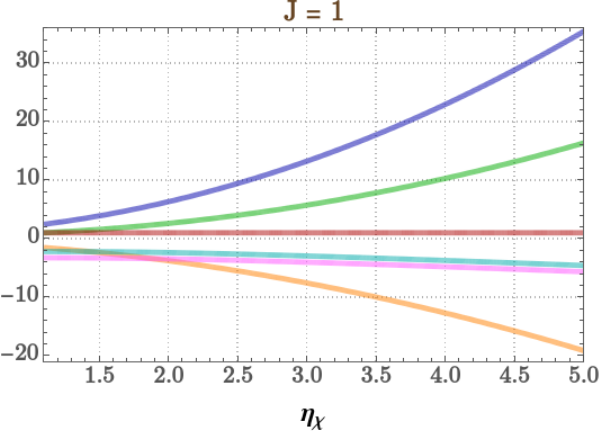}}
\subfigure{\includegraphics[width=0.32 \textwidth]{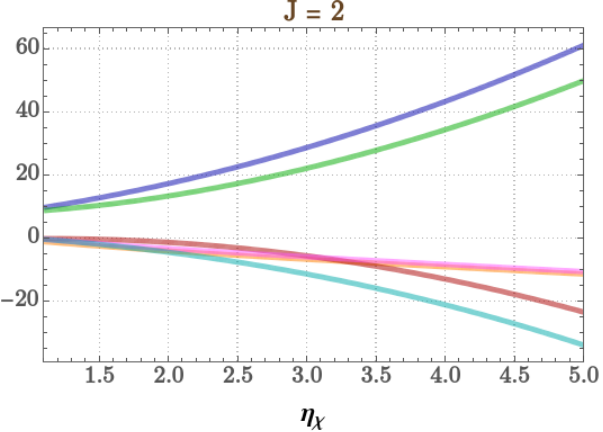}\hspace{0.1 cm}
\includegraphics[width=0.32 \textwidth]{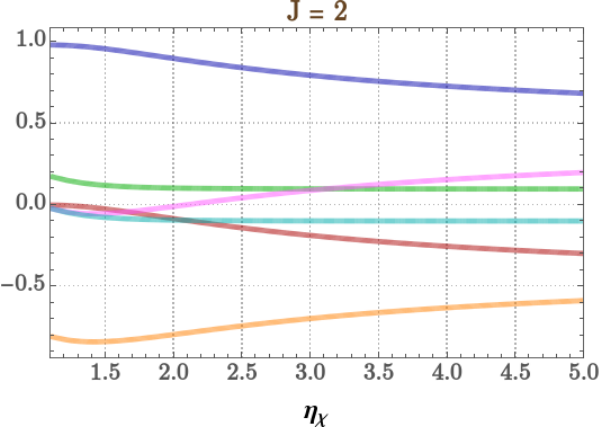}\hspace{0.1 cm}
\includegraphics[width=0.315 \textwidth]{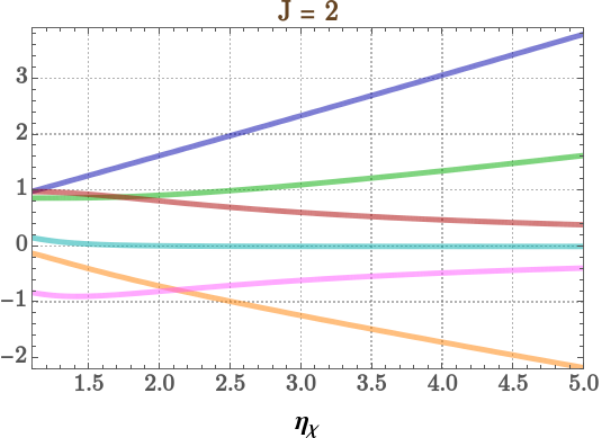}} 
\subfigure{\includegraphics[width=0.32 \textwidth]{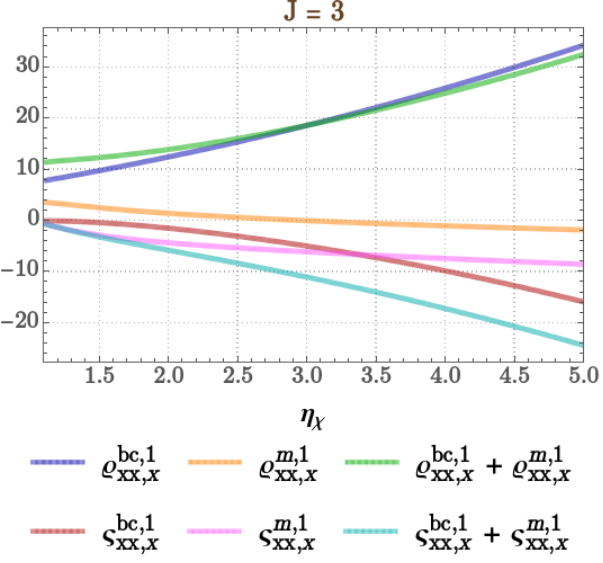}\hspace{0.1 cm}
\includegraphics[width=0.32 \textwidth]{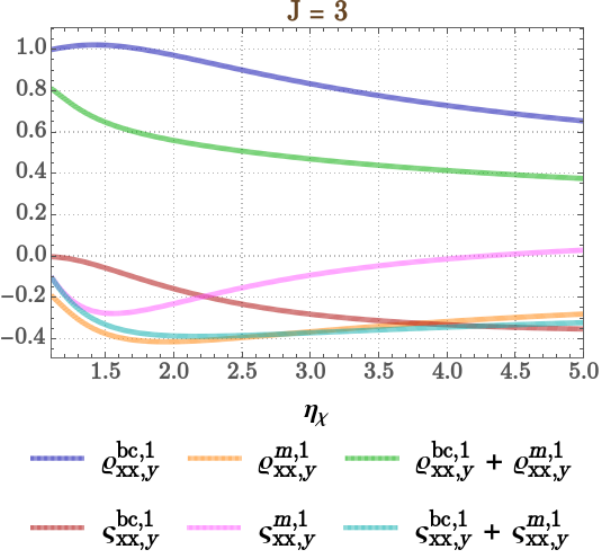} \hspace{0.1 cm}
\includegraphics[width=0.33 \textwidth]{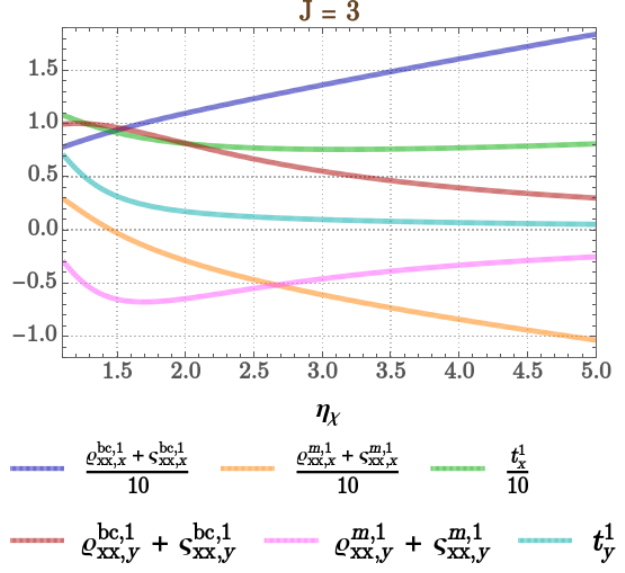}} 
\caption{
Longitudinal response for the type-II phase in set-up I: Comparison of the various parts for $J=1$, $J= 2$, and $J=3$ [cf. Eqs.~\eqref{eqtype2set1xx}, \eqref{eqxxj1}, \eqref{eqxxj2_1}, \eqref{eqxxj2_2}, \eqref{eqxxj3_1}, \eqref{eqxxj3_2}, \eqref{eqxxj3_3}, and \eqref{eqxxj3_4}]. The symbol $t^1_x$ ($t^1_y$) in the plot-legends denotes the total over all the terms accompanying $B_x^2$ ($B_y^2$).
\label{figset1long_type2}}
\end{figure*}

In the type-II phase, both the conduction and valence bands contribute for any given $ \mu $.
The net contributions are divided up into BC-only and OMM parts as
\begin{align}
\label{eqtype2set1xx}
& \sigma_{xx}^{ (\chi, bc) } =
 \frac {e^4 \, J^3 \, \tau  \, v_z  } 
 {128 \, \pi^2} 
\left ( \frac {\alpha_J} {\mu} \right)^{\frac{2} {J}} 
 \left[  B_x^2 \left( \varrho^{bc, 1}_{xx,x} + \varsigma^{bc, 1}_{xx,x}   \right)
 +  B_y^2 \left( \varrho^{bc, 1}_{xx,y}  + \varsigma^{bc, 1}_{xx,y}   \right) \right  ],
\nn &
\sigma_{xx}^{ (\chi, m) } = 
\frac {e^4 \, J^3 \, \tau  \, v_z  } 
 {128 \, \pi^2} 
\left (\frac {\alpha_J} {\mu} \right)^{\frac{2} {J}} 
 \left[  B_x^2 \left( \varrho^{m, 1}_{xx,x} 
 + \varsigma^{m, 1}_{xx,x}   \right)
 + 
 B_y^2 \left( \varrho^{m, 1}_{xx,y} 
 + \varsigma^{m, 1}_{xx,y}   \right) \right  ].
\end{align}
The symbols used above indicate the following: (1) $\varrho^{bc, 1}_{xx,x} $ and $ \varrho^{bc, 1}_{xx,y}$ ($ \varsigma^{bc, 1}_{xx,x} $ and $ \varsigma^{bc, 1}_{xx,y} $) represent the BC-only parts proportional to $B_x^2 $ and $B_y^2 $, respectively, coming from the $s=1$ ($s=2$) band, and (2) $\varrho^{m, 1}_{xx,x} $ and $ \varrho^{m, 1}_{xx,y}$ ($ \varsigma^{m, 1}_{xx,x} $ and $ \varsigma^{m, 1}_{xx,y}$) represent the OMM parts proportional to $B_x^2 $ and $B_y^2 $, respectively, coming from the $s=1$ ($s=2$) band. The final expressions are shown in Appendix~\ref{appset1_longtype2}, from where we find that there are no terms proportional to $\ln \Lambda $ or a positive power of $\Lambda $, implying that the integrals converge without the need of an ultraviolet-cutoff scale. Since the expressions are cumbersome, the overall behaviour is best deciphered with the help of plots. Fig.~\ref{figset1long_type2} serves the purpose, in conjunction with Fig.~\ref{fignetset1xx} (in Appendix~\ref{appset1_longtype2}). Defining $t^1_x = \varrho^{bc, 1}_{xx,x} + \varrho^{m, 1}_{xx,x} + \varsigma^{bc, 1}_{xx,x} + \varsigma^{m, 1}_{xx,x} $ and $t^1_y = \varrho^{bc, 1}_{xx,y} + \varrho^{m, 1}_{xx,y} + \varsigma^{bc, 1}_{xx,y} + \varsigma^{m, 1}_{xx,y} $, the latter displays exclusively the net response by magnifying the corresponding curves. We infer that, while $t^1_x$ turns out to be positive for all $J$-values, $t^1_y$ is negative for $J=1$, with $|t^1_x| \gg |t^1_y|$ for all the curves.

\subsection{Set-up I: In-plane transverse components}
\label{set1_inplane_trs}

The $J$-dependent expressions for the starting integrals are shown in Eq.~\eqref{eqappset1_inplane_trs} of Appendix~\ref{appset1_inplane_trs}. 

\subsubsection{\underline{Results for the type-I phase for $\mu>0$}}
\label{set1_inplane_trs_type1}

For $\mu>0$, only the conduction band contributes in the type-I phase.
The contributions are divided up into BC-only and OMM parts as
\begin{align}
\label{eqsetup1yxtype1}
 \sigma_{yx}^{ (\chi, bc) } = \frac{e^4  \, \tau  \, v_z}  {64 \, \pi^{\frac{3} {2}} \, J}
\left (\frac {\alpha_J} {\mu} \right)^{\frac{2} {J}} 
\, B_x \, B_y \, \ell^{bc}_{yx,1} \,,\quad
\sigma_{yx}^{ (\chi, m) } = 
 \frac{e^4  \, \tau  \, v_z}  {64 \, \pi^{\frac{3} {2}} \, J}
\left (\frac {\alpha_J} {\mu} \right)^{\frac{2} {J}} 
\, B_x \, B_y \, \ell^{m}_{yx,1} \, .
\end{align}
From the explicit final expressions shown in Appendix~\ref{appset1_inplane_trs_type1} for all the three values of $J$, we observe the following behaviour:
\begin{center}
\begin{tabular}{ |c||c|c|c|}
\hline
 & $ \ell^{bc}_{yx,1} $ &  $\ell^{m}_{yx,1}$ & $ \ell^{bc}_{yx,1} + \ell^{m}_{yx,1}$ \\ \hline
 $J=1$ & $ \frac{8 \, \left( 7 + 13 \, \eta_{\chi }^2\right)} {15 \, \sqrt{\pi }} $
& $ \frac{ -\, 8\, \left( 1 + 8 \, \eta_\chi^2 \right)} {15 \, \sqrt{\pi }}$ 
& $\frac{8 \, \left( 6 + 5  \, \eta_\chi^2 \right)} {15 \, \sqrt{\pi }} $ \\ \hline
 $J=2$ & $ \sqrt{\pi }\, \left( 13 + 8 \,\eta_\chi^2 \right)$
 & $4\, \sqrt \pi $  & $ \sqrt{\pi }\, \left( 17 + 8 \,\eta_\chi^2 \right)$\\ \hline
 $J=3 $ &  $>0$  &  $>0$ &  $>0$ \\ \hline
\end{tabular}
\end{center}
For $J=3$, since the expressions involve hypergeometric functions, the individual and overall behaviour is shown in Fig.~\ref{figell1}(b). From there, we find that $ \ell^{bc}_{yx,1} > 0$ and $ \ell^{m}_{yx,1} > 0$.

\subsubsection{\underline{Results for the type-II phase for $\mu>0$}}
\label{set1_inplane_trs_type2}

\begin{figure}[t!]
\centering 
\subfigure{\includegraphics[width= 0.32 \textwidth]{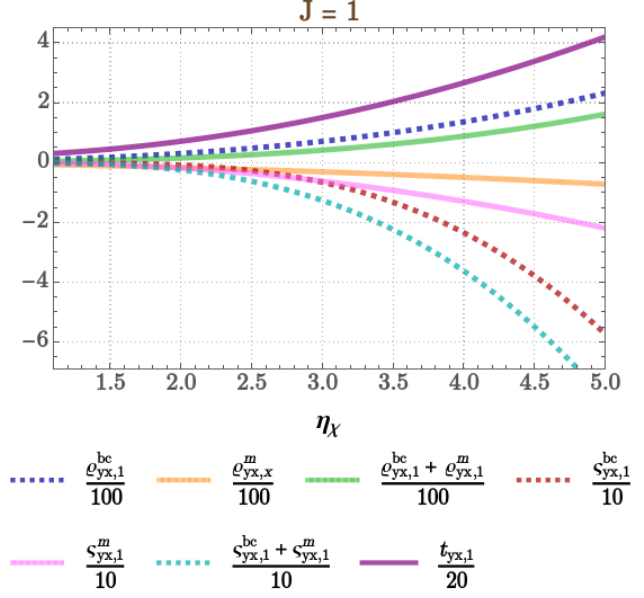}}\hspace{0.1 cm}
\subfigure{\includegraphics[width= 0.325 \textwidth]{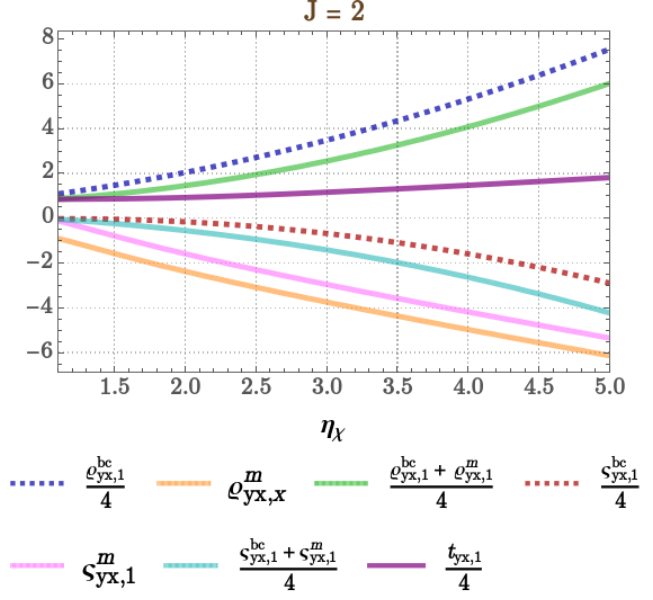}}\hspace{0.1 cm}
\subfigure{\includegraphics[width= 0.32 \textwidth]{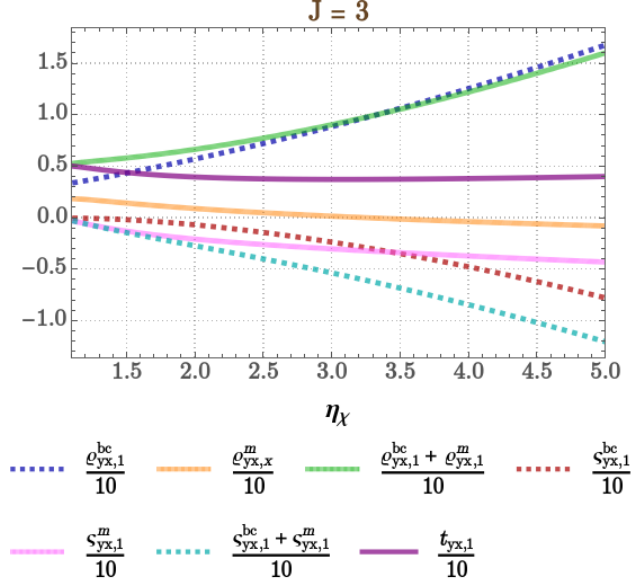}}
\caption{In-plane transverse response for the type-II phase in set-up I: Comparison of the various parts for $J=1$, $J= 2$, and $J=3$ [cf. Eqs.~\eqref{eqtype2set1yx}, \eqref{eqyxj1}, \eqref{eqyxj2}, \eqref{eqyxj3_1}, and \eqref{eqyxj3_2}]. The symbol $t_{yx, 1}$ in the plot-legends denotes the total, $\varrho^{bc}_{yx,1} + \varrho^{m}_{yx,1} + \varsigma^{bc}_{yx,1} + \varsigma^{m}_{yx,1}$, accompanying $B_x \, B_y$.
\label{fig5}}
\end{figure}

In the type-II phase, both the conduction and valence bands contribute for any given $ \mu $. The contributions are divided up into BC-only and OMM parts as
\begin{align}
\label{eqtype2set1yx}
 \sigma_{yx}^{ (\chi, bc) } =
 \frac {e^4 \, J\, \tau  \, v_z  } 
 { 64 \, \pi^2} 
\left (\frac {\alpha_J} {\mu} \right)^{\frac{2} {J}} 
\, B_x \, B_y 
\left ( \varrho^{bc}_{yx,1} +  \varsigma^{bc}_{yx,1} \right ),\quad
\sigma_{yx}^{ (\chi, m) } = 
 \frac {e^4 \, J\, \tau  \, v_z  } 
 { 64 \, \pi^2} 
\left (\frac {\alpha_J} {\mu} \right)^{\frac{2} {J}} 
\, B_x \, B_y 
\left ( \varrho^{m}_{yx,1} +  \varsigma^{m}_{yx,1} \right ) .
\end{align}
The symbols used above indicate the following: (1) $\varrho^{bc}_{yx,1} $ ($ \varsigma^{bc}_{yx,1} $) represents the BC-only part proportional to $B_x\, B_y $, arising from the $s=1$ ($s=2$) band, and (2) $\varrho^{m}_{yx,1} $ ($ \varsigma^{m}_{yx,1} $) represents the OMM part proportional to $B_x\, B_y $, arising from the $s=1$ ($s=2$) band.
The final expressions are shown in Appendix~\ref{appset1_inplane_trs_type2}, from where we find that there are no terms proportional to $\ln \Lambda $ or a positive power of $\Lambda $, implying that the integrals converge without the need of an ultraviolet-cutoff scale. Since the expressions are cumbersome, the overall behaviour is best understood by explicit plots, which are provided in Fig.~\ref{fig5}. Defining $t_{yx,1} =  \varrho^{bc}_{yx,1} + \varrho^{m}_{yx,1} + \varsigma^{bc}_{yx,1}+ \varsigma^{m}_{yx,1} $, the total response is also presented there. For $s=1$ ($s=2$), the net response is positive (negative) for $J$-values. In the end, summing over the two bands, $t_{yx,1}$ turns out to be positive for all $J$-values.

\subsection{Set-up I: Out-of-plane transverse components}
\label{secset1_outplane}

\begin{figure}[t!]
\centering 
\includegraphics[width= 0.65 \textwidth]{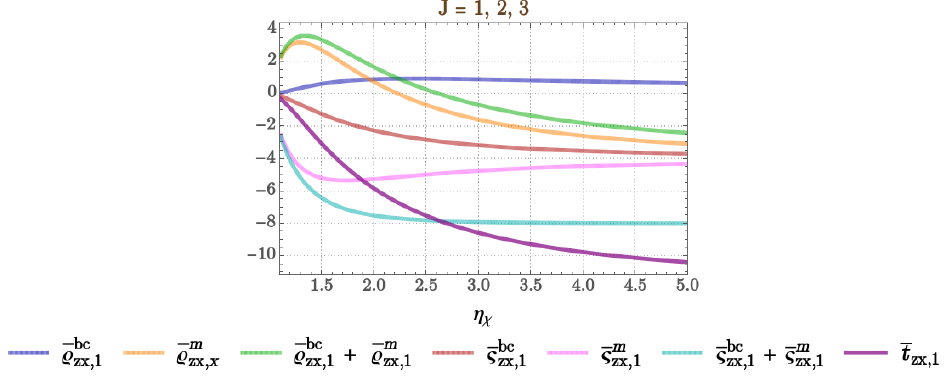}
\caption{Out-of-plane transverse response for the type-II phase in set-up I: Comparison of the various parts for $J=1$, $J= 2$, and $J=3$ [cf. Eqs.~\eqref{eqtype2set1zx} and \eqref{eqappzxset1}]. Here, the barred-over variables denote the coefficients of the $\ln (\Lambda/\mu) $ terms of the corresponding unbarred ones, and ${\bar t}_{yx, 1} \equiv  {\bar \varrho}^{bc}_{yx,1} 
+ \bar{\varrho}^{m}_{yx,1} + \bar{\varsigma}^{bc}_{yx,1} + \bar{\varsigma}^{m}_{yx,1}$ in the plot-legends denotes the total, which accompanies $ \chi B_x $.
\label{figzxset1}}
\end{figure}

The expressions for the integrals are shown in Eq.~\eqref{eqappset1_outplane} of Appendix~\ref{appset1_outplane}. We note that the terms turn out to be exclusively linear-in-$B$, with the $\mathcal{O}(B^2)$ terms vanishing altogether. The resulting magnetoelectric current, varying linearly with $B$, is caused by a nonzero current proportional to $\left( \mathbf E \cdot \mathbf B \right) \eta_\chi \, {\mathbf{\hat z}} $ (in agreement with Ref.~\cite{amit-magneto}).

\subsubsection{\underline{Results for the type-I phase for $\mu>0$}}
\label{set1_offplane_type1}

The contributions are divided up into BC-only and OMM parts as [cf. Appendix~\ref{appset1_offplane_type1}]
\begin{align}
 \sigma_{zx}^{ (\chi, bc) } =
 \frac{3 \,e^3 \, \tau\, J\,v_z  \,\eta_\chi^4} {512 \,\pi^2} 
 \, \chi \, B_x \,\ell^{bc}_{zx,1} \,,\quad
\sigma_{zx}^{ (\chi, m) } =  \frac{3 \,e^3 \, \tau\, J\,v_z  \,\eta_\chi^4} {512 \,\pi^2} 
\, \chi \, B_x \, \ell^{m}_{zx,1} \, ,
\end{align}
where
\begin{align}
\label{eqsetup1zxtype1}
 \ell^{bc}_{zx,1} 
=  - \, 6 \, \eta_\chi^5 + 5 \,\eta_\chi^3-3 \,\eta_\chi
+ 3 \left( 1- \eta_\chi^2 \right)^2 \tanh^{-1}\eta_{\chi} \,, \quad
 \ell^{m}_{zx,1} = 
 - \,13 \,\eta_\chi^3+15 \eta_\chi
 -3 \left(\eta_\chi^4-6 \,\eta_\chi^2+5\right) \tanh^{-1} \eta_{\chi}  \,.
\end{align}
Here, $ \ell^{bc}_{zx,1}$ and $\ell^{m}_{zx,1}$ are $J$-independent, with the $J$-dependence only appearing as an overall factor in $\sigma^{\chi, s}_{zx}$. To make sense of the tilt-dependence, we illustrate the behaviour via Fig.~\ref{figell1}(c). We find that  while $\ell^{bc}_{zx,1} > 0$, $\ell^{m}_{zx,1}$ goes from negative to positive. But the net response remains negative in the entire range of $ 0\leq \eta_\chi \leq 1 $.

\subsubsection{\underline{Results for the type-II phase for $\mu>0$}}
\label{set1_offplane_type2}

In the type-II phase, both the conduction and valence bands contribute for any given $ \mu $. The contributions are further divided up into BC-only and OMM parts as
\begin{align}
\label{eqtype2set1zx}
 \sigma_{zx}^{ (\chi, bc) } =
 \frac {e^3\, J\, \tau\,  v_z}
{64\, \pi^2}
\, \chi \, B_x 
\left ( \varrho^{bc}_{zx,1} +  \varsigma^{bc}_{zx,1} \right ),\quad
\sigma_{zx}^{ (\chi, m) } = 
\frac {e^3\, J\, \tau\,  v_z}
{64\, \pi^2}
\, \chi \, B_x 
\left ( \varrho^{m}_{zx,1} +  \varsigma^{m}_{zx,1} \right ) .
\end{align}
The symbols used above indicate the following: (1) $ \varrho^{bc}_{zx,1} $ ($ \varsigma^{bc}_{zx,1} $) represents the BC-only part proportional to $\chi \, B_x $, arising from the $s=1$ ($s=2$) band, and (2) $\varrho^{m, 1}_{xx,x} $ ($ \varsigma^{m, 1}_{xx,x} $) represents the OMM part proportional to $ \chi \,B_x $, arising from the $s=1$ ($s=2$) band. All these four coefficints are $J$-independent, with the $J$-dependence only appearing as an overall factor in $\sigma^{\chi, s}_{zx}$.

The final expressions are shown in Appendix~\ref{appset1_offplane_type2}, all of which are independent of the values of $J$. These are all divergent in $\Lambda$ logarithmically. Observing that the singular terms will dominate, we demonstrate in Fig.~\ref{figzxset1} the coefficients of the $\ln (\Lambda/\mu) $ terms for all the parts. We find that the net response is negative.

\section{Set-up II: $ \mathbf E = E \, {\mathbf{\hat x}} $, $\mathbf B = B_x \, {\mathbf{\hat x}} + B_z \, {\mathbf{\hat z}} $}
\label{secset2}

In set-up II, as shown in Fig.~\ref{figsetup}(b), the tilt-axis is perpendicular to $\mathbf E $, but not to $\mathbf B $. We choose $ \hat{\mathbf r}_E =  {\mathbf{\hat x}}  $ and $\hat{\mathbf r}_B = \cos \theta \, {\mathbf{\hat x}} 
+ \sin \theta \, {\mathbf{\hat z}}  $, such that $\mathbf E = E \, {\mathbf{\hat x}}$ and $\mathbf B = B_x \, {\mathbf{\hat x}} 
+ B_z \, {\mathbf{\hat z}} \equiv B  \,\hat{\mathbf r}_B  $.
The details of the generic forms of the integrals are shown in Appendix~\ref{appset2}.
Therein, Appendices~\ref{appset2long}, \ref{appset2_inplane_trs}, and \ref{appset2_outplane} deal with the longitudinal, in-plane transverse, and out-of-plane transverse components, respectively.

\subsection{Set-up II: Longitudinal components}
\label{set2long}

The expressions for the integrals are shown in Eq.~\eqref{eqappset2_long} of Appendix~\ref{appset2long}.
We find that the conductivity contains terms which are linear-in-$B$ as well those which are quadratic-in-$ B $. The former are caused by nonzero currents $ \propto \left( \mathbf B \cdot  \eta_\chi \, {\mathbf{\hat z}} \right ) \mathbf E $ (in agreement with Ref.~\cite{amit-magneto}).

\subsubsection{\underline{Results for the type-I phase for $\mu>0$}}
\label{set2long_type1}

\begin{figure*}[t!]
\centering 
\subfigure{\includegraphics[width=0.32 \textwidth]{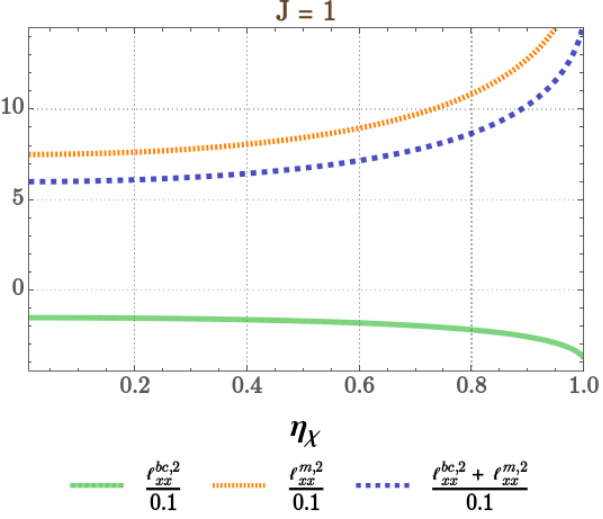} \hspace{0.1 cm}
\includegraphics[width=0.325 \textwidth]{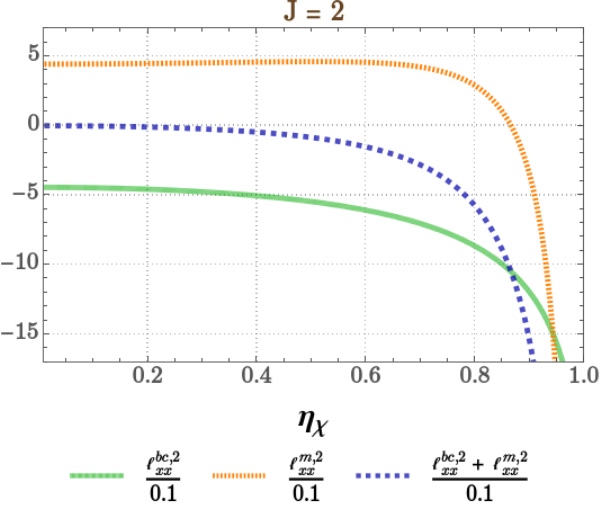} \hspace{0.1 cm}
\includegraphics[width=0.325 \textwidth]{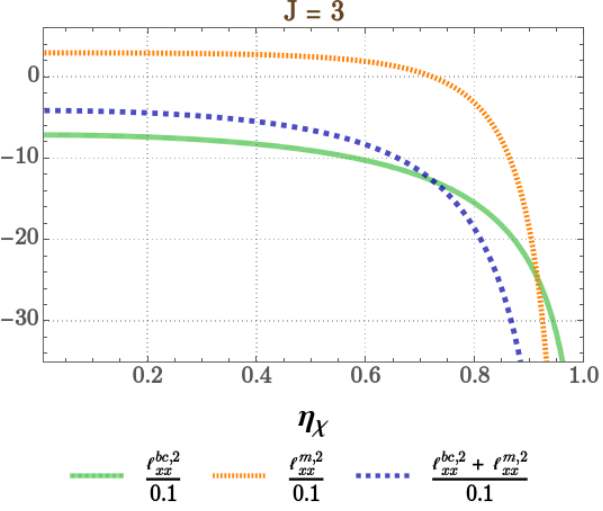}}
\subfigure{\includegraphics[width=0.33 \textwidth]{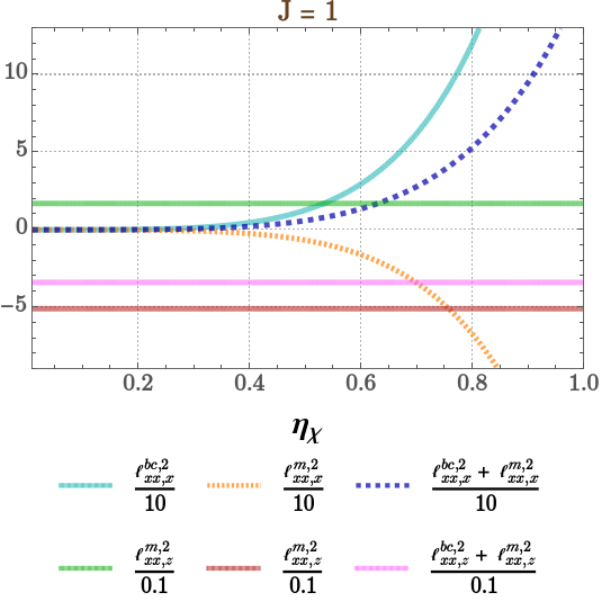} \hspace{0.1 cm}
\includegraphics[width=0.32 \textwidth]{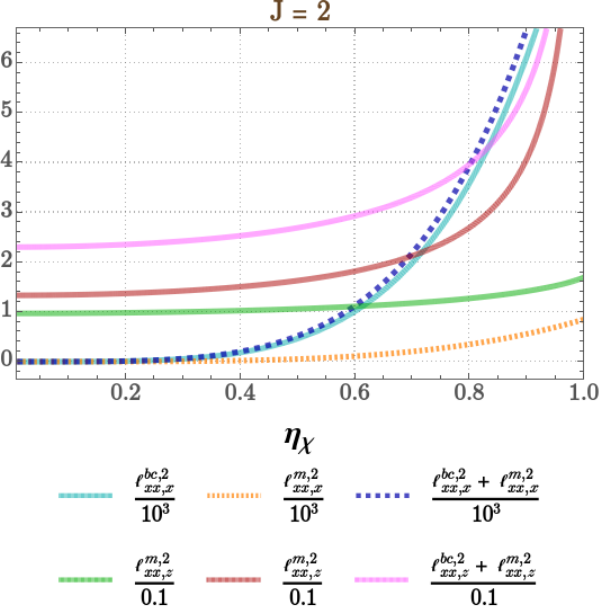} \hspace{0.1 cm}
\includegraphics[width=0.32 \textwidth]{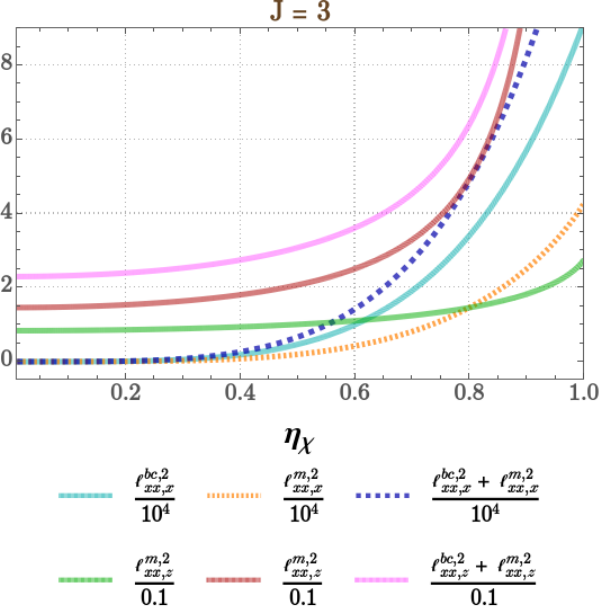}}
\caption{
Longitudinal response for the type-I phase in set-up II: Comparison of the various parts for $J=1$, $J= 2$, and $J=3$ [cf. Eq.~\eqref{eqsigxxset2type1}]. The top panel corresponds to Eq.~\eqref{ellxxset2linear}. The bottom panel illustrates the behaviour of the functions contained in Eqs.~\eqref{ellxxset2Bx} and \eqref{ellxxset2Bz}.
\label{figxxset2type1}}
\end{figure*}

For $\mu>0$, only the conduction band contributes in the type-I phase.
All the contributions are divided up as shown below:
\begin{align}
\label{eqsigxxset2type1}
& \sigma_{xx}^{ (\chi, bc) } =
\frac{e^3 \, \tau \, J^2 \,\mu^2  \,\eta_\chi \, \chi  \, B_z  }
{16 \, \pi^{\frac{3} {2} } \, v_z} \left(\frac{\alpha_J}{\mu }\right)^{\frac{2} {J} }
 \,\ell^{bc, 2}_{xx} 
+ \frac{e^4\, \tau \, v_z} {3840 \, \pi^{\frac{3} {2}} \, J^3  \,\eta_\chi^4} 
\left(\frac{\alpha_J}{\mu }\right)^{\frac{2} {J} }  B_x^2 \,\ell^{bc,2}_{xx,x}  
+ 
\frac{e^4  \, \tau \, J^5  \, \mu^2  } {64  \, \pi  \,  v_z} 
\left(\frac{\alpha_J} {\mu }\right)^{\frac{4} {J} }
B_z^2 \, \ell^{bc,2}_{xx,z} \,,
\nn & \sigma_{xx}^{ (\chi, m) } = 
\frac{e^3 \, \tau \, J^2 \,\mu^2  \,\eta_\chi \, \chi  \, B_z  }
{16 \, \pi^{\frac{3} {2} } \, v_z} \left(\frac{\alpha_J}{\mu }\right)^{\frac{2} {J} }
\,\ell^{m, 2}_{xx} 
 + \frac{e^4\, \tau \, v_z} {3840 \, \pi^{\frac{3} {2} }\, J^3  \,\eta_\chi^4} 
\left(\frac{\alpha_J}{\mu }\right)^{\frac{2} {J} }   B_x^2 
 \,\ell^{m,2}_{xx,x}  
+ 
\frac{e^4  \, \tau \, J^5  \, \mu^2  } {64  \, \pi  \,  v_z}
\left(\frac{\alpha_J}{\mu }\right)^{\frac{4} {J} } B_z^2 \, \ell^{m,2}_{xx,z} \,.
\end{align}
Here, $\ell^{bc, 2}_{xx}$, $\ell^{bc,2}_{xx,x}$, and $\ell^{bc,2}_{xx,z}$
designate the BC-only parts accompanying $\chi \, B_z $, $B_x^2 $, and $ B_z^2 $, respectively. Similarly, $\ell^{m}_{xx, 23} $, $\ell^{m,2}_{xx,x}$, and $\ell^{m,2}_{xx,z}$ demarcate the OMM parts proportional to $\chi \, B_z $, $B_x^2 $, and $ B_z^2 $, respectively.

More details about the generic-$J$ expressions can be found in Appendix~\ref{appset2long_type1}, which are quite involved. We tabulate below the final forms explicitly for the cases which are simple enough to decipher:
\begin{center}
\begin{tabular}{ |c||c|c|c|}
\hline
 & $\ell^{bc, 2}_{xx} $ & $\ell^{m, 2}_{xx}$ & $\ell^{bc, 2}_{xx} + \ell^{m, 2}_{xx}$ \\ \hline
 $J=1$ &   $\frac{ - \,4 \; \, _2F_1\left(1,\frac{3}{2};\frac{7}{2};\eta_\chi^2\right)}
 {15 \, \sqrt{\pi }}$ &
$ \frac{10 \,  \left[ 3 \,\left(\eta_\chi^2-1\right)  
-2 \,\eta_\chi^3 + 3 \,\eta_\chi 
\tanh^{-1} \eta_\chi \right] } {3  \, \sqrt{\pi }  \, \eta_\chi^5}$ &
$ \frac{8  \, \left[ 3 \,\left(\eta_\chi^2-1\right)
-2  \, \eta_\chi^3+3  \, \eta_\chi 
\tanh^{-1} \eta_\chi \right ]} 
{3  \, \sqrt{\pi }  \, \eta_\chi^5}$
 \\ \hline
 $J=2$ & $   \frac{ - \, 2  \, \sqrt{\pi } \,  
 \left[ 4 \, + \, \left(\sqrt{1-\eta_\chi^2}-3\right) 
 \eta_\chi^2-4  \, \sqrt{1-\eta_\chi^2} \, \right ] } 
 {\eta_\chi^6}$ &
 $  \frac{ - \,2  \, \sqrt{\pi }  \, \left(\eta_\chi^2+2  \, \sqrt{1-\eta_\chi^2}-2\right) 
 \,  \left(2  \, \eta_\chi^2+3  \, \sqrt{1-\eta_\chi^2}-3\right)}
 {\eta_\chi^6 \sqrt{1-\eta_\chi^2}}$ &
$ \frac{- \, 2  \, \sqrt{\pi }  \, \left[ 8 \, + \, 
\eta_\chi^4 
+ 4 \, \left(\sqrt{1-\eta_\chi^2}-2\right) 
\eta_\chi^2-8  \, \sqrt{1-\eta_\chi^2} \,\right] } 
{\eta_\chi^6  \, \sqrt{1-\eta_\chi^2}}  $ 
  \\ \hline
 $J=3 $ & $<0$ & 
 \makecell{\small{transitions from positive}\\
\small{to negative at $\eta_\chi \simeq 0.724 $}} 
& $<0 $\\ \hline
\end{tabular}
\end{center}
\begin{center}
\begin{tabular}{ |c||c|c|c||c|c|c|}
\hline
 & $ \ell^{bc, 2}_{xx,x} $ &
 $\ell^{m, 2}_{xx,x}$ & $ \ell^{bc, 2}_{xx,x} + \ell^{m, 2}_{xx,x}$
 & $ \ell^{bc, 2}_{xx,z} $ &  $\ell^{m, 2}_{xx,z}$ & 
 $ \ell^{bc, 2}_{xx,z} + \ell^{m, 2}_{xx,z}$ \\ \hline
 $J=1$ & $ \frac{ 32 \,\eta_\chi^4  \, \left( 8 \, + \, 13  \, \eta_\chi^2 \right)}
 {\sqrt{\pi }}$ &
 $ \frac{ - \,128  \,\eta_\chi^4 \, \left( 1 \, + \,2  \, \eta_\chi^2  \right)}
 {\sqrt{\pi }}$ &
$ \frac{32  \, \eta_\chi^4 \left( 4\, + \, 5 \,  \eta_\chi^2 \right)}{\sqrt{\pi }} $ &
$ \frac{8} {15 \, \pi } $ & $\frac{-\,8}{5 \, \pi }$ & $ \frac{-\,16}{15 \, \pi } $
 \\ \hline
 $J=2$ & 
 \makecell{
 \footnotesize{$ 120 \,  \sqrt{\pi } \,  \eta_\chi^4$} \\  
\footnotesize{ $ \times \left( 31 + 16  \, \eta_\chi^2 \right)$}} &
\footnotesize{$ 480  \, \sqrt{\pi }  \, \eta_\chi^4 $} &
\makecell{
\footnotesize{$ 120 \,  \sqrt{\pi }  \, \eta_\chi^4$}\\   
\footnotesize{$ \times \left( 35 + 16  \, \eta_\chi^2 \right)$}}&
  $ \frac{32 \;_2F_1\left(1,\frac{3}{2};\frac{9}{2};\eta_\chi^2\right)}
  {105 \, \pi } $ &
 \makecell{
$ \frac{4  \, \left(\eta_\chi^2-2\right)^2 \tanh^{-1} \eta_\chi}
 {\pi  \,  \eta_\chi^7}$ 
\\ $ - \,\frac{   4 \, \left( 60 \, + \,7 \, \eta_\chi^4 - 40  \, \eta_\chi^2 \right)}
{15  \, \pi  \,  \eta_\chi^6}$
 }  &
\makecell{ $ \frac{4  \, \left( 5 + 2  \, \eta_\chi^4  - 6  \, \eta_\chi^2 \right) 
\tanh^{-1}\eta_\chi} {\pi \,   \eta_\chi^7}$ \\
$  -\,\frac{  4\, \left( 15 \, + \, 3  \, \eta_\chi^4-13  \, \eta_\chi^2 \right)}
{3  \, \pi   \, \eta_\chi^6}$
} \\ \hline
 $J=3 $ & $>0$ & $>0$ & $>0$ & $>0$  & $>0$& $>0$ \\ \hline
\end{tabular} 
\end{center}
Additionally, we comprehend the overall behaviour for all the cases with the help of Fig.~\ref{figxxset2type1}.
For $J=1$, the overall (i.e., BC+OMM) coefficients of $\chi B_z $ and $B_x^2$ are positive, while that for  $B_z^2$ is negative.
For $J=2$ and $J=3$, the net coefficients of $\chi B_z$ are negative, while those for $B_x^2$ and  $B_z^2$ are positive.

\subsubsection{\underline{Results for the type-II phase for $\mu>0$}}
\label{set2long_type2}

\begin{figure*}[h!]
\centering 
\includegraphics[width=0.32 \textwidth]{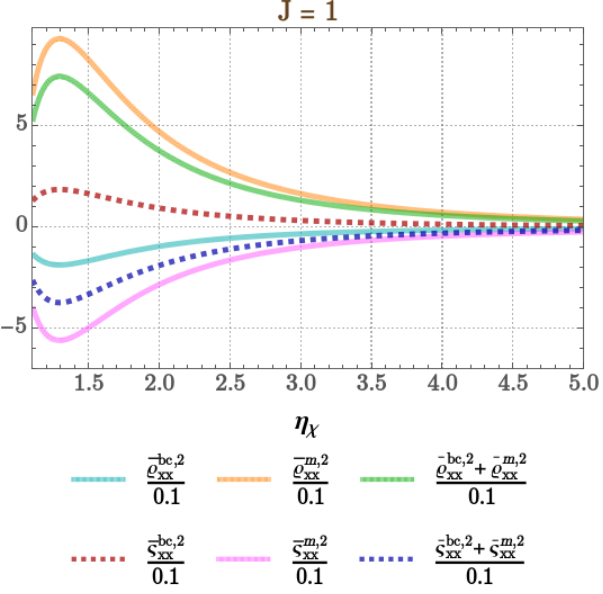} \hspace{0.1 cm}
\includegraphics[width=0.32 \textwidth]{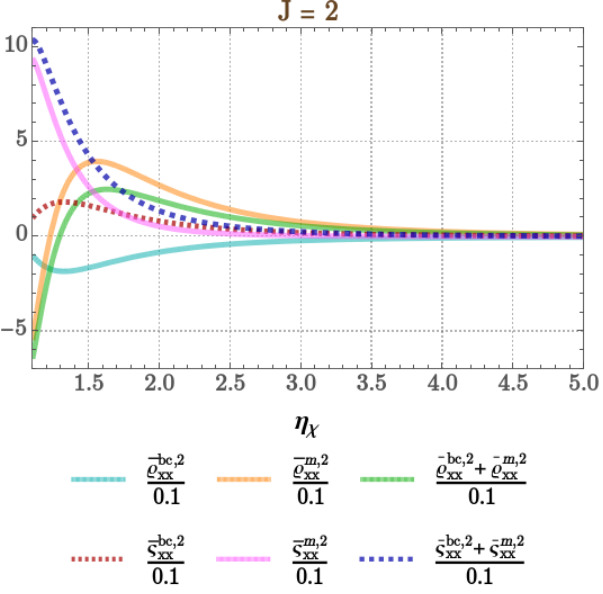} \hspace{0.1 cm}
\includegraphics[width=0.32 \textwidth]{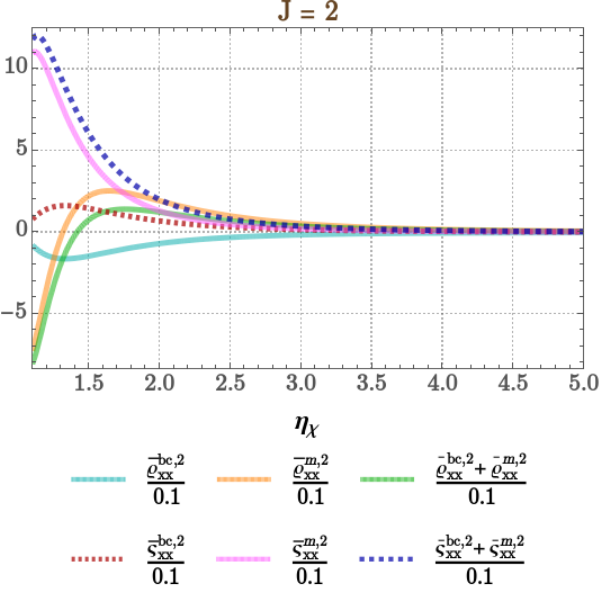}
\caption{
Longitudinal response for the type-II phase in set-up II: Comparison of the $\propto \chi B_z$-parts for $J=1$, $J= 2$, and $J=3$ [cf. Eqs.~\eqref{eqset2long_type2}, \eqref{eqlong_type2_lin_j1}, \eqref{eqlong_type2_lin_j2}, and \eqref{eqlong_type2_lin_j3}.
\label{figset2long_type2_lin}}
\end{figure*}

In the type-II phase, both the conduction and valence bands contribute for any given $ \mu $. The contributions are divided up into BC-only and OMM parts as
\begin{align}
\label{eqset2long_type2}
 \sigma_{xx}^{ (\chi, bc) } & =
\frac{e^3 \, \tau \, J^2 \,\mu^2  \,\eta_\chi \, \chi  \, B_z  }
{16 \, \pi^{2} \, v_z} \left(\frac{\alpha_J}{\mu }\right)^{\frac{2} {J} }
 \left( \varrho^{bc, 2}_{xx} + \varsigma^{bc, 2}_{xx} \right)
+ \frac{e^4\, \tau \, J^5\,v_z} {3840 \, \pi^2} 
\left(\frac{\alpha_J}{\mu }\right)^{\frac{2} {J} }  B_x^2 
\left( \varrho^{bc,2}_{xx,x}  +  \varsigma^{bc,2}_{xx,x} \right )
 \nn &  \,\quad
+ \frac{e^4 \,\tau\, J^9 \, \mu^2 } {1920 \,\pi^2 \,v_z}
\left(\frac{\alpha_J} {\mu }\right)^{\frac{4} {J} } B_z^2 
 \left( \varrho^{bc,2}_{xx,z}  +  \varsigma^{bc,2}_{xx,z} \right ) ,
\nn  \sigma_{xx}^{ (\chi, m) } & = 
\frac{e^3 \, \tau \, J^2 \,\mu^2  \,\eta_\chi \, \chi  \, B_z  }
{16 \, \pi^{2} \, v_z} \left(\frac{\alpha_J}{\mu }\right)^{\frac{2} {J} }
\left( \varrho^{m, 2}_{xx} + \varsigma^{m, 2}_{xx} \right)
 + \frac{e^4\, \tau \, J^5\,v_z} {3840 \, \pi^2}
\left(\frac{\alpha_J}{\mu }\right)^{\frac{2} {J} }   B_x^2 
\left( \varrho^{m,2}_{xx,x}  +  \varsigma^{m,2}_{xx,x} \right )
 \nn &  \,\quad 
+ \frac{e^4 \,\tau\, J^9 \, \mu^2 } {1920 \,\pi^2 \,v_z}
\left(\frac{\alpha_J} {\mu }\right)^{\frac{4} {J} } B_z^2 
 \left( \varrho^{m,2}_{xx,z}  +  \varsigma^{m,2}_{xx,z} \right ).
\end{align}
The symbols used above indicate the following: (1) $ \varrho^{bc,2}_{xx} $ ($ \varsigma^{bc,2}_{xx} $) represents the BC-only part proportional to $ \chi \, B_z $, arising from the $s=1$ ($s=2$) band; (2) $ \varrho^{bc,2}_{xx,x} $ ($ \varsigma^{bc,2}_{xx,x} $) represents the BC-only part proportional to $ B_x^2 $, arising from the $s=1$ ($s=2$) band; (3) $ \varrho^{bc,2}_{xx,z} $ ($ \varsigma^{bc,2}_{xx,z} $) represents the BC-only part proportional to $ B_z^2 $, arising from the $s=1$ ($s=2$) band; (4) $\varrho^{m,2}_{xx} $ ($ \varsigma^{m,2}_{xx} $) represents the OMM part proportional to $ \chi \,B_z $, arising from the $s=1$ ($s=2$) band; (5) $\varrho^{m,2}_{xx,x} $ ($ \varsigma^{m,2}_{xx,x} $) represents the OMM part proportional to $ B_x^2 $, arising from the $s=1$ ($s=2$) band; and
(6) $\varrho^{m,2}_{xx,z} $ ($ \varsigma^{m,2}_{xx,z} $) represents the OMM part proportional to $ B_z^2 $, arising from the $s=1$ ($s=2$) band.

\begin{figure*}[t!]
\centering 
\subfigure{\includegraphics[width=0.32 \textwidth]{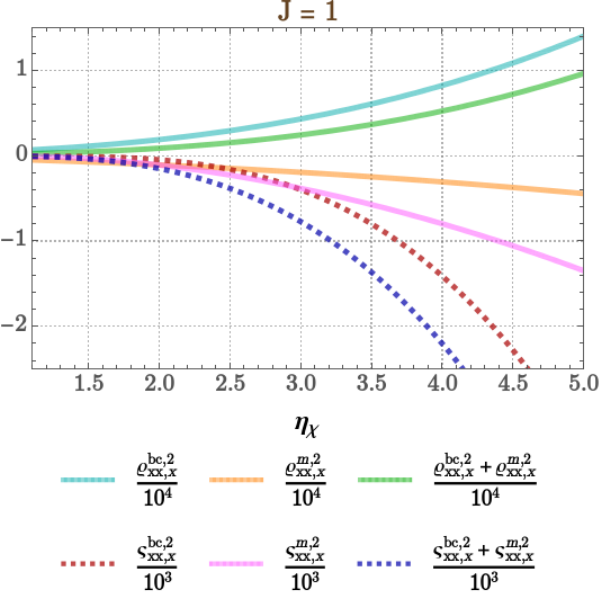} \hspace{0.1 cm}
\includegraphics[width=0.32 \textwidth]{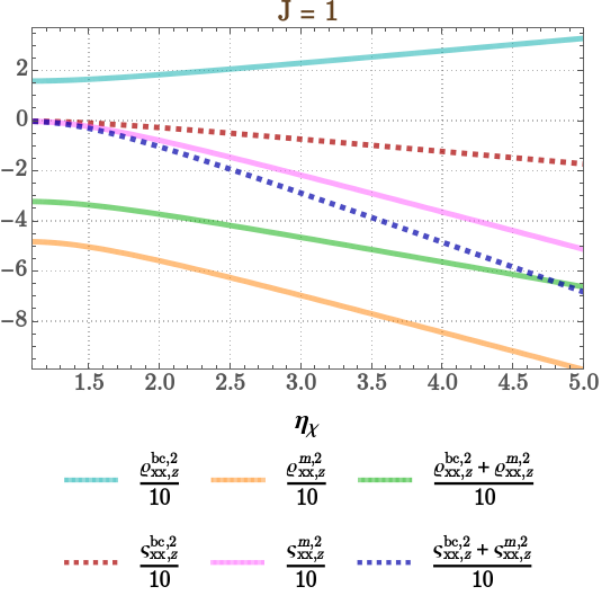} \hspace{0.1 cm}
\includegraphics[width=0.32 \textwidth]{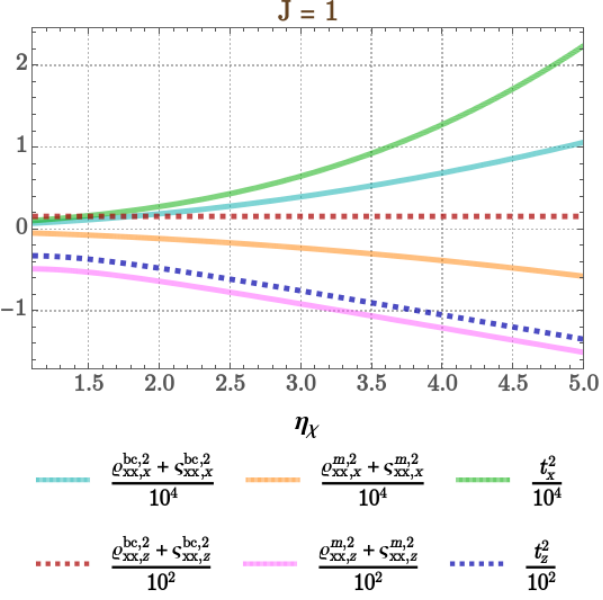}}
\subfigure{\includegraphics[width=0.32 \textwidth]{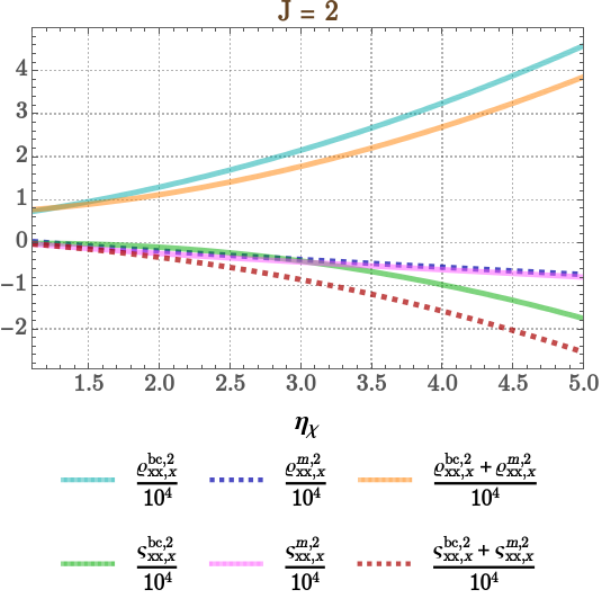}\hspace{0.1 cm}
\includegraphics[width=0.325 \textwidth]{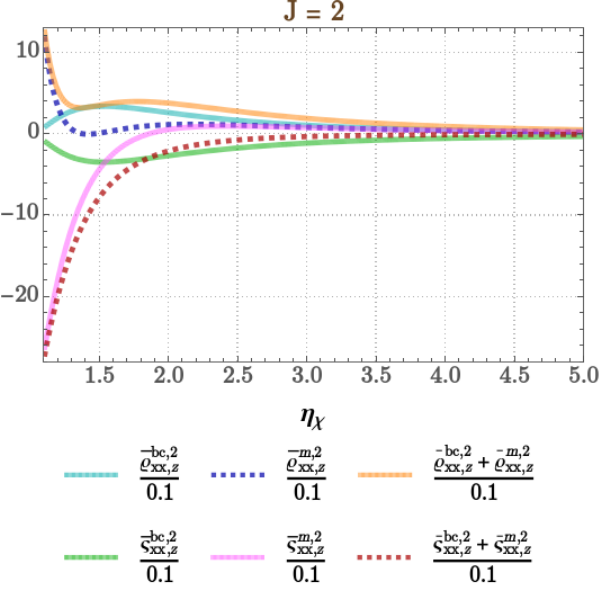}\hspace{0.1 cm}
\includegraphics[width=0.325 \textwidth]{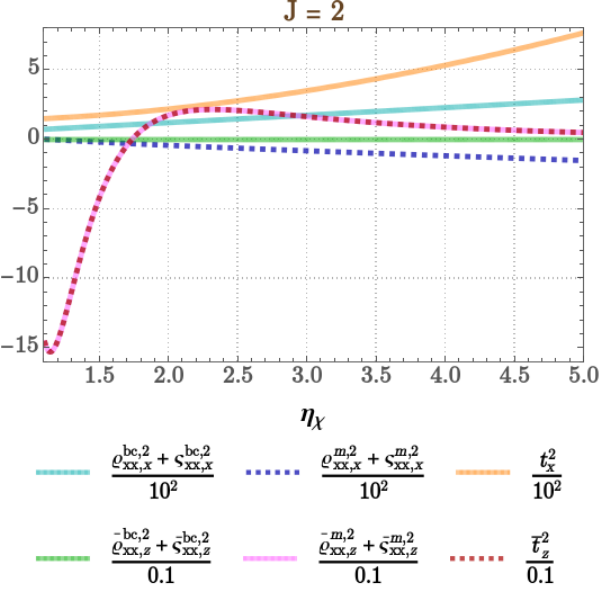}} 
\subfigure{\includegraphics[width=0.32 \textwidth]{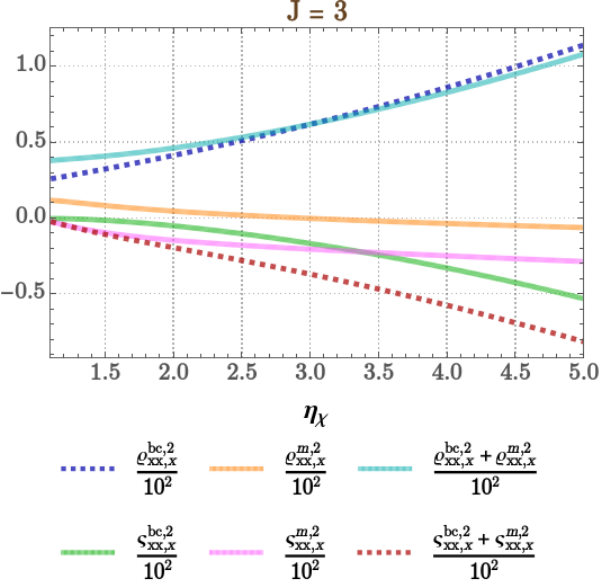}\hspace{0.1 cm}
\includegraphics[width=0.32 \textwidth]{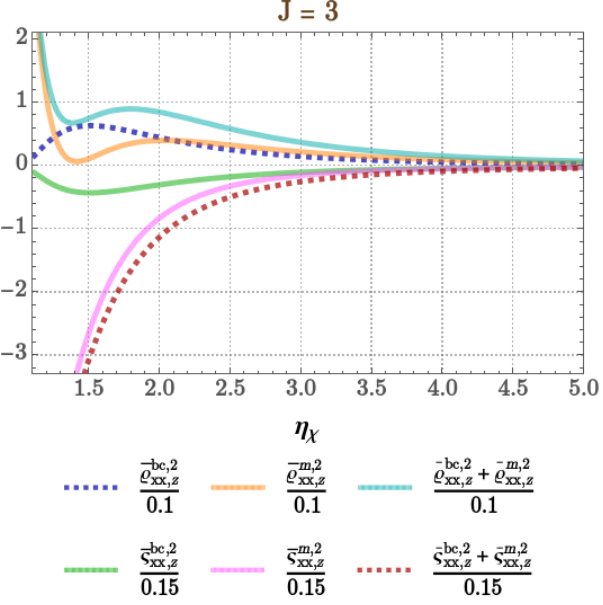} \hspace{0.1 cm}
\includegraphics[width=0.33 \textwidth]{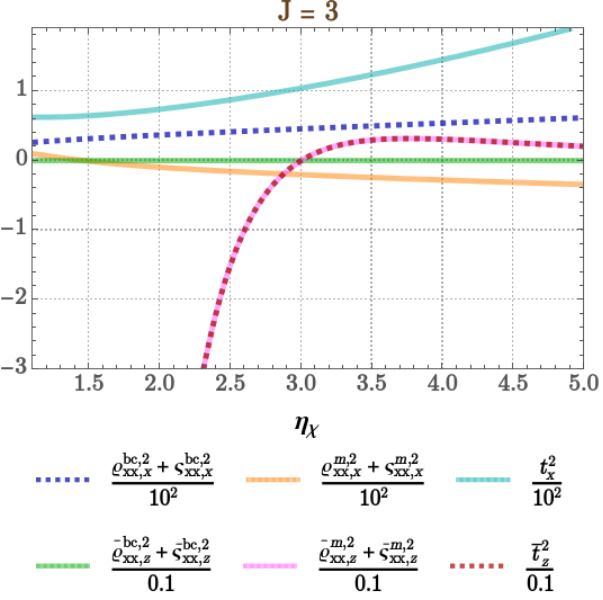}} 
\caption{
Longitudinal response for the type-II phase in set-up II: Comparison of the quadratic-in-$B$ parts for $J=1$, $J= 2$, and $J=3$ [cf. Eqs.~\eqref{eqset2long_type2}, \eqref{eqxx_type2_Bx_j1}, \eqref{eqxx_type2_Bx_j2}, \eqref{eqxx_type2_Bx_j3}, \eqref{eqxx_type2_Bz_j1}, and \eqref{eqxx_type2_Bz_j2}, and \eqref{eqxx_type2_Bz_j3}]. The symbol $t^2_x$ ($t^2_z$ or $\bar t^2_z $) in the plot-legends denotes the total over all the terms accompanying $B_x^2$ ($B_z^2$).
\label{figset2long_type2}}
\end{figure*}

The final expressions are shown in Appendix~\ref{appset2long_type2}, from where we find that different terms have different convergent/divergent behaviour. For the terms possessing singularities, the barred-over variables denote the coefficients of the most singular terms of the corresponding unbarred ones. While the $\propto B_x^2$-terms are completely convergent, the $\propto \chi   B_z$ ones are divergent in $\Lambda/\mu $, with the degree of divergence of the leading-order singularity increasing with $J$ [viz., $\sim \ln ( \Lambda / \mu)$, $\sim \Lambda / \mu$, and $\sim (\Lambda / \mu)^{4/3 }$ for $J$ equalling $1$, $ 2$, and $3$, respectively].
Interestingly, for all $J $'s, we find that $\bar  \varrho^{bc, 2}_{xx} +  \bar  \varsigma^{bc, 2}_{xx} = 0$, such that the net $\chi B_z$-dependent contribution arises solely from $\bar  \varrho^{m, 2}_{xx} +  \bar  \varsigma^{m, 2}_{xx}$.
For the $\propto B_z^2$-terms, although a $J=1$ node leads to convergent contributions, the $J=2 $ and $J=3 $ nodes produce singularities $ \sim \ln ( \Lambda / \mu) $ and $\sim (\Lambda / \mu)^{2/3 }$, respectively, at the leading order in $\Lambda$.

Since the expressions are cumbersome, the overall behaviour is best understood with the help of plots. Figs.~\ref{figset2long_type2_lin} and \ref{figset2long_type2} serve the purpose, depicting the behaviour of all kinds of terms. In the plots, $t^2_x = \varrho^{bc, 2}_{xx,x} + \varrho^{m, 2}_{xx,x} + \varsigma^{bc, 2}_{xx,x} + \varsigma^{m, 2}_{xx,x} $, $t^2_z = \varrho^{bc, 2}_{xx,z} + \varrho^{m, 2}_{xx,z} + \varsigma^{bc, 2}_{xx,z} + \varsigma^{m, 2}_{xx,z} $, and $\bar t^2_z = \bar \varrho^{bc, 2}_{xx,z} +\bar  \varrho^{m, 2}_{xx,z} +\bar  \varsigma^{bc, 2}_{xx,z} +\bar  \varsigma^{m, 2}_{xx,z} $. For each $J$-value, $t^2_x > 0$, and it represents the summation of completely-convergent integrals. On the contrary, $t^2_z $ is independent of UV-cutoff for $J=1$ (remaining always negative) and dependent on the cutoff for $J\geq 2 $. The latter scenario necessitates the usage of $\bar t^2_z $ for depicting the dominant response, with each $\bar t^2_z $-curve starting out with a negative value, but transitioning to a positive-valued one eventually. We note that $ \varrho^{bc, 2}_{xx,z} +   \varsigma^{bc, 2}_{xx,z} =16 $ for $J=1$, and $ \bar \varrho^{bc, 2}_{xx,z} +  \bar \varsigma^{bc, 2}_{xx,z} = 0$ for $J=2 \text{ and } 3$.

\subsection{Set-up II: In-plane transverse components}
\label{set2_inplane_trs}

The expressions for the integrals are shown in Eq.~\eqref{eqappset2_zx} of Appendix~\ref{appset2_inplane_trs}.
We find that the conductivity contains terms which are linear-in-$B$ as well those which are quadratic-in-$ B $. The former are caused by a nonzero current proportional to $\left( \mathbf E \cdot \mathbf B \right)  \eta_\chi \, {\mathbf{\hat z}} $.

\subsubsection{\underline{Results for the type-I phase for $\mu>0$}}
\label{set2_inplane_type1}

\begin{figure*}[t!]
\centering 
\subfigure[]{\includegraphics[width=0.32 \textwidth]{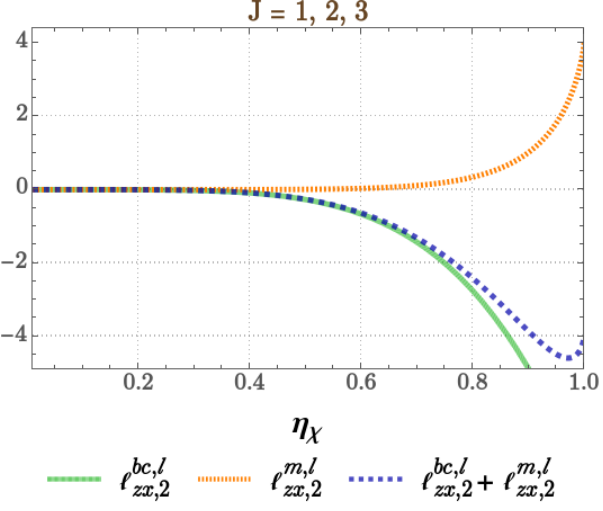}} \hspace{2 cm}
\subfigure[]{\includegraphics[width=0.325 \textwidth]{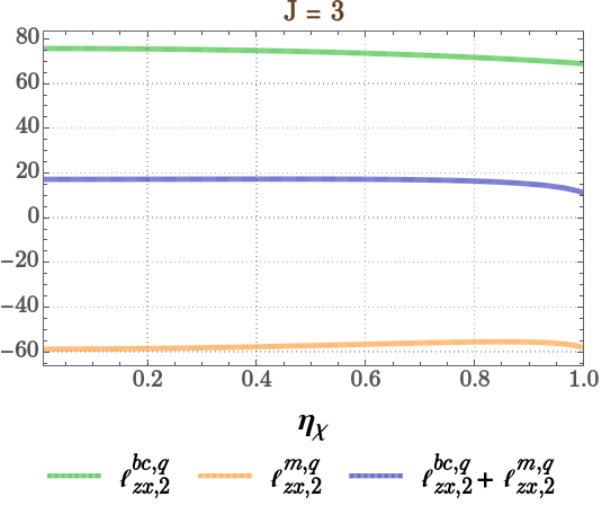}}
\caption{
Planar Hall response for the type-I phase in set-up II: Comparison of the various parts shown in Eqs.~\eqref{eqzx_set2_type1}, \eqref{eqzx_set2_type1_lin}, and \eqref{eqzx_set2_type1_q}.
\label{figzxset2type1}}
\end{figure*}

For $\mu>0$, only the conduction band contributes in the type-I phase.
The  contributions are divided up as shown below:
\begin{align}
\label{eqzx_set2_type1}
 & \sigma_{zx}^{ (\chi, bc) } = 
 \frac{e^3 \, \tau \, J\, v_z} {48 \,\pi^2 \, \eta_\chi^4} \, \chi \,B_x\, \ell^{bc, l}_{zx,2}
 + \frac{e^4 \,\tau\, J \, v_z} {32\, \pi^{ \frac{3} {2} } }
\left (\frac {\alpha_J} {\mu} \right)^{\frac {2} {J}} B_x \, B_z\, \ell^{bc, q}_{zx,2}\,,
\nn &
\sigma_{zx}^{ (\chi, m) } = 
 \frac{e^3 \, \tau \, J\, v_z} {48 \,\pi^2 \, \eta_\chi^4} \, \chi \,B_x\, \ell^{m, l}_{zx,2}
 +
\frac{e^4 \,\tau\, J \, v_z} {32\, \pi^{ \frac{3} {2} } }
\left (\frac {\alpha_J} {\mu} \right)^{\frac {2} {J}}
B_x \, B_z\,\ell^{m, q}_{zx,2} \, .
\end{align}
Here, $ \ell^{bc, l}_{zx,2}$ and $ \ell^{bc, q}_{zx,2} $ represent the BC-only parts proportional to $\chi  B_x $ and $B_x B_z $, respectively. Similarly, $ \ell^{m, l}_{zx,2} $ and $ \ell^{m, q}_{zx,2} $ represent the OMM parts proportional to $\chi B_x $ and $B_x B_z $, respectively.

Appendix~\ref{appset2_inplane_type1} contains the final expressions. From there, we observe that $ \ell^{bc, l}_{zx,2}$ and $ \ell^{m, l}_{zx,2}$ are $J$-independent, as well as $\ell^{bc, l}_{zx,2} <0$, $\ell^{m, l}_{zx,2} >0$, and $\ell^{bc, l}_{zx,2} + \ell^{m, l}_{zx,2}<0 $. The numerical values can be inferred from the curves in Fig.~\ref{figzxset2type1}(a).
For the coefficients accompanying the $B^2$ parts, we observe the following behaviour:
\begin{center}
\begin{tabular}{ |c||c|c|c|}
\hline
 & $ \ell^{bc, q}_{zx,2} $ &  $\ell^{m, q}_{zx,2}$ & $ \ell^{bc, q}_{zx,2} + \ell^{m, q}_{zx,2} $ \\ \hline
 $J=1$ & $ \frac{4 \,\left( 7 \, + \,3 \, \eta_\chi^2 \right)} {15  \,\sqrt{\pi }} $ &
$ \frac{ \,4 \, \left( 1 \, + \,7  \,\eta_\chi^2 \right)}{15  \,\sqrt{\pi }} $ &
$ \frac{8 \, \left(3-2 \, \eta_\chi^2\right)}{15  \,\sqrt{\pi }} $ \\ \hline
 $J=2$ & 
$ 2  \,\sqrt{\pi } $ & $ \frac{5 \, \sqrt{\pi }}{4} $ & $ \frac{3  \,\sqrt{\pi }}{4} $\\ \hline
 $J=3 $ & $> 0$ & $<0$ & $ > 0$ \\ \hline
\end{tabular}
\end{center}
For $J=3$, since the expressions involve hypergeometric functions, the individual and overall behaviour is shown in Fig.~\ref{figzxset2type1}(b). For each $J$-value, $B_x B_z $ has a positive coefficient overall.

\subsubsection{\underline{Results for the type-II phase for $\mu>0$}}
\label{set2_inplane_type2}

In the type-II phase, both the conduction and valence bands contribute for any given $ \mu $. The contributions are divided up as shown below:
\begin{align}
 & \sigma_{zx}^{ (\chi, bc) } = 
 \frac{e^3 \, \tau \, J\, v_z} {48 \,\pi^2 \, \eta_\chi^4} \, \chi \,B_x 
 \left(  \varrho^{bc, l}_{zx,2} + \varsigma^{bc, l}_{zx,2} \right)
 +
 \frac {e^4  \, J^2 \,\tau  \, v_z} {32 \, \pi^2} 
\left (\frac {\alpha_J} {\mu} \right)^{\frac {2} {J} }
B_x \, B_z
\left ( \varrho^{bc, q}_{zx,2} + \varsigma^{bc, q}_{zx,2} \right) ,
\nn & \sigma_{zx}^{ (\chi, m) } = 
\frac{e^3 \, \tau \, J\, v_z} {48 \,\pi^2 \, \eta_\chi^4}\, \chi \,B_x
 \left(  \varrho^{m, l}_{zx,2} + \varsigma^{m, l}_{zx,2} \right)
 +
\frac {e^4  \, J^2 \,\tau  \, v_z} {32 \, \pi^2}
\left (\frac {\alpha_J} {\mu} \right)^{\frac {2} {J}}
B_x \, B_z
\left ( \varrho^{m, q}_{zx,2} + \varsigma^{m, q}_{zx,2} \right) .
\end{align}
The symbols used above indicate the following: (1) $ \varrho^{bc, l}_{zx,2} $ ($ \varsigma^{bc, l}_{zx,2} $) represents the BC-only part proportional to $ \chi \, B_x $, arising from the $s=1$ ($s=2$) band; (2) $ \varrho^{m, l}_{zx,2} $ ($ \varsigma^{m, l}_{zx,2} $) represents the OMM part proportional to $ \chi \, B_x $, arising from the $s=1$ ($s=2$) band; (3) $\varrho^{bc, q}_{zx,2} $ ($ \varsigma^{bc, q}_{zx,2} $) represents the BC-only part proportional to $ B_x\, B_z $, arising from the $s=1$ ($s=2$) band; and (4) $\varrho^{m, q}_{zx,2} $ ($ \varsigma^{m, q}_{zx,2} $) represents the OMM part proportional to $ B_x B_z $, arising from the $s=1$ ($s=2$) band.

The coefficients accompanying $J \chi  B_x $ are $J$-independent and are logarithmically divergent.
Defining
$$ \left \lbrace \varrho^{bc, l}_{zx,2}, \, \varrho^{m, l}_{zx,2} , \,
\varsigma^{bc, l}_{zx,2}, \, \varsigma^{m, l}_{zx,2} \right \rbrace 
= \left \lbrace 
\bar \varrho^{bc, l}_{zx,2}, \,\bar \varrho^{m, l}_{zx,2} , \,
\bar \varsigma^{bc, l}_{zx,2}, \, \bar \varsigma^{m, l}_{zx,2} \right \rbrace  \ln(\Lambda/ \mu) 
+ \order{ \left( {\mu}/ {\Lambda}  \right)^0} \,, $$
we have
\begin{align}
\label{eqzx_set2_type2_lin}
& {\bar \varrho}^{bc, l}_{zx,2} = 3 \left(\eta_\chi^2-1\right)^2\,,\quad
{\bar \varrho}^{m, l}_{zx,2} = 3 \left(5-\eta_\chi^2\right) \left(\eta_\chi^2-1\right) \quad 
\Rightarrow  \quad
{\bar \varrho}^{bc, l}_{zx,2}  +  {\bar \varrho}^{m, l}_{zx,2}  = 2 \left(\eta_\chi^2-1\right),
\nn & {\bar \varsigma}^{bc, l}_{zx,2} = -3 \left(\eta_\chi^2-1\right)^2\,,\quad
{\bar \varsigma}^{m, l}_{zx,2} = -3 \left(\eta_\chi^2-1\right) \left(\eta_\chi^2+3\right)\quad 
\Rightarrow  \quad
{\bar \varsigma}^{bc, l}_{zx,2}  +  {\bar \varsigma}^{m, l}_{zx,2}  = 
6\left (1- \eta_\chi^4 \right) .
\end{align}
We note that, since $\bar \varsigma^{bc, l}_{zx,2} + \,\bar \varrho^{bc, l}_{zx,2} =0 $, on summing over the two bands, the OMM parts yield the net contribution, which takes a negative value of $-\,6\left (\eta_\chi^2 - 1 \right)^2$.

The $B_x B_z$-coefficients are $J$-dependent and divergent, with the degree of divergence increasing with the value of $J$. The final expressions are discussed in Appendix~\ref{appset2_inplane_type2}. In summary, the overall response is dominated by
\begin{align}
\begin{cases}
 \frac{\mu^2 \left(\eta_\chi^2-1\right)^2}  {\eta_\chi^5}  
  \ln \bigg( \frac{\Lambda}{\mu} \bigg)
 & \text { for } J=1 \\
\frac{4 \,\mu^2 \left(\eta_\chi^2-1\right)^{ \frac{5} {2} }}{\eta_\chi^7} \,\frac{\Lambda}{\mu}
+ \frac{4 \, \mu \, \left(\eta_\chi^2-1\right)^{ \frac{3} {2} } 
\left [\mu \, \left(\eta_\chi^2-5\right)-6 \right ]}
{\eta_\chi^6} \ln \bigg( \frac{\Lambda}{\mu} \bigg) & \text { for } J=2 \\
\frac{27 \,\mu^2 \left(\eta_\chi^2-1\right)^{ \frac{8} {3} }}
{4 \, \eta_\chi^{\frac{23} {3} } }     \bigg( \frac{\Lambda}{\mu} \bigg)^{ \frac{4} {3} }
+
\frac{9 \, \mu  \left(\eta_\chi^2-1\right)^{ \frac{5} {3} } 
\left [(3 \, \mu -1) \,\eta_\chi^2-2\, (8 \,\mu + 9 )\right] }
{\eta_\chi^{ \frac{20} {3} }} \, \bigg( \frac{\Lambda}{\mu} \bigg)^{ \frac{1} {3} } 
& \text { for } J=3
\end{cases} \,.
\end{align}
All these singular terms originate from $\varsigma^{bc, q}_{zx,2} $, as the rest of the contributions are non-divergent.

\subsection{Set-up II: Out-of-plane transverse components}
\label{set2_outplane}

All the out-of-plane components of the conductivity tensor vanish identically. This follows from the fact that, with $ \mathbf E = E \, {\hat{{\mathbf x}}} $, $\mathbf B = B_x \, {\hat{{\mathbf x}}} + B_z \, {\hat{{\mathbf z}}} $, the currents sourced by $\chi \left( \mathbf E \cdot \mathbf B \right)   {\mathbf{\hat z}} $ or $ \chi \left( \mathbf B \cdot   {\mathbf{\hat z}} \right ) \mathbf E $ cannot furnish a nonzero $\sigma_{yx}$.

\section{Set-up III: $\mathbf E = E \, {\mathbf{\hat z}}$, $\mathbf B = B_x \, {\mathbf{\hat x}} + B_z \, {\mathbf{\hat z}}  $}
\label{secset3}

In set-up III, as shown in Fig.~\ref{figsetup}(b), the tilt-axis is parallel to $\mathbf E $, but not to $\mathbf B $. We choose $ \hat{\mathbf r}_E =  {\mathbf{\hat z}}  $ and $\hat{\mathbf r}_B = \cos \theta \, {\mathbf{\hat x}} 
+ \sin \theta \, {\mathbf{\hat z}}  $, such that $\mathbf E = E \, {\mathbf{\hat z}}$ and $\mathbf B = B_x \, {\mathbf{\hat x}} 
+ B_z \, {\mathbf{\hat z}} \equiv B  \,\hat{\mathbf r}_B  $.
The details of the generic forms of the integrals are shown in Appendix~\ref{appset3}.
Therein, Appendices~\ref{appset3long}, \ref{appset3_inplane_trs}, and \ref{appset3_outplane} deal with the longitudinal, in-plane transverse, and out-of-plane transverse components, respectively.

\subsection{Set-up III: Longitudinal components}
\label{set3long}

The expressions for the integrals are shown in Eq.~\eqref{eqappset3_long} of Appendix~\ref{appset3long}, which corroborate the existence of terms varying linearly with $B$. Since $\mathbf E $ is parallel to the tilt-axis for this set-up, we can have nonzero currents proportional to $\chi \left( \mathbf E \cdot \mathbf B \right) {\hat{{\mathbf z}}} $ and/or $ \chi \left( \mathbf B \cdot   {\hat{{\mathbf z}}} \right ) \mathbf E $.

\subsubsection{\underline{Results for the type-I phase for $\mu>0$}}
\label{set3long_type1}

\begin{figure*}[t!]
\centering 
\subfigure[]{\includegraphics[width=0.32 \textwidth]{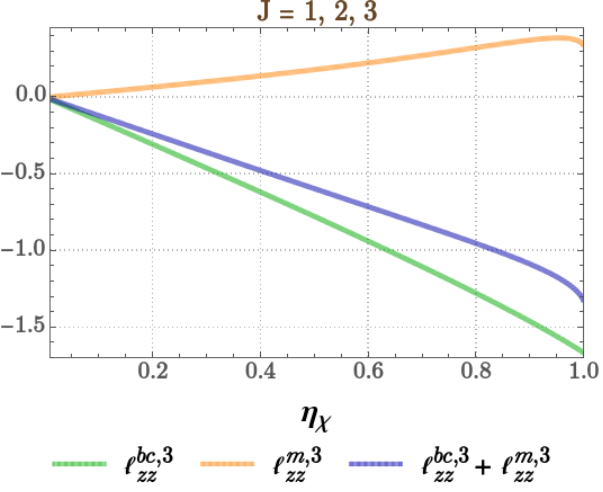}} \hspace{2 cm}
\subfigure[]{\includegraphics[width=0.325 \textwidth]{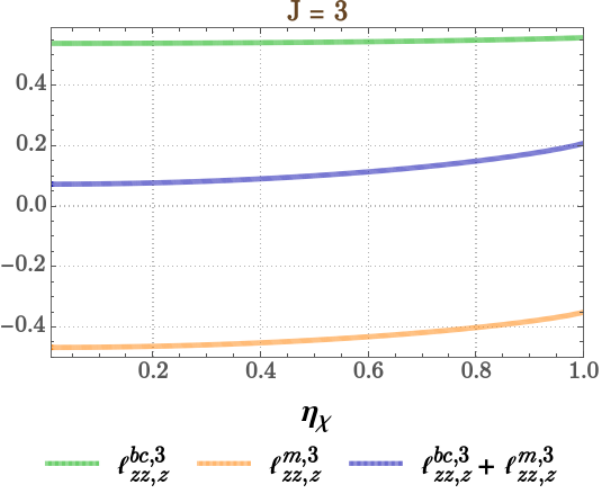} }
\caption{Longitudinal response for the type-I phase in set-up III: Comparison of the various parts shown in Eqs.~\eqref{eqzzset3type1}, \eqref{eqzz_set3_type1_lin}, and \eqref{eqzz_set3_type1_q}.
\label{figzzset3type1}}
\end{figure*}

For $\mu>0$, only the conduction band contributes in the type-I phase. The contributions are divided up as shown below:
\begin{align}
\label{eqzzset3type1}
 & \sigma_{zz}^{ (\chi, bc) } =  \frac {e^3 \, J\, \tau \, v_z}
{ 4 \, \pi^2} \, \chi  \, B_z \,\ell^{bc, 3}_{zz}
+
 \frac {e^4 \, \tau \, J   \, v_z^3}
{ 128 \, \pi^2\, \mu^2} \, B_x^2 \,\ell^{bc, 3}_{zz, x}
+ \frac {e^4 \, \tau \, J^3\, v_z} {16\, \pi^{\frac{3}{2}} } 
\left (\frac {\alpha_J} {\mu} \right)^{\frac {2}  {J}}
 B_z^2 \,\ell^{bc, 3}_{zz, z}\,,
\nn & \sigma_{zz}^{ (\chi, m) } = \frac {e^3 \, J\, \tau \, v_z}
{ 4 \, \pi^2} \, \chi  \, B_z \,\ell^{m, 3}_{zz}
+
 \frac {e^4 \, \, \tau\, J \, \tau  \, v_z^3}
{ 128 \, \pi^2\, \mu^2} \, B_x^2 \,\ell^{m, 3}_{zz, x}
+ \frac {e^4 \, \tau \, J^3\, v_z} {16\, \pi^{\frac{3}{2}} } 
 \left (\frac {\alpha_J} {\mu} \right)^{\frac {2}  {J}}
B_z^2 \,\ell^{m, 3}_{zz, z} \,.
\end{align}
Here, $ \ell^{bc, 3}_{zz} $, $ \ell^{bc, 3}_{zz, x} $, and $ \ell^{bc, 3}_{zz, z} $
represent the BC-only parts proportional to $\chi B_z $, $B_x^2 $, and $ B_z^2 $, respectively. Similarly, $ \ell^{m, 3}_{zz} $, $ \ell^{m, 3}_{zz, x} $, and $ \ell^{m, 3}_{zz, z} $ represent the OMM parts proportional to $\chi B_z $, $B_x^2 $, and $ B_z^2 $, respectively.

The final expressions are demonstrated in Appendix~\ref{appset3long_type1}. From there, we observe that $ 
\ell^{bc, 3}_{zz} $, $ \ell^{m, 3}_{zz} $, $ \ell^{bc, 3}_{zz, x} $, and $ \ell^{m, 3}_{zz, x} $ are $J$-independent.
Furthermore, $ \ell^{bc, 3}_{zz} <0 $, $ \ell^{m, 3}_{zz} > 0$, and $ \ell^{bc, 3}_{zz} + \ell^{m, 3}_{zz} < 0 $.
The coefficients accompanying $ B_x^2 $ take the simple forms of
\begin{align}
& \ell^{bc, 3}_{zz, x} =  \frac{16 \left( 1 + 7 \, \eta_\chi^2 \right)}{15} \,,\quad
\ell^{m, 3}_{zz, x} =  \frac{ - \, 16 \left( 3 + \eta_\chi^2 \right)}{15} 
\quad
\Rightarrow \quad
\ell^{bc, 3}_{zz, x}  + \ell^{m, 3}_{zz, x} = 
\frac{- \,32\left( 1 - 3 \, \eta_\chi^2 \right ) }  {15}  \,.
\end{align}
Therefore, their sum goes from negative to positive at $\eta_\chi = 1/\sqrt 3$.
The numerical values of the linear-in-$B$ coefficients can be inferred from the curves demonstrated in Fig.~\ref{figzzset3type1}(a).
For the coefficients accompanying the $B_z^2$ parts, we observe the following behaviour:
\begin{center}
\begin{tabular}{ |c||c|c|c|}
\hline
 & $ \ell^{bc, 3}_{zz, z} $ & $ \ell^{m, 3}_{zz, z} $ & 
 $ \ell^{bc, 3}_{zz, z} + \ell^{m, 3}_{zz, z} $\\ \hline
 $J=1$ & $\frac{16}{15  \, \sqrt{\pi }}$ &
 $ \frac{ -\,2 \,\left( 4 \,+\, 7  \, \eta_\chi^2 \right)}  {15  \,  \sqrt{\pi }}$ &
$ \frac{2  \, \left(4-\,7 \, \eta_\chi^2\right)}{15 \, \sqrt{\pi }} $ 
 \\ \hline
 $J=2$ &  $ \frac{5  \, \sqrt{\pi }}{16} $ & 
 $  \frac{ - \,\sqrt{\pi }}{4} $  & $ \frac{ \sqrt{\pi }}{16} $\\ \hline
 $J=3 $ & $>0$ & $<0$ & $>0$  \\ \hline
\end{tabular} 
\end{center}
For $J=3$, since the expressions involve hypergeometric functions, the individual and overall behaviour is shown in Fig.~\ref{figzzset3type1}(b). While the net $B_z^2$-dependent response remains positive for $J=2 \text{ and } 3$, for $J=1$, it crosses over from positive to negative at $ \eta_\chi = 2/\sqrt 7$.

\subsubsection{\underline{Results for the type-II phase for $\mu>0$}}
\label{set3long_type2}

In the type-II phase, both the conduction and valence bands contribute for any given $ \mu $. The contributions are divided up into BC-only and OMM parts as
\begin{align}
\label{eqzz_type2}
 \sigma_{zz}^{ (\chi, bc) } & = \frac{e^3 \, \tau \,J \,  v_z}
 {4 \, \pi^2} \, \chi  \, B_z
 \left( \varrho^{bc, 3}_{zz} + \varsigma^{bc, 3}_{zz} \right)
+ \frac{e^4 \, \tau \, J\, v_z^3} {128\, \pi^2 \,\mu^2}\, B_x^2 
\left( \varrho^{bc,3}_{zz,x}  +  \varsigma^{bc,3}_{zz,x} \right )
+ \frac{e^4 \, \tau \,J \, v_z} {128 \, \pi^2}
\left(\frac{\alpha_J} {\mu }\right)^{\frac{2} {J} } B_z^2 
 \left( \varrho^{bc,3}_{zz,z}  +  \varsigma^{bc,3}_{zz,z} \right ) ,
\nn  \sigma_{zz}^{ (\chi, m) } & = \frac{e^3 \, \tau \,J \,  v_z}
 {4 \, \pi^2} \, \chi  \, B_z
 \left( \varrho^{m, 3}_{zz} + \varsigma^{m, 3}_{zz} \right)
+ \frac{e^4 \, \tau \, J\, v_z^3} {128\, \pi^2 \,\mu^2}\, B_x^2 
\left( \varrho^{m,3}_{zz,x}  +  \varsigma^{m,3}_{zz,x} \right )
+ \frac{e^4 \, \tau \,J \, v_z} {128 \, \pi^2}
\left(\frac{\alpha_J} {\mu }\right)^{\frac{2} {J} } B_z^2 
 \left( \varrho^{m,3}_{zz,z}  +  \varsigma^{m,3}_{zz,z} \right ).
\end{align}
The symbols used above indicate the following: (1) $ \varrho^{bc,3}_{zz} $ ($ \varsigma^{bc,3}_{zz} $) represents the BC-only part proportional to $ \chi B_z $, arising from the $s=1$ ($s=2$) band; (2) $ \varrho^{bc,3}_{zz,x} $ ($ \varsigma^{bc,3}_{zz,x} $) represents the BC-only part proportional to $ B_x^2 $, arising from the $s=1$ ($s=2$) band; (3) $ \varrho^{bc,3}_{zz,z} $ ($ \varsigma^{bc,3}_{zz,z} $) represents the BC-only part proportional to $ B_z^2 $, arising from the $s=1$ ($s=2$) band; (4) $\varrho^{m,3}_{zz} $ ($ \varsigma^{m,3}_{zz} $) represents the OMM part proportional to $ \chi \,B_z $, arising from the $s=1$ ($s=2$) band; (5) $\varrho^{m,3}_{zz,x} $ ($ \varsigma^{m,3}_{zz,x} $) represents the OMM part proportional to $ B_x^2 $, arising from the $s=1$ ($s=2$) band; and
(6) $\varrho^{m,3}_{zz,z} $ ($ \varsigma^{m,3}_{zz,z} $) represents the OMM part proportional to $ B_z^2 $, arising from the $s=1$ ($s=2$) band. The final expressions are shown in Appendix~\ref{appset3long_type2}. From there, we see that the coefficients accompanying $ J \chi  B_z $ are $J$-independent and are logarithmically divergent.

\begin{figure}[t!]
\centering 
\subfigure[]{\includegraphics[width= 0.335 \textwidth]{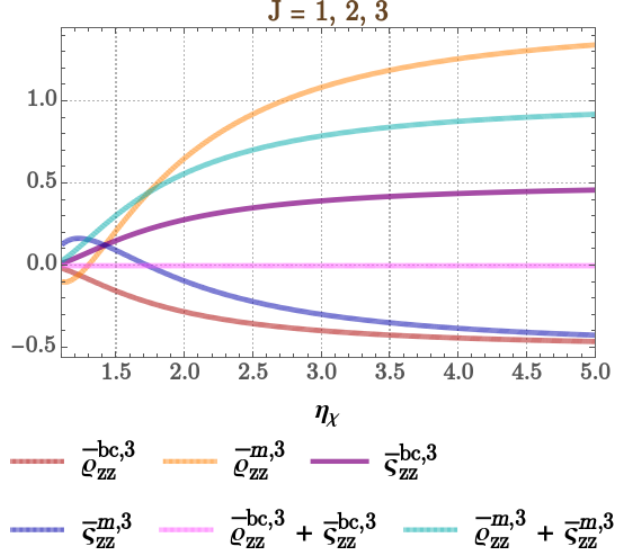}}\hspace{1 cm}
\subfigure[]{\includegraphics[width= 0.32 \textwidth]{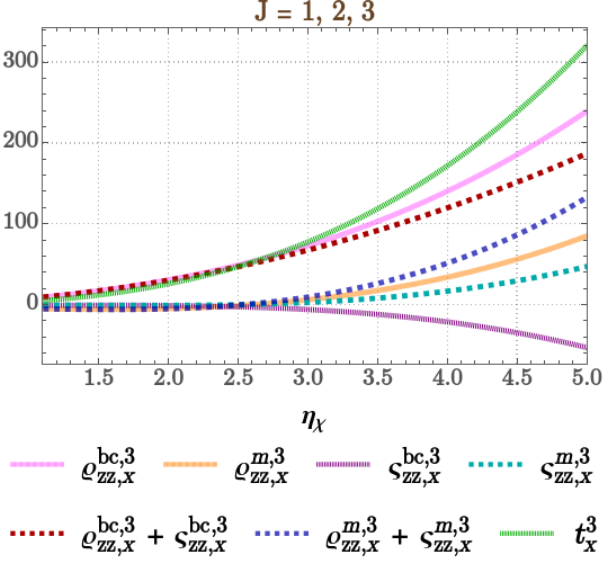}}\\
\subfigure[]{\includegraphics[width= 0.32 \textwidth]{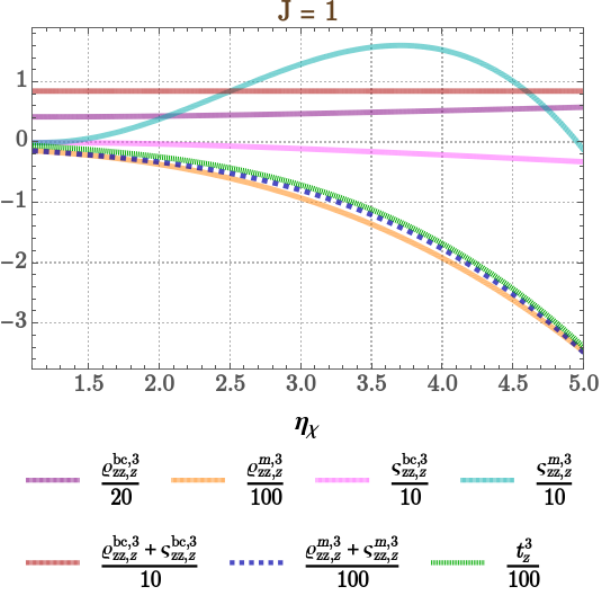}\hspace{0.1 cm}
\includegraphics[width= 0.32 \textwidth]{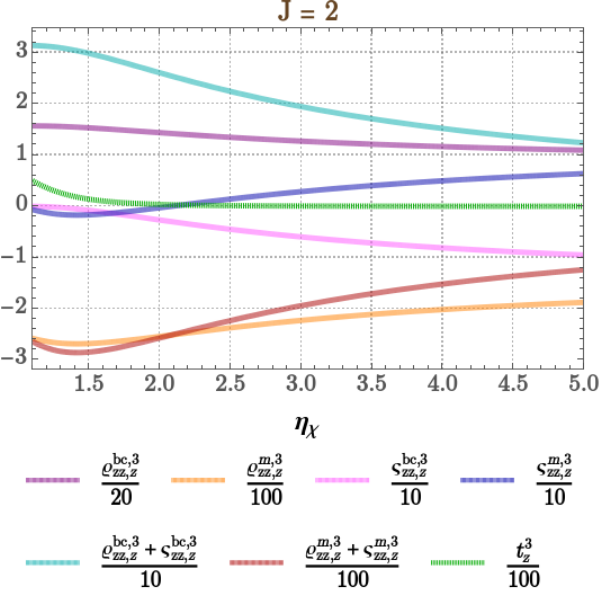}\hspace{0.1 cm}
\includegraphics[width= 0.32 \textwidth]{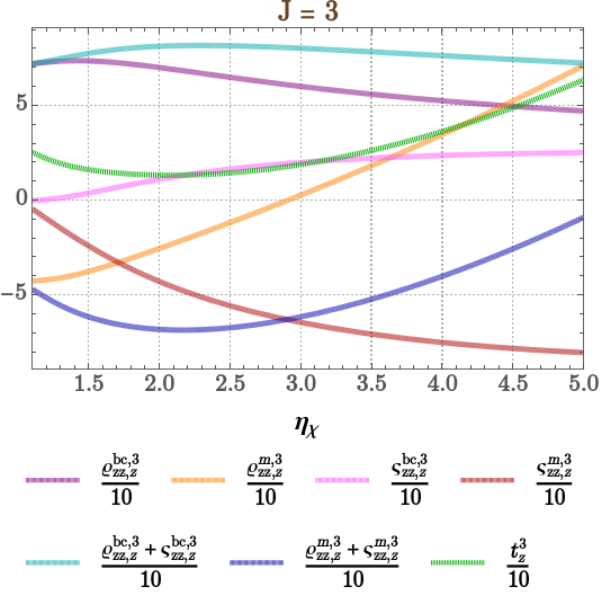}}
\caption{Longitudinal response for the type-II phase in set-up III: Comparison of the various parts for $J=1$, $J= 2$, and $J=3$ [cf. Eqs.~\eqref{eqzz_type2}, \eqref{eqzx_set2_type2_lin}, \eqref{eqzx_set2_type2_Bx}, \eqref{eqzx_type2_Bz1}, \eqref{eqzx_type2_Bz2}, \eqref{eqzx_type2_Bz31}, and \eqref{eqzx_type2_Bz32}], accompanying (a) $\chi B_z$, (b) $B_x^2$, and (c) $B_z^2$. Here, $t^3_x $ and $t^3_z $ represent the net response, after summing over the two bands.
\label{figzzset3type2}}
\end{figure}

Defining
$$ \left \lbrace \varrho^{bc, 3}_{zz}, \, \varrho^{m, 3}_{zz} , \,
\varsigma^{bc, 3}_{zz}, \, \varsigma^{bc, 3}_{zz} \right \rbrace 
= 
\left \lbrace 
\bar \varrho^{bc, 3}_{zz}, \, \bar \varrho^{m, 3}_{zz} , \,
\bar \varsigma^{bc, 3}_{zz}, \,  \bar \varsigma^{bc, 3}_{zz} 
\right \rbrace  \ln(\Lambda/ \mu) 
+ \order{ \left( {\mu}/ {\Lambda}  \right)^0} \,, $$
we have
\begin{align}
\label{eqzx_set2_type2_lin}
& \bar \varrho^{bc, 3}_{zz} = \frac{ - \, \left(\eta_\chi^2-1\right)^2}
{2 \, \eta_\chi^4} \,,\quad
\bar \varrho^{m, 3}_{zz} = \frac{3}{2}-\frac{4}{\eta_\chi^2}+\frac{5}{2 \, \eta_\chi^4}
\quad \Rightarrow  \quad
\bar \varrho^{bc, 3}_{zz}  +  \bar \varrho^{m, 3}_{zz} =
1-\frac{3}{\eta_\chi^2}+\frac{2}{\eta_\chi^4} \, ,
\nn & \bar \varsigma^{bc, 3}_{zz} = \frac{\left(\eta_\chi^2-1\right){}^2}
{2 \, \eta_\chi^4}\,,\quad
\bar \varsigma^{bc, 3}_{zz} = -\,\frac{1}{2}
+\frac{2}{\eta_\chi^2}-\frac{3}{2 \, \eta_\chi^4}
\quad \Rightarrow  \quad
\bar \varsigma^{bc, 3}_{zz} +  {\bar \varsigma}^{m, l}_{zz,2}  = 
\frac{\eta_\chi^2-1}{\eta_\chi^4}\, .
\end{align}
Clearly, $ \bar \varrho^{bc, 3}_{zz} + \bar \varsigma^{bc, 3}_{zz}  =  0$, such that the OMM parts yield the net contribution of $ {\left(\eta_\chi^2-1\right)^2} /{\eta_\chi^4}$, which is positive. Fig.~\ref{figzzset3type2}(a) illustrates the corresponding curves.
We refer the reader to Appendix~\ref{appset3long_type2} for the explicit expressions of the quadratic-in-$B$ parts.
Subfigures (b) and (c) of Fig.~\ref{figzzset3type2} represent their characteristics. In the plots, $t^3_x = \varrho^{bc, 3}_{zz,x} + \varrho^{m, 3}_{zz,x} + \varsigma^{bc, 3}_{zz,x} + \varsigma^{m, 3}_{zz,x} $ and $t^3_z = \varrho^{bc, 3}_{zz,z} + \varrho^{m, 3}_{zz,z} + \varsigma^{bc, 3}_{zz,z} + \varsigma^{m, 3}_{zz,z} $ stand for the net response, after summing over the two bands.

\subsection{Set-up III: In-plane transverse components}
\label{set3_inplane_trs}

The expressions for the integrals are detailed in Appendix~\ref{appset3_inplane_trs}, which show the presence of terms which are linear-in-$B$ as well those which are quadratic-in-$ B $.
The former are caused by a nonzero current $\propto  \chi \left( \mathbf E \cdot  {\hat{{\mathbf z}}} \right ) B_x \, {\hat{{\mathbf x}}} $. In the end, we find that the final results are the same as those for the $zx$-components obtained for set-up II. Hence, the behaviour outlined in Sec.~\ref{set2_inplane_type2} applies here.

\subsection{Set-up III: Out-of-plane transverse components}
\label{set3_outplane}

All the out-of-plane components of the conductivity tensor vanish identically. This follows from the fact that, with $ \mathbf E = E \, {\mathbf{\hat z}} $, $\mathbf B = B_x \, {\mathbf{\hat x}} + B_z \, {\mathbf{\hat z}} $, the currents sourced by $\chi \left( \mathbf E \cdot \mathbf B \right)  {\mathbf{\hat z}} $, $ \chi \left( \mathbf B \cdot   {\mathbf{\hat z}} \right ) \mathbf E $, or $ \chi \left( \mathbf E \cdot {\mathbf{\hat z}} \right)  \mathbf B $ cannot furnish a nonzero $\sigma_{yz}$.


\section{Discussions, summary, and conclusion}
\label{secsum}

In this paper, we have derived all the relevant components of the magnetoelectric conductivity (including the in-plane and out-of-plane components) in planar Hall set-ups involving WSMs and mWSMs, which constitute a complete description incorporating the effects of both the BC and the OMM systematically. Hence, the results here supplement the studies in Ref.~\cite{ips-rahul-tilt}, where the OMM was ignored, and type-II phases were not considered. In particular, we have considered both the undertilted and overtilted regimes of the dispersion, and have chalked out how a nonzero OMM affects the final response. The results show that, in various situations, the OMM-contributed parts turn out to be comparable to or even greater than the BC-only parts. In the latter case, if the BC-only and the OMM parts are of opposite signs, the sign of the overall response is opposite to the BC-only part. Hence, we have demonstrated that the conclusions regarding the nature of the response is prone to be erroneous if the OMM is neglected, emphasizing on the importance of treating all effects of topological origin on equal footing.

Let us summarize the relation of the contents of this paper with the earlier works as follows:
\begin{enumerate}
\item In Ref.~\cite{ips-rahul-ph}, we computed the planar Hall response at a finite $T$ (in the limit $T \ll \mu $) for WSMs/mWSMs, considering only the set-up I of Fig.~\ref{figsetup}. There, we ignored the OMM and considered untilted nodes.
\item In Ref.~\cite{ips-rahul-tilt}, we computed the planar Hall response at a finite $T$ (in the limit $T \ll \mu $) for WSMs/mWSMs, considering all the set-ups shown in Fig.~\ref{figsetup}. There, we ignored the OMM and considered tilted nodes
in the type-I phases only.
\item In Ref.~\cite{ips-medel}, we computed the planar Hall response at a finite $T$ (in the limit $T \ll \mu $) for WSMs/mWSMs, considering only the set-up I of Fig.~\ref{figsetup}. There, we included the effects from OMM, but considered only untilted nodes.
\item In Ref.~\cite{ips-spin1-ph}, we computed the planar Hall response at a finite $T$ (in the limit $T \ll \mu $) for a pseudospin-1 version of the WSMs/mWSMs, considering only set-up I of Fig.~\ref{figsetup}. There, we included OMM, but considered untilted nodes only. These systems comprise three bands (hence, sometimes dubbed as triple-point semimetals or TSMs) meeting at each nodal point, with the Hamiltonian carrying the spin-1 representation of the SO(3) group. One of the three bands is flat (i.e., dispersionless) and, hence, does not contribute to transport. Although the dispersion of the non-flat bands show a behaviour similar to the $s=1$ and $s=2$ bands of the WSMs/mWSMs discussed here, their topological properties are different, arising from the three-component structure of the eigenspinors. In fact, the TSM-nodes carry Chern numbers with values $2 J \chi $, which are twice the values of $J \chi $ (harboured by the systems considered here).
\item In Ref.~\cite{ips-rsw-ph}, we dealt with the planar Hall response of Rarita-Schwinger-Weyl nodes, which are the fourfould cousins of the Weyl nodes. We restricted ourselves to untilted nodes, which are isotropic. There, all the three set-ups give the same results due to the 3d rotational symmetry. For definiteness, we chose the orientations of the electromagnetic fields of set-up I. Because they are pinned at high-symmetry points of the Brillouin zone of chiral crystals (possessing discernable spin-orbit couplings), a tilt-term proportional to the identity matrix is prohibited by symmetry. Hence, tilting is not at all relevant for these systems.
\end{enumerate}

A comprehensive summary of the characteristics of the response in the three distinct set-ups (see Fig.~\ref{figsetup}) can be found in Table~\ref{table_sets}. It also provides a glimpse into the distinctions between the type-I and type-II phases. We have found that tilting gives rise to terms linear-in-$B$, depending on the relative orientation of the $\mathbf E $-$\mathbf B$ plane with respect to the tilt-axis. For the type-II phases, due to the existence of open Fermi pockets arising from the effective continuum Hamiltonian, some of the integrals are divergent, which are regularized by introducing a UV cutoff $\Lambda $. Although we have shown the results for $\mu > 0$ and $ \eta_\chi \geq 0$, the corresponding expressions for the $\mu<0 $ and/or $ \eta_\chi < 0 $ cases can be obtained by following the same procedure. In particular, for the type-II phases, we have to implement the correct limits of integration for the $\gamma $-integrals \cite{amit-magneto} [cf. Eq.~\eqref{eqint}]. Finally, when we add up the contributions coming from a pair of conjugate nodes (with chiralities $\chi $ and $-\chi $), we need to consider the distinct values of the chemical potential and the tilt parameter for the two nodes (which need not be of the same sign). 

One way to do away with the cutoff for regularizing the integrals in the type-II phase, which turn to be divergent in $\Lambda $, is to add suitable terms to the effective Hamiltonian. These are subleading terms which are higher-order in momentum, as outlined in Refs.~\cite{marcus-emil, ips_optical_cond}, and are naturally expected to be present in a realistic bandstructure. The additional terms lead to closed Fermi pockets in the type-II regime, capturing the actual/physical scenarios, thus eliminating the need for using a seemingly \textit{ad hoc} UV cutoff. However, such terms will substantially complicate the already cumbersome computations. Hence, we leave it for a follow-up work, remembering that one way to simplify the calculations is to obtain the relevant characteristics numerically.

In the future, it will be worthwhile to investigate the cases when the tilting is taken with respect to the $x$- or $y$-axis for the mWSMs.\footnote{For the WSMs, the choice of the tilt-axis does not matter, because the untilted system is isotropic.} This will significantly increase the complexity of the integrals because the integrands will then depend on the azimuthal angle $ \phi $. Another direction is to recompute the response after the inclusion of internode scatterings in the collision integrals, which appear in the Boltzmann equations \cite{das19_linear2, ips-internode}. We would like to emphasize that, since we have used the methodology based on the relaxation-time approximation, it is our aim to gain a better understanding by going beyond this approximation by an exact computation of the relevant collision integrals \cite{timm}.
Yet another avenue to be explored is to go beyond the weak-magnetic-field limit, and determine the response in the presence of the quantized Landau levels caused by the applied magnetic field \cite{goerbig-LL-weyl, ips-kush, fu22_thermoelectric, marcus-emil, ips_optical_cond}.
While all the above scenarios involve noninteracting Hamiltonians, the response arising in the presence of disorder and/or strong interactions will essentially involve employing many-body techniques \cite{ips-seb, ips_cpge, ips-biref, ips-klaus, rahul-sid, ips-rahul-qbt, ips-qbt-sc, ips-plasmons, ips-jing-plasmons, ips-hermann-review}.

\section*{Acknowledgments}
We thank Firdous Haidar for useful discussions.


\appendix
\section{Set-up I --- $\mathbf E = E \, {\mathbf{\hat x}}$, $\mathbf B = B_x \, {\mathbf{\hat x}} + B_y \, {\mathbf{\hat y}}  $}
\label{appset1}

In set-up I, as shown in Fig.~\ref{figsetup}(a), the tilt-axis is perpendicular to the plane spanned by $\mathbf E $ and $\mathbf B $. Due to the rotational symmetry of the dispersion of each semimetallic node within the $xy$-plane, the exact directions of $\mathbf E $ and $\mathbf B $ does not matter --- the only physically relevant parameter is the angle between $ \hat{\mathbf r}_E$ and $ \hat{\mathbf r}_B$. Hence, without any loss of generality, we choose $ \hat{\mathbf r}_E = {\mathbf{\hat x}}  $ and $\hat{\mathbf r}_B = \cos \theta \, {\mathbf{\hat x}} + \sin \theta \, {\mathbf{\hat y}}  $, such that $\mathbf E = E \, {\mathbf{\hat x}}$ and $\mathbf B = B_x \, {\mathbf{\hat x}} + B_y \, {\mathbf{\hat y}} \equiv B  \,\hat{\mathbf r}_B  $.
In the following, we will include a prefactor of $\zeta $ ($\upsilon $) for each factor of a component of BC (OMM). This helps us distinguish whether the term originates from BC, OMM, or both.

\subsection{Set-up I: Longitudinal components}
\label{appset1long}

The starting expression is captured by
\begin{align}  
\label{eqappset1long}
  \sigma_{xx}^{\chi, s}  & = \sigma^{\chi, s}_{{\rm Drude}, xx}  +
\frac{e^4\,\tau\, J^2 \,\alpha_J^4 \,v_z^2  }   {64 }
\int \frac{d \epsilon \, d \gamma } {(2\,\pi)^2 } \,
\frac{  k_\perp^{4 J-4}}  {\epsilon^8}
\left(\zeta^2 \, t^1_{1xx} +  \upsilon^2 \,t^1_{2xx} + 
\zeta \,\upsilon \,t^1_{3xx} \right)
\mathcal J  ,\nn 
t^1_{1xx}  & = 
 \left [8 \,\epsilon \, B_x^2 \,\alpha_J^2\,  k_\perp^{2 J}
  \left \lbrace (-1)^s \,\eta_\chi\, k_z \,v_z-\epsilon \right \rbrace
  +\alpha_J^4 \left(3 \, B_x^2+B_y^2\right)  k_\perp^{4 J}
  +8  \,\epsilon^2 \,B_x^2 
  \left \lbrace -2 \,(-1)^s \,\epsilon\, \eta_\chi\, k_z\, v_z
  +\eta_\chi^2 \,k_z^2\, v_z^2+\epsilon^2 \right \rbrace
 \right ] \nn &    \quad \times
2 \,J^2 \,\delta (\epsilon -\mu ) \,,\nn
t^1_{2xx}  & = 
8  \left [ J^2\, k_z^4 \, v_z^4 \left(3 \,B_x^2+B_y^2\right)
-2\, J \,\epsilon^2\, k_z^2\, v_z^2
   \left(B_x^2 + B_y^2\right)+\epsilon^4 \left(B_x^2+B_y^2\right)
\right] \delta (\epsilon -\mu )  \nn &  \quad
+ 8  \, J \, (-1)^s  \,\epsilon  \, \alpha_J^2  \, k_\perp^{2 J}  
 \left [
 \epsilon^2 \left(B_x^2+B_y^2\right)-J\,  k_z^2  \,v_z^2 \left(3  \,B_x^2+B_y^2\right)
 \right ]   \delta^\prime (\epsilon -\mu ) 
+ J^2 \,\epsilon^2\, \alpha_J^4 \, k_\perp^{4 J} \left(3\, B_x^2+B_y^2\right)  
\delta^{\prime \prime} (\epsilon -\mu ) \,,\nn
\frac{ t^1_{3xx} } {2\, J } & =  16 \,J\, \epsilon \, B_x^2 \,k_z^2 \,v_z^2 
 \left [ \,(-1)^s \,\eta_\chi\, k_z\, v_z-\epsilon 
 \right ] \delta (\epsilon -\mu )
- J \,(-1)^s \, \epsilon \, \alpha_J^4 
\left(3\, B_x^2 + B_y^2\right)  k_\perp^{4 J}  \,\delta^\prime (\epsilon -\mu ) 
\nn & \quad
+ 4 \,\alpha_J^2 \, k_\perp^{2 J} 
\Big [
 \, \Big \lbrace
J \,k_z^2 \,v_z^2 \left(3\, B_x^2+B_y^2\right) \delta (\epsilon -\mu )
-2 \,  J \, \epsilon^2 \, B_x^2 \,\eta_\chi\, k_z \,v_z \, \delta^\prime (\epsilon -\mu )
\nn & \hspace{ 2.5 cm }
+\epsilon^2 \left(B_x-B_y\right)  \left(B_x+B_y\right) \delta (\epsilon -\mu ) 
\Big \rbrace
+ 2 \,J \, (-1)^s \,\epsilon^3\, B_x^2\, \delta^\prime (\epsilon -\mu )
 \Big ] .
\end{align}
We find that there exists no term with a linear-in-$B$ dependence, showing that the inclusion of the OMM does not lead to an $\order{B}$ term.

\subsubsection{Results for the type-I phase for $\mu>0$}
\label{appset1_longtype1}

For $\mu>0$, only the conduction band contributes in the type-I phase.
The contributions are divided up into BC-only (i.e., $t^1_{1xx} $-contributed) and OMM [i.e., $ ( t^1_{2xx} + t^1_{3xx} ) $-contributed] parts as
\begin{align}
& \sigma_{xx}^{ (\chi, bc) } =
 \frac {e^4 \,  \tau  \, J^3\, v_z  } 
 {128 \, \pi^{ \frac{3} {2}} } 
\left (\frac {\alpha_J} {\mu} \right)^{\frac{2} {J}} 
 \left(  B_x^2 \,\ell^{bc, 1}_{xx,x}  +   B_y^2 \, \ell^{bc, 1}_{xx,y} \right )
\text{ and } \sigma_{xx}^{ (\chi, m) } = 
\frac {e^4 \,  \tau  \,J^3\,  v_z  } 
 {128 \, \pi^{ \frac{3} {2}} } 
\left (\frac {\alpha_J} {\mu} \right)^{\frac{2} {J}} 
 \left( 
B_x^2 \, \ell^{m, 1}_{xx,x} + B_y^2\, \ell^{m, 1}_{xx,y} \right ) .
\end{align}
Here, $ \ell^{bc, 1}_{xx,x} $ and $\ell^{m, 1}_{xx,x}$ ($ \ell^{bc, 1}_{xx,y} $ and $\ell^{m, 1}_{xx,y}$)
represent the parts proportional to $B_x^2 $ ($ B_y^2 $).
The final expressions turn out to be
\begin{align}
   \frac{ \ell^{bc, 1}_{xx,x} }
{ \frac{\Gamma \left(2-\frac{1}{J}\right)}{30\, J^4\, \eta_{\chi }^4} } & =
J \left[ 30 \,J^3+120 \,J^2 \,\eta_\chi^6-97\, J^2
+J \left \lbrace J \,(122\, J-115) + 207 \right \rbrace
\eta_\chi^4-4 \,(J-2) \,(2 \,J-1) \,(9\, J-7) 
\,\eta_\chi^2 + 89 \,J+\frac{4}{J}-32\right ]
\nn & \qquad \times
   \,_2\tilde{F}_1\left(\frac{1}{2}-\frac{1}{J},\frac{J-1}{J};\frac{5}{2}-\frac{1}{J};
   \eta_{\chi }^2\right)
\nn & \quad
+  J \, (J-2)  \left(\eta_\chi^2-1\right)
 \left[ 8 \,J \, (2\, J+9) \,\eta_\chi^4 -30 \,J^2 
 + \left \lbrace J \,(54\, J-77)+25 \right \rbrace 
 \eta_\chi^2 + 37\, J+\frac{2}{J}-15\right ]
\nn & \qquad \times
 \,_2\tilde{F}_1\left(\frac{3}{2}-\frac{1}{J},\frac{J-1}{J};\frac{5}{2}-\frac{1}{J};
 \eta_{\chi}^2 \right)  
\end{align}
\begin{align}
 \ell^{bc, 1}_{xx,y} = 
 \Gamma \left(4-\frac{1}{J}\right) \, 
_2\tilde{F}_1\left(\frac{1}{2}-\frac{1}{J},\frac{J-1}{J};\frac{9}{2}-\frac{1}{J};
 \eta_{\chi }^2\right),
\end{align}
\begin{align}
 \frac {\ell^{m, 1}_{xx,x} } {\frac{\Gamma \left(2-\frac{1}{J}\right)}
 {30 \, J^6 \, \eta_\chi^4}}  & =  
\Big [   \frac{(J-2)^2 \,(2 \,J-1) (3\, J-1) (4 \,J-1) (5 \,J-2)}{J^2}
+(J \,(J \,(2 \,J \,(32 \,J+57)-461)+59)+30) \,\eta_\chi^4
\nn & \qquad 
-\frac{4 \,(J-2) \,(2 \,J-1) \,(3 \,J-2) \,(J \,(2 \,J-9)+3) \,\eta_\chi^2}{J}
\Big] \, J^2 \,
\,_2\tilde{F}_1\left(\frac{3}{2}-\frac{1}{J},\frac{J-1}{J};\frac{5}{2}-\frac{1}{J};
   \eta_\chi^2 \right)
\nn & \quad + \Big [
 J \,\big \lbrace -120 \, J^2+32\, (\,(J-7) J+2) \,\eta_\chi^4
 +(157-2 \,J (12 \,J+25)) \,\eta_\chi^2 + 418 \,J-453 \big \rbrace
 \nn & \qquad \quad
+  \frac{4}{J^2} 
+ \frac{6 \left(3 \,\eta_\chi^2-8\right)} {J}
+ 217-99 \,\eta_\chi^2 
\Big] \, J^2 \,
(J-2) \left(1-\eta_\chi^2\right) \,
  _2\tilde{F}_1\left(\frac{3}{2}-\frac{1}{J},\frac{J-1}{J};\frac{5}{2}-\frac{1}{J};
   \eta_\chi^2\right),
\end{align}   
\begin{align}
&  \frac {  \ell^{m, 1}_{xx,y}}
{ \frac{\Gamma \left(2-\frac{1}{J}\right)}
{4 \,J^4\, \Gamma \left(\frac{9}{2}-\frac{1}{J}\right)} }   =
(J+2)\, (7 J-2) \left [ 2\, J \,\,
_3F_2\left(\frac{3}{2},\frac{1}{2}-\frac{1}{J},1-\frac{1}{J};
\frac{1}{2},\frac{7}{2}-\frac{1}{J};\eta_\chi^2\right)+(2-5 \,J) 
 \,  \,_2F_1\left(\frac{1}{2}-\frac{1}{J},\frac{J-1}{J};\frac{5}{2}-\frac{1}{J};
   \eta_{\chi}^2\right) \right ]
\nn & \hspace{ 2.5 cm } 
  +3 \,(J-2) \,(4 \,J-1)\, J^2 \,\,
  _3F_2\left(\frac{5}{2},\frac{1}{2}-\frac{1}{J},1-\frac{1}{J};
   \frac{1}{2},\frac{9}{2}-\frac{1}{J};\eta_\chi^2\right) .
\end{align}
Here, $_2\tilde {F}_1\left ( a, b; c ;\eta_\chi^2 \right)$ is the regularized hypergeometric function, $_2 {F}_1\left (  a, b; c ;\eta_\chi^2\right ) / \Gamma(c)$, and $_3 F_2\left (a_1, a_2, a_2; b_1, b_2, b_3 ; \eta_\chi^2 \right) $ represents the generalized hypergeometric function. The resulting behaviour is discussed in Sec.~\ref{set1_longtype1} of the main text. The $J=3$ case is captured by Fig.~\ref{figell1}(a).

\subsubsection{Results for the type-II phase for $\mu>0$}
\label{appset1_longtype2}

In the type-II phase, both the conduction and valence bands contribute for any given $ \mu $.
The contributions are divided up into BC-only and OMM parts as
\begin{align}
& \sigma_{xx}^{ (\chi, bc) } =
 \frac {e^4 \, J^3 \, \tau  \, v_z  } {128 \, \pi^2} 
\left ( \frac {\alpha_J} {\mu} \right)^{\frac{2} {J}} 
 \left[  B_x^2 \left( \varrho^{bc, 1}_{xx,x} + \varsigma^{bc, 1}_{xx,x}   \right)
 +  B_y^2 \left( \varrho^{bc, 1}_{xx,y}  + \varsigma^{bc, 1}_{xx,y}   \right) \right  ],
\nn &
\sigma_{xx}^{ (\chi, m) } = 
\frac {e^4 \, J^3 \, \tau  \, v_z  } 
 {128 \, \pi^2} 
\left (\frac {\alpha_J} {\mu} \right)^{\frac{2} {J}} 
 \left[  B_x^2 \left( \varrho^{m, 1}_{xx,x} 
 + \varsigma^{m, 1}_{xx,x}   \right)
 + B_y^2 \left( \varrho^{m, 1}_{xx,y} 
 + \varsigma^{m, 1}_{xx,y}   \right) \right  ].
\end{align}
The symbols used above indicate the following: (1) $\varrho^{bc, 1}_{xx,x} $ and $ \varrho^{bc, 1}_{xx,y}$ ($ \varsigma^{bc, 1}_{xx,x} $ and $ \varsigma^{bc, 1}_{xx,y} $) represent the BC-only parts proportional to $B_x^2 $ and $B_y^2 $, respectively, coming from the $s=1$ ($s=2$) band, and (2) $\varrho^{m, 1}_{xx,x} $ and $ \varrho^{m, 1}_{xx,y}$ ($ \varsigma^{m, 1}_{xx,x} $ and $ \varsigma^{m, 1}_{xx,y}$) represent the OMM parts proportional to $B_x^2 $ and $B_y^2 $, respectively, coming from the $s=1$ ($s=2$) band.

Here, the integrals turn out to be quite complicated and, in order to extract the answers, we need to perform them separately for each value of $J$. Finally, we find that neither of them is divergent in $\Lambda$. The final expressions and their behaviour are obtained as discussed below:
\begin{enumerate}

\item $J=1$:
\begin{align}
\label{eqxxj1}
\varrho^{bc, 1}_{xx,x}  & =
\frac{64}{15} + 2 \,\eta_\chi^3+\frac{104 \,\eta_\chi^2}{15}
+\frac{17 \,\eta_\chi}{2}+\frac{5}{6 \,\eta_\chi}
-\frac{7}{30 \,\eta_{\chi }^3}+\frac{1}{10 \,\eta_\chi^5} \,,
\quad \varrho^{bc, 1}_{xx,y}   =
\frac{8}{15}  + \frac{\eta_\chi}{6}+\frac{1}{2 \,\eta_\chi}
-\frac{1}{6\, \eta_\chi^3}
+\frac{1}{30\, \eta_\chi^5} \,,
\nn \varrho^{m, 1}_{xx,x}  & = - \,\frac{32}{15}
-\frac{64  \,\eta_\chi^2}{15}-\frac{15  \,\eta_\chi}{2}
+\frac{5}{6  \,\eta_\chi}
+\frac{17}{30 \, \eta_{\chi}^3}-\frac{3}{10  \,\eta_\chi^5}\,,
\quad \varrho^{m, 1}_{xx,y}   = - \, \frac{8}{5}
-\frac{\eta_\chi}{2}-\frac{3}{2  \,\eta_\chi}
+\frac{1}{2  \,\eta_\chi^3}
-\frac{1}{10  \,\eta_\chi^5} \,,\nn
\varsigma^{bc, 1}_{xx,x}  & = \frac{64}{15}-2 \,\eta_\chi^3
+\frac{104 \,\eta_\chi^2}{15}-\frac{17\, \eta_\chi}{2}
-\frac{5}{6 \,\eta_\chi}+\frac{7}{30 \,\eta_{\chi }^3}
-\frac{1}{10 \,\eta_\chi^5} \,,
\quad \varsigma^{bc, 1}_{xx,y}  =
\frac{8}{15} -\frac{\eta_\chi}{6}-\frac{1}{2 \,\eta_\chi}
+\frac{1}{6 \,\eta_\chi^3}
-\frac{1}{30 \,\eta_\chi^5} \,,
\nn \varsigma^{m, 1}_{xx,x}  & =
\frac{112}{15} -\frac{32 \,\eta_\chi^2}{15}+\frac{\eta_\chi}{2}
-\frac{43}{6 \,\eta_\chi}+\frac{49}{30 \,\eta_\chi^3}
-\frac{3}{10 \,\eta_\chi^5} \,,
\quad \varsigma^{m, 1}_{xx,y}  = 
\frac{8}{5} -\frac{\eta_\chi}{2}-\frac{3}{2 \,\eta_\chi}
+\frac{1}{2 \,\eta_\chi^3}-\frac{1}{10\, \eta_\chi^5}  \,.
\end{align}

\item $J=2$:
\begin{align}
\label{eqxxj2_1}
\varrho^{bc, 1}_{xx,x}  & =
\frac{\sqrt{\eta_\chi^2-1} \left(656 \,\eta_\chi^6
+223 \,\eta_\chi^4-294\, \eta_\chi^2+120\right)}
{240 \,\eta_{\chi}^6}+\left(2 \,\eta_\chi^2 + \frac{31}{8}\right) \cot^{-1}
   \left(\frac{\eta_\chi-1}{\sqrt{\eta_\chi^2-1}}\right) ,
\nn \varrho^{bc, 1}_{xx,y} & = 
\frac{1}{96} \left [\frac{2 \,\sqrt{\eta_\chi^2-1} 
\left(33\, \eta_\chi^4-26 \,\eta_\chi^2+8 \right)}{\eta_{\chi}^6}
+30 \csc^{-1} \eta_\chi +15 \,\pi \right ],
\nn \varrho^{m, 1}_{xx,x}  & =
\frac{\sqrt{\eta_\chi^2-1} \left(   42  -128 \,\eta_\chi^4
+101 \,\eta_\chi^2 \right)} {60 \,\eta_\chi^4}
+\frac{\tan^{-1}\left(\eta_\chi-\sqrt{\eta_\chi^2-1}\right)}{2} 
-\frac{3\, \pi }{8} \,,
\nn \varrho^{m, 1}_{xx,y}  & = 
\frac{\left(2-3 \,\eta_\chi^2\right) 
\sqrt{\eta_\chi^2-1}} {4 \,\eta_\chi^4}
-\frac{1}{2} \cot^{-1}\left(\frac{\eta_{\chi }-1}{\sqrt{\eta_\chi^2-1}}\right) ,
\end{align}
\begin{align}
\label{eqxxj2_2}
\varsigma^{bc, 1}_{xx,x}  & = 
\frac{\sqrt{\eta_\chi^2-1} \left(656 \,\eta_\chi^6
+223 \,\eta_\chi^4-294 \,\eta_\chi^2+120\right)}{240 \,\eta_{\chi}^6}
-\left(2\, \eta_\chi^2+\frac{31}{8}\right) 
\cot^{-1}\left(\frac{\eta_\chi+1}
{\sqrt{\eta_{\chi}^2-1} }\right), \nn
\varsigma^{bc, 1}_{xx,y} & =
\frac{\sqrt{\eta_\chi^2-1} \left(33 \,\eta_\chi^4
-26 \,\eta_\chi^2+8\right)}{48 \,\eta_\chi^6}
-\frac{5}{16} \sec^{-1} \eta_\chi \, ,
\nn \varsigma^{m, 1}_{xx,x}  & =
\frac{\sqrt{\eta_\chi^2-1} 
\left( 42 -128 \,\eta_\chi^4 + 101 \,\eta_\chi^2 \right)}
{60 \,\eta_\chi^4}-\frac{1}{2}
   \cot^{-1}\left(\eta_\chi-\sqrt{\eta_\chi^2-1}\right) +\frac{\pi }{8} \,,
\nn \varsigma^{m, 1}_{xx,y} & = \frac{\left(2-3 \,\eta_\chi^2\right) 
\sqrt{\eta_\chi^2-1}} {4 \,\eta_\chi^4}
+\frac{1}{2} \cot^{-1}\left(\frac{\eta_\chi+1}{\sqrt{\eta_\chi^2-1}}\right) .
\end{align}

\begin{figure*}[t!]
\centering 
\subfigure{\includegraphics[width= 0.32 \textwidth]{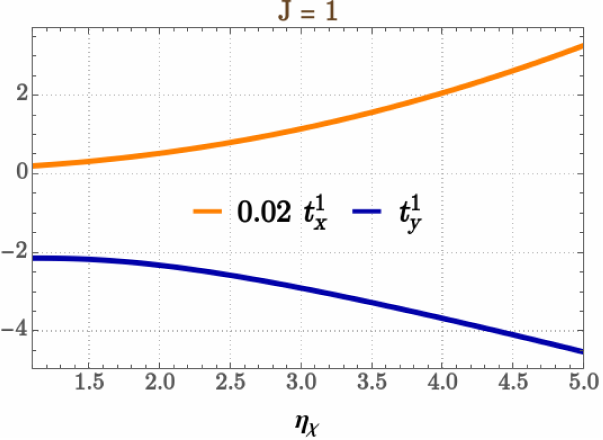}} \hspace{0.1 cm}
\subfigure{\includegraphics[width= 0.32 \textwidth]{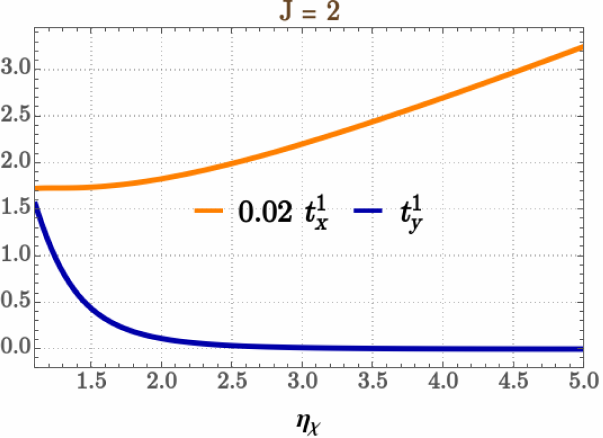}} \hspace{0.1 cm}
\subfigure{\includegraphics[width= 0.315 \textwidth]{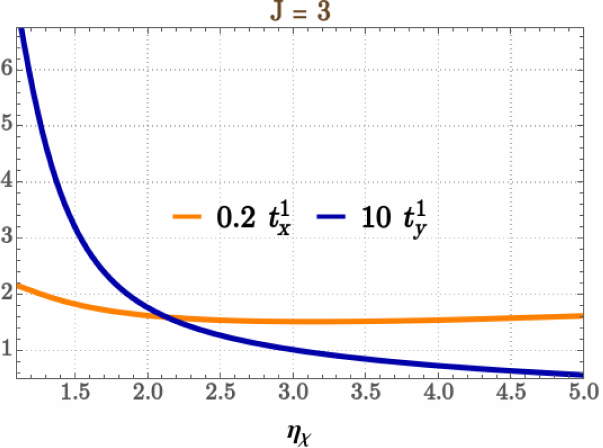}} 
\caption{
Longitudinal response for the type-II phase in set-up I: The total over all the terms accompanying $B_x^2$ ($B_y^2$).
Here, $t^1_x= \varrho^{bc, 1}_{xx,x} + \varrho^{m, 1}_{xx,x} + \varsigma^{bc, 1}_{xx,x} + \varsigma^{m, 1}_{xx,x} $ and $t^1_y = \varrho^{bc, 1}_{xx,y} + \varrho^{m, 1}_{xx,y} + \varsigma^{bc, 1}_{xx,y} + \varsigma^{m, 1}_{xx,y} $, used in the plot-legends.
\label{fignetset1xx}}
\end{figure*}

\item $J=3$:
\begin{align}
\label{eqxxj3_1}
\varrho^{bc, 1}_{xx,x}  & = \frac{4 \,\sqrt{\pi } \,
\Gamma \left(\frac{5}{3}\right) \,\left(\eta_\chi+1\right)^{ \frac{2} {3}}}
{729 \,\eta_\chi^{ \frac{17} {3}} \,  \sqrt[3]{\eta_\chi^2-1}} \,
\Big [
\frac{\sqrt[3]{\left(\eta_\chi-1\right)^2 \eta_\chi} 
\left \lbrace 9 \left(27 \eta_\chi^4+20 \, \eta_\chi^2-26\right)
   \eta_\chi^2+91 \right  \rbrace }
   {\Gamma \left(\frac{7}{6}\right)} \nn &
\hspace{ 4.5 cm } + 
\frac{2 \, \sqrt{3 \, \pi } 
\left \lbrace  3 \, \eta_\chi^2 
\left(81\,  \eta_\chi^6 + 135\,  \eta_\chi^4
-90 \, \eta_{\chi}^2 + 91\right)-91 \right \rbrace
\,_2F_1\left(\frac{1}{3},\frac{2}{3};\frac{4}{3};\frac{\eta_\chi+1} 
{1-\eta_{\chi}} \right)}
   {\Gamma^2 \left(\frac{1}{3}\right)}   
\Big ], \nn
\varrho^{bc, 1}_{xx,y}  & =
\frac{8\, \pi \, \Gamma \left(\frac{5}{3}\right)}{729\, \sqrt{3}\,
 \Gamma^2 \left(\frac{1}{3}\right)\,\sqrt[3]{\eta_\chi^{19}
   \left(\eta_\chi^2-1\right)}}\,
\Big  [
\sqrt[3]{2} \left \lbrace  \left(\eta_\chi^2-1\right)^{ \frac{2} {3}} 
\left(99\, \eta_\chi^2-92\right) \eta_\chi^{ \frac{7} {3}}
+91\,   \sqrt[3] \, {\eta_\chi \left(\eta_\chi^2-1\right)^2}
\right \rbrace
\nn & \hspace{ 5.5 cm }
+ \left(\eta_\chi+1\right)^{ \frac{2} {3}} \left(45 \,\eta_\chi^2 
\left(9 \,\eta_\chi^4-9\, \eta_\chi^2+7\right)-91\right)
   \,_2F_1\left(\frac{1}{3},\frac{2}{3};\frac{4}{3};\frac{\eta_\chi+1}{1-\eta_\chi}\right)
\Big ]  ,
\end{align}
\begin{align}
\label{eqxxj3_2}
\varrho^{m, 1}_{xx,x}  & = \frac{8 \,\pi  \,\Gamma \left(\frac{5}{3}\right) \left(\eta_\chi+1\right)^{ \frac{2} {3}}
}
{2187 \,\sqrt{3} \,
\Gamma \left(\frac{1}{3}\right)^2 \eta_\chi^{ \frac{17} {3}}
\, \sqrt[3]{\eta_\chi^2-1}} \,
\Big [
\sqrt[3]{2} \left \lbrace 
3 \left(\eta_\chi-1\right)^{ \frac{2} {3}} 
\left ( -432 \,\eta_\chi^4
+432 \,\eta_\chi^2+59\right) \eta_{\chi }^{ \frac{7} {3}}
+ 1001\, \sqrt[3]{\left(\eta_\chi-1\right)^2 \eta_\chi}\right \rbrace
\nn & \hspace{ 5.5 cm }
 + \left  \lbrace
 9 \,\eta_\chi^2 \left(468\, \eta_\chi^4-123 \,\eta_\chi^2+28\right)-1001 
 \right \rbrace \,
  _2F_1\left(\frac{1}{3},\frac{2}{3};\frac{4}{3};\frac{\eta_\chi+1}{1-\eta_\chi}
   \right ) \Big ],\nn
\varrho^{m, 1}_{xx,y}  & =
\frac{4 \,\sqrt{\frac{\pi }{3}}\, \Gamma \left(-\frac{1}{6}\right)}
{59049 \,\Gamma \left(\frac{1}{3}\right)}
\,\Big [
\frac{2 \, \sqrt[3]{\eta_\chi^2-1} 
\left(1917  \,\eta_\chi^4-744  \,\eta_\chi^2-1001\right)}
{\eta_\chi^6}
\nn & \hspace{ 3 cm } +
\frac{2^{ \frac{2} {3}} \left  \lbrace
9 \left(153 \,\eta_\chi^4-261\, \eta_\chi^2+ 35\right) 
\eta_\chi^2+1001 \right \rbrace \,
  _2F_1\left(\frac{1}{3},\frac{2}{3};\frac{4}{3};\frac{\eta_\chi+1}{1-\eta_\chi}\right)}
   {\sqrt[3]{\frac{\left(\eta_{\chi }-1\right) \eta_\chi^{19}}{\eta_\chi+1}}}
 \Big ],
\end{align}
\begin{align}
\label{eqxxj3_3}
\varsigma^{bc, 1}_{xx,x}  & =
\frac{4 \,\Gamma \left(\frac{5}{3}\right) \,\Gamma \left(\frac{11}{6}\right)}
{405\, \sqrt{\pi }}
\, \Big [
\frac{2 \,\sqrt[3]{\eta_\chi^2-1} 
\left \lbrace 9 \left(27 \,\eta_\chi^4+20\, \eta_\chi^2-26\right) 
\eta_{\chi}^2+91\right \rbrace }
{\eta_\chi^6}
\nn & \hspace{ 3 cm } +
2^{ \frac{2} {3}} \,\sqrt[3]{\frac{\eta_\chi-1}{\eta_\chi^{19}
 \left(\eta_\chi+1\right)}} 
\left \lbrace 91-3 \,\eta_\chi^2
   \left(81 \,\eta_\chi^6+135 \,\eta_\chi^4-90 \,\eta_\chi^2+91\right)
   \right \rbrace \,
  _2F_1\left(\frac{1}{3},\frac{2}{3};\frac{4}{3};\frac{2}{\eta_\chi+1}-1\right)
\Big ], \nn
\varsigma^{bc, 1}_{xx,y}  & = \frac{2 \, \Gamma \left(\frac{5}{3}\right)}{2187}
\, \Big [
\frac{\Gamma \left(\frac{-1}{6}\right)
 \sqrt[3]{\eta_\chi^2-1} 
\left(-297 \,\eta_\chi^4+276 \,\eta_{\chi}^2-91\right)}
{\sqrt{\pi } \,\eta_\chi^6}
\nn & \hspace{ 2 cm } +  
\frac{2 \,\Gamma \left(\frac{2}{3}\right) 
\,\sqrt[3]{\frac{\eta_\chi-1}{\eta_\chi^{19} 
\left(\eta_\chi+1\right)}}
   \left \lbrace
   91-45\, \eta_\chi^2 \left(9 \,\eta_\chi^4-9 \,\eta_\chi^2+7\right)
   \right \rbrace \,
  _2F_1\left(\frac{1}{3},\frac{2}{3};\frac{4}{3};\frac{2}{\eta_\chi+1}-1\right)}
   {\Gamma \left(\frac{4}{3}\right)}
\Big ],
\end{align}
\begin{align}
\label{eqxxj3_4}
\varsigma^{m, 1}_{xx,x}  & = \frac{\Gamma \left(\frac{-1}{6}\right)
\, \Gamma \left(\frac{5}{3}\right)}{6561 \,\sqrt{\pi }}
\, \Big [
\frac{2 \,\sqrt[3]{\eta_\chi^2-1} 
\left(2592 \,\eta_\chi^6-378 \,\eta_\chi^4-2877\, \eta_\chi^2+637\right)}
{\eta_{\chi }^6}
\nn & \hspace{ 3 cm } +
2^{ \frac{2} {3}} \,\sqrt[3]{\frac{\eta_\chi-1}
{\eta_\chi^{19} \left(\eta_\chi+1\right)}} 
\left \lbrace 
 \left(198 \,\eta_{\chi}^4+87 \,\eta_\chi^2-350\right) \eta_\chi^2
 +637\right \rbrace \,
  _2F_1\left(\frac{1}{3},\frac{2}{3};\frac{4}{3};\frac{2}{\eta_\chi+1}-1\right)
\Big ], \nn
\varsigma^{m, 1}_{xx,y}  & =
\frac{\Gamma \left(\frac{- 1}{6}\right) \,\Gamma \left(\frac{5}{3}\right)}
{19683 \,\sqrt{\pi }} \, \Big [
\frac{14  \,\sqrt[3]{\eta_\chi^2-1} \left(459  \,\eta_\chi^4
-546  \,\eta_\chi^2+91\right)}{\eta_\chi^6}
\nn & \hspace{ 3 cm } +
2^{ \frac{2} {3}}  \,\sqrt[3]{\frac{\eta_\chi-1}{\eta_\chi^{19}
 \left(\eta_\chi+1\right)}} 
 \left\lbrace  637-9  \,\eta_\chi^2
   \left(153 \, \eta_\chi^4-531  \, \eta_\chi^2+455\right)
   \right \rbrace \,
  _2F_1\left(\frac{1}{3},\frac{2}{3};\frac{4}{3};\frac{2}{\eta_\chi+1}-1\right)
\Big ].
\end{align}

\end{enumerate}
In the above expressions, $F_1 (a; b_1, b_2; c; z_1, z_2) $ stands for the the Appell hypergeometric function of two variables, $z_1$ and $z_2$.  The resulting behaviour is discussed in Sec.~\ref{set1_longtype2} with the help of representative plots [cf. Fig.~\ref{figset1long_type2}]. In addition to that, we also show here the net response in Fig.~\ref{fignetset1xx}, which supplements Fig.~\ref{figset1long_type2} by magnifying the corresponding curves.

\subsection{Set-up I: In-plane transverse components}
\label{appset1_inplane_trs}

The starting expression is captured by
\begin{align}  
\label{eqappset1_inplane_trs}
  \sigma_{yx}^{\chi, s}
& = 
\frac{e^4\,\tau \, J^3 \, \alpha_J^4 \,v_z^2 \, B_x \, B_y }       {32} 
\int \frac{d \epsilon \, d \gamma } {(2\,\pi)^2 } \,
\frac{  k_\perp^{4 J-4}}  {\epsilon^8}
\left(\zeta^2 \, t^1_{1yx} +  \upsilon^2 \,t^1_{2yx} + \zeta \,\upsilon \,t^1_{3yx} \right)
\mathcal J  ,\nn 
 t^1_{1yx}  & = 
 2 \,J
 \left [ \alpha_J^2 \, k_\perp^{2 J}-2 \,\epsilon 
  \left \lbrace \epsilon -(-1)^s \,\eta_\chi \,k_z\,v_z\right \rbrace
 \right ]^2 \delta (\epsilon -\mu ) \,,
\nn   t^1_{2yx}  & = 
8\, J \,k_z^4 \,v_z^4 \, \delta (\epsilon -\mu )
-8\, J \,(-1)^s\, \epsilon \, \alpha_J^2 \,k_z^2\, v_z^2\,  k_\perp^{2 J}\,  \delta^\prime (\epsilon -\mu )
+ J \,\epsilon^2 \,\alpha_J^4\,  k_\perp^{4 J} \, \delta^{\prime \prime}(\epsilon -\mu ) \,,
\nn    t^1_{3yx} & = 
8 
 \left [\alpha_J^2\,  k_\perp^{2 J} 
 \left(J \,k_z^2 \,v_z^2+\epsilon^2\right)+2 \,J \,\epsilon \, k_z^2
   v_z^2 \left((-1)^s\, \eta_\chi \,k_z \,v_z-\epsilon \right)
 \right ] \delta (\epsilon -\mu )
\nn & \quad
-2\,  J \, (-1)^s \, \epsilon \,  \alpha_J^2 \,  k_\perp^{2 J} 
\left [\alpha_J^2  k_\perp^{2 J} +4\,  (-1)^s\,  \epsilon \,  \eta_\chi\,  k_z \, v_z
\right ] \delta^\prime (\epsilon -\mu ) \,.
\end{align}
We find that there exists no term with a linear-in-$B$ dependence, showing that the inclusion of the OMM does not lead to an $\order{B}$ term.

\subsubsection{Results for the type-I phase for $\mu>0$}
\label{appset1_inplane_trs_type1}

For $\mu>0$, only the conduction band contributes in the type-I phase.
The contributions are divided up into BC-only (i.e., $t^1_{1yx} $-contributed) and OMM [i.e., $ ( t^1_{2yx} + t^1_{3yx} ) $-contributed] parts as
\begin{align}
 \sigma_{yx}^{ (\chi, bc) } = \frac{e^4  \, \tau  \, v_z}  {64 \, \pi^{\frac{3} {2}} \, J}
\left (\frac {\alpha_J} {\mu} \right)^{\frac{2} {J}} 
\, B_x \, B_y \, \ell^{bc}_{yx,1}  \text{ and }
\sigma_{yx}^{ (\chi, m) } = 
 \frac{e^4  \, \tau  \, v_z}  {64 \, \pi^{\frac{3} {2}} \, J}
\left (\frac {\alpha_J} {\mu} \right)^{\frac{2} {J}} 
\, B_x \, B_y \, \ell^{m}_{yx,1} \, .
\end{align}
Here, the final expressions turn out to be
\begin{align}
\frac{ \ell^{bc}_{yx,1}} {\frac{\Gamma \left(2-\frac{1}{J}\right)}{90 \,\eta_\chi^4}}
& = J \,\Big [
72 \, \eta_\chi^2 +  \frac{4}{J}-32 
+ J \,\Big \lbrace
30 \,J^2+180 \,J \,\eta_\chi^6 + 3 \,(J (46 \,J-45)+101) \,\eta_\chi^4
-36\, (J\, (2 \,J-7)+7) \eta_\chi^2-97\, J+89
\Big \rbrace
\Big] \nn
& \qquad \times \,_2\tilde{F}_1\left(\frac{1}{2}-\frac{1}{J},\frac{J-1}{J};\frac{5}{2}-\frac{1}{J};
\eta_\chi^2\right) \nn &
\quad + J\,
(J-2)  \,\left(\eta_\chi^2-1\right) 
\Big [(33 \, \eta_\chi^2+J \left(12 \, (2  \,J+9)  \,\eta_\chi^4
+3  \, (18 \, J-31)  \,\eta_{\chi}^2-30 J+ 37\right)+\frac{2}{J}-15 \Big ]\nn
& \qquad \times
\,_2\tilde{F}_1\left(\frac{3}{2}-\frac{1}{J},\frac{J-1}{J};\frac{5}{2}-\frac{1}{J};\eta_\chi^2\right),
\end{align}
\begin{align}
\frac{ J\, \ell^{m}_{yx,1} } 
{\frac{\Gamma \left(2-\frac{1}{J}\right)} {4  \, \Gamma \left(\frac{9}{2}-\frac{1}{J}\right)}}
& =
\frac{(7 \, J-2) 
\left[ \left(20 \,J^2-13 \,J+2\right) \,
  _2F_1\left(\frac{1}{2}-\frac{1}{J},\frac{J-1}{J};\frac{5}{2}-\frac{1}{J};\eta_\chi^2\right)
 -4\, (J-2) \,J \,\eta_{\chi}^2 \,_2F_1\left(\frac{3}{2}-\frac{1}{J},\frac{J-1}{J};\frac{7}{2}-\frac{1}{J};\eta_\chi^2\right) \right ]}
   {J}
\nn & \quad +
3 \,J \,(4\, J-1) \,_3F_2\left(\frac{5}{2},\frac{1}{2}-\frac{1}{J},1-\frac{1}{J};\frac{1}{2},\frac{9}{2}-\frac{1}{J};
\eta_{\chi}^2\right)   
\nn & \quad + 12\, J \,(3\, J-2) \,\eta_\chi^2 \,
  _3F_2\left(\frac{5}{2},1-\frac{1}{J},\frac{3}{2}-\frac{1}{J};\frac{3}{2},\frac{9}{2}-\frac{1}{J};
   \eta_\chi^2\right).
\end{align}
The resulting behaviour is discussed in Sec.~\ref{set1_inplane_trs_type1} of the main text. The $J=3$ case is captured by Fig.~\ref{figell1}(b).

\subsubsection{Results for the type-II phase for $\mu>0$}
\label{appset1_inplane_trs_type2}

In the type-II phase, both the conduction and valence bands contribute for any given $ \mu $. The contributions are divided up into BC-only and OMM parts as
\begin{align}
 \sigma_{yx}^{ (\chi, bc) } =
 \frac {e^4 \, J\, \tau  \, v_z  } 
 { 64 \, \pi^2} 
\left (\frac {\alpha_J} {\mu} \right)^{\frac{2} {J}} \, B_x \, B_y 
\left ( \varrho^{bc}_{yx,1} +  \varsigma^{bc}_{yx,1} \right ),\quad
\sigma_{yx}^{ (\chi, m) } = 
 \frac {e^4 \, J\, \tau  \, v_z  } 
 { 64 \, \pi^2} 
\left (\frac {\alpha_J} {\mu} \right)^{\frac{2} {J}} \, B_x \, B_y 
\left ( \varrho^{m}_{yx,1} +  \varsigma^{m}_{yx,1} \right ) .
\end{align}
The symbols used above indicate the following: (1) $\varrho^{bc}_{yx,1} $ ($ \varsigma^{bc}_{yx,1} $) represents the BC-only part proportional to $B_x\, B_y $, arising from the $s=1$ ($s=2$) band; (2) $\varrho^{m}_{yx,1} $ ($ \varsigma^{m}_{yx,1} $) represents the OMM part proportional to $B_x\, B_y $, arising from the $s=1$ ($s=2$) band.

The results for the integrals are extracted by performing them separately for each value of $J$, and neither of them is divergent in $\Lambda$. The final expressions and their behaviour are obtained as discussed below:
\begin{enumerate}

\item $J=1$:
\begin{align}
\label{eqyxj1}
& \varrho^{bc}_{yx,1} = \frac{28}{15} + \eta_\chi^3 +
\frac{52 \,  \eta_\chi^2}{15}+\frac{25 \,  \eta_\chi }{6}
+\frac{1}{6 \,  \eta_\chi }-\frac{1}{30\,  \eta_\chi^3}
+\frac{1}{30 \,  \eta_\chi^5} \,,
\quad \varrho^{m}_{yx,1} = \frac{ - \,4}{15} -\frac{32 \,  \eta_\chi^2}{15}
-\frac{7 \,  \eta_\chi }{2}+\frac{7}{6 \,  \eta_\chi }
+\frac{1}{30 \,  \eta_\chi^3}-\frac{1}{10 \,  \eta_\chi^5}
\,, 
\nn & \varsigma^{bc}_{yx,1}
 = \frac{28}{15} - \eta_\chi^3
+\frac{52 \,  \eta_\chi^2}{15}-\frac{25 \,  \eta_\chi }{6}
-\frac{1}{6  \, \eta_\chi }+\frac{1}{30 \,  \eta_\chi^3} 
-\frac{1}{30 \,  \eta_\chi^5}\,, 
\quad \varsigma^{m}_{yx,1} = \frac{44}{15}-\frac{16 \,  \eta_\chi^2}{15}
+\frac{ \eta_\chi }{2}-\frac{17}{6  \, \eta_\chi }
+\frac{17}{30 \,  \eta_\chi^3}-\frac{1}{10 \,  \eta_\chi^5} \,.
\end{align}

\item $J=2$:
\begin{align}
\label{eqyxj2}
& \varrho^{bc}_{yx,1} =
\frac{\sqrt{ \eta_\chi^2-1}
 \left(328   \, \eta_\chi^6+29   \, \eta_\chi^4-82  \,  \eta_\chi^2+40\right)}
 {240  \, \eta_\chi^6}+
 \left( \eta_\chi^2+\frac{13}{8}\right) \cot^{-1}\left(\frac{ \eta_\chi -1}
   {\sqrt{ \eta_\chi^2-1}}\right),\nn
& \varrho^{m}_{yx,1} = \frac{\sqrt{ \eta_\chi^2-1} 
\left(-64  \,  \eta_\chi^4+73 \,   \eta_\chi^2+6\right)}
{60   \, \eta_\chi^4}+\frac{1}{2} \tan^{-1}\left( \eta_\chi -\sqrt{ \eta_\chi^2-1}\right)
-\frac{3  \, \pi }{8}  \,,\nn
& \varsigma^{bc}_{yx,1} =
\frac{\sqrt{ \eta_\chi^2-1} 
\left(328  \,  \eta_\chi^6+29 \,   \eta_\chi^4-82  \,  \eta_\chi^2+40\right)}
{240  \eta_\chi^6}-\left( \eta_\chi^2+\frac{13}{8}\right) \cot^{-1}\left(\frac{ \eta_\chi +1}
{\sqrt{ \eta_\chi^2-1}}\right),\nn
& \varsigma^{m}_{yx,1} =
\frac{\sqrt{ \eta_\chi^2-1} 
\left(-64  \,  \eta_\chi^4+73  \,  \eta_\chi^2+6\right)}
{60  \eta_\chi^4}-\frac{1}{2} \cot^{-1}\left( \eta_\chi -\sqrt{ \eta_\chi^2-1}\right)
+\frac{\pi }{8} \,.
\end{align}

\item $J=3$:
\begin{align}
\label{eqyxj3_1}
& \varrho^{bc}_{yx,1} = \frac{2  \,  \pi  \,  \Gamma \left(\frac{5}{3}\right)}
{729 \, \sqrt{3} \,  \Gamma \left(\frac{1}{3}\right)^2  \,  \eta_\chi^{ \frac{19} {3} }
   \sqrt[3]{ \eta_\chi^2-1}} \,\Big [
2  \,  \sqrt[3]{ 2\,\eta_\chi  \left( \eta_\chi^2-1\right)^2} 
\left(729  \, \eta_\chi^6 + 243   \, \eta_\chi^4-426  \,  \eta_\chi^2+182\right) \nn
&  \hspace{2 cm } + \left( \eta_\chi +1\right)^{ \frac{2} {3} } 
\left \lbrace   \eta_\chi^2 \left(81  \,  \eta_\chi^6 + 90  \,  \eta_\chi^4
-45  \, \eta_{\chi}^2+56\right)-182\right \rbrace  \,
   _2F_1\left(\frac{1}{3},\frac{2}{3};\frac{4}{3};\frac{ \eta_\chi +1}{1- \eta_\chi }\right)   
\Big ] \,,   
\nn & \varrho^{m}_{yx,1} = \frac{4 \,\pi \, 
\Gamma \left(\frac{5}{3}\right) \left( \eta_\chi +1\right)^{ \frac{2} {3} }}
{6561  \, \sqrt{3}  \, \Gamma \left(\frac{1}{3}\right)^2 
\sqrt[3]{ \eta_\chi^{19} \left( \eta_\chi^2-1\right)}}
\,\Big[
\sqrt[3]{2} \left  \lbrace   3 \left( \eta_\chi -1\right)^{ \frac{2} {3} }  
\eta_\chi^{\frac{7}{3}} 
\left(1296   \, \eta_\chi^4-1935  \,  \eta_\chi^2+71\right)-
 2002 \,  \sqrt[3]{\left( \eta_\chi -1\right)^2  \eta_\chi }\right \rbrace
\nn & \hspace{ 2 cm } +
\left(2002-9  \,  \eta_\chi^2 \left(1557  \,  \eta_\chi^4
-630   \, \eta_\chi^2+119\right)\right) \,
  _2F_1\left(\frac{1}{3},\frac{2}{3};\frac{4}{3};\frac{ \eta_\chi +1}{1- \eta_\chi }\right) 
\Big ]\,,  
\end{align}
\begin{align}
\label{eqyxj3_2}
&  \varsigma^{bc}_{yx,1} = 
\frac{\Gamma \left(-\frac{1}{6}\right) \,  \Gamma \left(\frac{5}{3}\right)}
{4374  \, \sqrt{\pi }}
\,\Big[
\frac{ - \,2  \, \sqrt[3]{ \eta_\chi^2-1} 
\left(729  \,  \eta_\chi^6+243  \,  \eta_\chi^4-426  \,  
\eta_\chi^2+182\right)}  { \eta_\chi^6}
\nn & \hspace{ 2 cm } +
\sqrt[3]{\frac{ 4\, (\eta_\chi -1) }{ \eta_\chi^{19} 
\left( \eta_\chi +1\right)}} 
\left \lbrace 9  \,  \eta_\chi^2
 \left(81   \,  \eta_\chi^6+90   \, \eta_\chi^4-45  \,  \eta_\chi^2+56\right)
 -182\right \rbrace \,
  _2F_1\left(\frac{1}{3},\frac{2}{3};\frac{4}{3};\frac{2}{ \eta_\chi +1}-1\right)
\Big ]\,, \nn  
 \nn &  \varsigma^{m}_{yx,1} = \frac{\Gamma \left(-\frac{1}{6}\right) 
 \Gamma \left(\frac{5}{3}\right)}  {39366 \, \sqrt{\pi }}
 \,\Big[ 
 \frac{2 \, \sqrt[3]{ \eta_\chi^2-1} 
 \left(7776 \,  \eta_\chi^6 -4347   \, \eta_\chi^4-4809  \,  \eta_\chi^2+1274\right)}
 { \eta_\chi^6}
\nn & \hspace{ 2 cm } +
\sqrt[3]{\frac{ 4\, (\eta_\chi -1) }{ \eta_\chi^{19} 
\left( \eta_\chi +1\right)}} 
\left \lbrace
9 \left(747  \, \eta_\chi^4-270  \,  \eta_\chi^2-595\right)  \eta_\chi^2
+1274\right \rbrace \,
  _2F_1\left(\frac{1}{3},\frac{2}{3};\frac{4}{3};\frac{2}{ \eta_\chi +1}-1\right)
\Big ]\,. 
\end{align}

\end{enumerate}
The resulting behaviour is discussed in Sec.~\ref{set1_inplane_trs_type2} of the main text. In particular, Fig.~\ref{fig5} provides a pictorial depiction of the relative and additive magnitudes of the various terms.

\subsection{Set-up I: Out-of-plane transverse components}
\label{appset1_outplane}

The starting expression is captured by
\begin{align}  
\label{eqappset1_outplane}
  \sigma_{yx}^{\chi, s}
& = 
\frac{ -\, e^3 \,\tau \, J^2 \, \alpha_J^2 \, v_z^2 }    {4}  \,  \chi \, B_x
\int \frac{d \epsilon \, d \gamma } {(2\,\pi)^2 } \,
\frac{k_\perp^{2 J-2}} {\epsilon^5}
\left(  \zeta  \, t^1_{1zx} + \upsilon  \, t^1_{2zx} \right) \mathcal J\,,
\nn &  t^1_{1zx} = 
\left [  \,(-1)^s \, k_z \, v_z-\epsilon  \, \eta_\chi\right ] 
\left [ \alpha_J^2  \, k_\perp^{2 J}-2 \epsilon 
   \left \lbrace \epsilon -(-1)^s  \,\eta_\chi  \,k_z  \,v_z
 \right \rbrace \right ]\delta (\epsilon -\mu )\,,
\nn &  t^1_{2zx} =
2\, k_z \,v_z \left [ k_z \,v_z 
\left \lbrace\, (-1)^s\, k_z\, v_z-\epsilon \, \eta_\chi\right \rbrace
-(-1)^s\, \alpha_J^2 \, k_\perp^{2 J}\right ] \delta (\epsilon -\mu ) 
 + \epsilon \, \alpha_J^2 \, k_\perp^{2 J}  
\left [\,(-1)^s \,\epsilon \, \eta_\chi-k_z\, v_z \right ]\delta^\prime (\epsilon -\mu )\,.
\end{align} 
We find that the only nonzero terms have a linear-in-$B$ dependence, with the $\mathcal{O}(B^2)$ terms vanishing altogether. The part of the magnetoelectric current, varying linearly with $B$, is $ \propto \chi \left( \mathbf E \cdot \mathbf B \right)  {\mathbf{\hat z}} $ (in agreement with Ref.~\cite{amit-magneto}).

\subsubsection{Results for the type-I phase for $\mu>0$}
\label{appset1_offplane_type1}

For $\mu>0$, only the conduction band contributes in the type-I phase.
The contributions are divided up into BC-only (i.e., $ t^1_{1zx}$-contributed) and OMM (i.e., $ t^1_{2zx}$-contributed) parts as
\begin{align}
 \sigma_{zx}^{ (\chi, bc) } =
 \frac{3 \,e^3 \, \tau\, J\,v_z  \,\eta_\chi^4} {512 \,\pi^2} 
 \, \chi \, B_x \,\ell^{bc}_{zx,1} \text{ and }
\sigma_{zx}^{ (\chi, m) } =  \frac{3 \,e^3 \, \tau\, J\,v_z  \,\eta_\chi^4} {512 \,\pi^2} 
\, \chi \, B_x \, \ell^{m}_{zx,1} \, ,
\end{align}
where
\begin{align}
\label{eqsigzx}
 \ell^{bc}_{zx,1} 
=  -6 \, \eta_\chi^5 + 5 \,\eta_\chi^3-3 \,\eta_\chi
+ 3 \left( 1- \eta_\chi^2 \right)^2 \tanh^{-1}\eta_{\chi} \text{ and }
 \ell^{m}_{zx,1} = 
 -13 \,\eta_\chi^3+15 \eta_\chi
 -3 \left(\eta_\chi^4-6 \,\eta_\chi^2+5\right) \tanh^{-1} \eta_{\chi}  .
\end{align}
We observe that the response goes to zero if $\eta_\chi \rightarrow 0 $.
The consequences of the above expressions are discussed in Sec.~\ref{set1_offplane_type1} of the main text. In particular, Fig.~\ref{figell1}(c) provides a pictorial depiction of the relative and additive magnitudes of the two terms.

\subsubsection{Results for the type-II phase for $\mu>0$}
\label{appset1_offplane_type2}

In the type-II phase, both the conduction and valence bands contribute for any given $ \mu $. The contributions are divided up into BC-only and OMM parts as
\begin{align}
 \sigma_{zx}^{ (\chi, bc) } =
 \frac {e^3\, J\, \tau\,  v_z}
{64\, \pi^2}
\, \chi \, B_x 
\left ( \varrho^{bc}_{zx,1} +  \varsigma^{bc}_{zx,1} \right ) \text{ and }
\sigma_{zx}^{ (\chi, m) } = 
\frac {e^3\, J\, \tau\,  v_z}
{64\, \pi^2}
\, \chi \, B_x 
\left ( \varrho^{m}_{zx,1} +  \varsigma^{m}_{zx,1} \right ) .
\end{align}
The symbols used above indicate the following: (1) $ \varrho^{bc}_{zx,1} $ ($ \varsigma^{bc}_{zx,1} $) represents the BC-only part proportional to $\chi \, B_x $, arising from the $s=1$ ($s=2$) band, and (2) $\varrho^{m, 1}_{xx,x} $ ($ \varsigma^{m, 1}_{xx,x} $) represents the OMM part proportional to $ \chi \,B_x $, arising from the $s=1$ ($s=2$) band. The corresponding integrals are $J$-independent and take the following forms:
\begin{align}
\label{eqappzxset1}
& \varrho^{bc}_{zx,1} = - \, 14 -8 \,\eta_\chi+\frac{20}{3 \,\eta_\chi}
+\frac{16} {\eta_\chi^2}-\frac{4} {\eta_\chi^3}-\frac{22}{3 \,\eta_{\chi}^4}
+ {\bar \varrho}^{bc}_{zx,1} 
\left[ \ln \bigg(\frac{\Lambda} {\mu} \bigg)
+ \ln \left( \frac {\eta_\chi+1} {\eta_\chi}  \right) \right ] , \quad
{\bar \varrho}^{bc}_{zx,1} = 
\frac{ - \,4 \left(\eta_\chi^2-1\right)^2 }{\eta_\chi^4}\,,\nn
& \varrho^{m}_{zx,1} = 6-\frac{52}{3 \,\eta_\chi}
-\frac{40}{\eta_\chi^2}+\frac{20}{\eta_\chi^3}
+\frac{110}{3 \,\eta_\chi^4}
+ {\bar \varrho}^{m}_{zx,1} \left[ \ln \bigg(\frac{\Lambda} {\mu} \bigg)
+ \ln \left( \frac {\eta_\chi+1} {\eta_\chi}  \right) \right ], \quad
{\bar \varrho}^{m}_{zx,1} = \frac{ - \,4 \left(\eta_\chi^4
-6 \,\eta_\chi^2+5\right) }   {\eta_\chi^4} \,,
\nn &  \varsigma^{bc}_{zx,1} = 14 -8 \,\eta_\chi+\frac{20}{3\, \eta_\chi}
- \frac{16}  {\eta_\chi^2} -\frac{4}{\eta_\chi^3}+\frac{22}{3\, \eta_{\chi}^4} 
+ {\bar \varsigma}^{bc}_{zx,1} \left[ \ln \bigg(\frac{\Lambda} {\mu} \bigg)
+ \ln \left( \frac {\eta_\chi+1} {\eta_\chi}  \right) \right ], \quad
{\bar \varsigma}^{bc}_{zx,1} = \frac{ - \,4 \left(\eta_\chi^2-1\right)^2 } {\eta_\chi^4} \,,
\nn &  \varsigma^{m}_{zx,1} = -\frac{4}{\eta_\chi}+ \frac{16}  {\eta_\chi^2} 
+\frac{12}{\eta_\chi^3}-\frac{22}{\eta_\chi^4}-2
+ {\bar \varsigma}^{m}_{zx,1} \left[ \ln \bigg(\frac{\Lambda} {\mu} \bigg)
+ \ln \left( \frac {\eta_\chi+1} {\eta_\chi}  \right) \right ] \quad
{\bar \varsigma}^{m}_{zx,1} = \frac{- \,4 \left(\eta_\chi^4
+2 \,\eta_\chi^2-3\right)} {\eta_{\chi }^4} \,.
\end{align}
Hence, all of them are logarithmically divergent. The consequences of the above expressions are discussed in Sec.~\ref{set1_offplane_type2} of the main text, and illustrated in Fig.~\ref{figzxset1}.

\section{Set-up II --- $ \mathbf E = E \, {\mathbf{\hat x}} $, $\mathbf B = B_x \, {\mathbf{\hat x}} + B_z \, {\mathbf{\hat z}} $}
\label{appset2}

In set-up II, as shown in Fig.~\ref{figsetup}(b), the tilt-axis is perpendicular to $\mathbf E $, but not to $\mathbf B $. We choose $ \hat{\mathbf r}_E =  {\mathbf{\hat x}}  $ and $\hat{\mathbf r}_B = \cos \theta \, {\mathbf{\hat x}} 
+ \sin \theta \, {\mathbf{\hat z}}  $, such that $\mathbf E = E \, {\mathbf{\hat x}}$ and $\mathbf B = B_x \, {\mathbf{\hat x}} 
+ B_z \, {\mathbf{\hat z}} \equiv B  \,\hat{\mathbf r}_B  $.
In the following, we will include a prefactor of $\zeta $ ($\upsilon $) for each factor of a component of BC (OMM). This helps us distinguish whether the term originates from BC or OMM or both.

\subsection{Set-up II: Longitudinal components}
\label{appset2long}

The starting expression is captured by
\begin{align}  
\label{eqappset2_long}
  \sigma_{xx}^{\chi, s}   & = \sigma^{\chi, s}_{{\rm Drude}, xx} +
\frac{e^3 \, \tau \,J^3  \, \alpha_J^4 \,v_z\, \chi \, B_z}  {4} 
\int \frac{d \epsilon \, d \gamma } {(2\,\pi)^2 } \,
\frac{k_z \, k_\perp^{4 J-4}} {\epsilon^5}
\left(\zeta \, l^2_{1xx}   + \upsilon \,l^2_{2xx}  \right ) \mathcal J
\nn & \quad  +
\frac{e^4\,\tau\, J^2 \,\alpha_J^4 \,v_z^2  }   {64 }
\int \frac{d \epsilon \, d \gamma } {(2\,\pi)^2 } \,
\frac{  k_\perp^{4 J-6 }}  {\epsilon^8}
\left(\zeta^2 \, t^2_{1xx} +  \upsilon^2 \,t^2_{2xx} + \zeta \,\upsilon \,t^2_{3xx} \right)
\mathcal J  ,\nn 
 l^2_{1xx} & = J \,  \alpha_J^2 \, k_\perp^{2 J} \, \delta (\epsilon -\mu ) \,,
\quad l^2_{2xx}  =  
4  \, (-1)^s   \left( \epsilon^2 - J  \, k_z^2 \,  v_z^2 \right)
\delta (\epsilon -\mu )
+ J \,  \epsilon  \,  \alpha_J^2 \,  k_\perp^{2 J} \delta^\prime (\epsilon -\mu )\,,\nn
t^2_{1xx}  & = 
2  \, J^2  \, B_x^2 \,  k_\perp^2 
\left [ 3  \, \alpha_J^4  \, k_\perp^{4 J} + 8  \,  \epsilon  \,  \alpha_J^2  \, k_\perp^{2 J}
   \left\lbrace (-1)^s  \, \eta_{\chi }  \, k_z  \, v_z-\epsilon \right \rbrace
+ 8  \, \epsilon^2 \left \lbrace 
\epsilon^2 -  2  \, (-1)^s  \, \epsilon \,   \eta_{\chi } \,  k_z  \, v_z
+\eta_{\chi }^2 \,  k_z^2  \, v_z^2 \right \rbrace
  \right ]  \delta (\epsilon -\mu ) 
\nn & \quad +
8 \, J^4  \, B_z^2  \, \alpha_J^4  \, k_z^2\, k_\perp^{4 J} \, \delta (\epsilon -\mu )\,,  \nn
t^2_{2xx}  & = 8  \, B_x^2  \, k_\perp^2  
\left(3 \,  J^2 \,  k_z^4  \, v_z^4 -2 \,  J  \, \epsilon^2 \,  k_z^2 \,  v_z^2
+\epsilon^4\right)
\delta (\epsilon -\mu ) 
+ 8  \, J \,  (-1)^s  \, \epsilon  \,  B_x^2  \, \alpha_J^2  \, k_\perp^{2 J+2} 
 \left(\epsilon^2-3  \, J  \, k_z^2 \,  v_z^2\right) \delta^\prime (\epsilon -\mu )
\nn & \quad
+  3  \, J^2  \, \epsilon^2  \, B_x^2  \, \alpha_J^4  \, k_\perp^{4 J+2} 
 \, \delta^{\prime \prime}(\epsilon -\mu )
+ 32  \, J^2 \,  B_z^2 \,  k_z^2 \left(\epsilon^2-J \,  k_z^2  \, v_z^2\right)^2 
\delta (\epsilon -\mu ) 
\nn & \quad
-32  \, J^3 \,  (-1)^s  \, \epsilon \,   B_z^2  \, \alpha_J^2  \, k_z^2  \, k_\perp^{2 J} 
\left(J  \, k_z^2  \, v_z^2-\epsilon^2\right)
 \delta^\prime(\epsilon -\mu )
 4  \, J^4 \,  \epsilon^2 \,  B_z^2  \, \alpha_J^4  \, k_z^2  \, k_\perp^{4 J} \,
 \delta^{\prime \prime} (\epsilon -\mu )\,,  \nn
t^2_{3xx}  & = 8  \, J  \, B_x^2 \,  k_\perp^2 
\left[ \alpha_J^2 \,  k_\perp^{2 J} \left(3  \, J \,  k_z^2 v_z^2+\epsilon^2\right)
+ 4 \,  J  \, \epsilon  \,  k_z^2 \,  v_z^2 
\left \lbrace (-1)^s  \, \eta_\chi  \,  k_z  \, v_z-\epsilon \right \rbrace
\right ] \delta (\epsilon -\mu ) 
\nn & \quad
-2  \, J^2 \,  (-1)^s  \, \epsilon  \,  B_x^2 \,  \alpha_J^2 \,  k_\perp^{2 J+2} 
 \left [ 3  \, \alpha_J^2  \, k_\perp^{2 J}
 -8  \,   \epsilon  \left [ \epsilon -(-1)^s  \, \eta_\chi  \,  k_z  \, v_z\right ]
\right ] \delta^\prime(\epsilon -\mu )
\nn & \quad
+ 32  \,  J^3 \,B_z^2  \, \alpha_J^2  \, k_z^2 \,k_\perp^{2 J}
 \left(J  \, k_z^2  \, v_z^2-\epsilon^2\right)
\delta (\epsilon -\mu ) 
-8  \, J^4 \,  (-1)^s  \, \epsilon   \, B_z^2  \, \alpha_J^4 
 \, k_z^2  \, k_\perp^{4 J}  \, \delta^\prime (\epsilon -\mu )\,.
\end{align}  
We find that $\sigma_{xx}^{ \chi, s } $ contains linear-in-$B$ as well as quadratic-in-$ B $ terms. 
The former are caused by a nonzero current proportional to $\chi \left( \mathbf B \cdot  {\mathbf{\hat z}} \right ) \mathbf E $ (in agreement with Ref.~\cite{amit-magneto}).

\subsubsection{Results for the type-I phase for $\mu>0$}
\label{appset2long_type1}

For $\mu>0$, only the conduction band contributes in the type-I phase. The net expression is divided up into BC
and OMM as shown below:
\begin{align}
& \sigma_{xx}^{ (\chi, bc) } =
\frac{e^3 \, \tau \, J^2 \,\mu^2  \,\eta_\chi \, \chi  \, B_z  }
{16 \, \pi^{\frac{3} {2} } \, v_z} \left(\frac{\alpha_J}{\mu }\right)^{\frac{2} {J} }
 \,\ell^{bc, 2}_{xx} 
+ \frac{e^4\, \tau \, v_z} {3840 \, \pi^{\frac{3} {2}} \, J^3  \,\eta_\chi^4} 
\left(\frac{\alpha_J}{\mu }\right)^{\frac{2} {J} }  B_x^2 \,\ell^{bc,2}_{xx,x}  
+ 
\frac{e^4  \, \tau \, J^5  \, \mu^2  } {64  \, \pi  \,  v_z} 
\left(\frac{\alpha_J} {\mu }\right)^{\frac{4} {J} }
B_z^2 \, \ell^{bc,2}_{xx,z} \,,
\nn & \sigma_{xx}^{ (\chi, m) } = 
\frac{e^3 \, \tau \, J^2 \,\mu^2  \,\eta_\chi \, \chi  \, B_z  }
{16 \, \pi^{\frac{3} {2} } \, v_z} \left(\frac{\alpha_J}{\mu }\right)^{\frac{2} {J} }
\,\ell^{m, 2}_{xx} 
 + \frac{e^4\, \tau \, v_z} {3840 \, \pi^{\frac{3} {2} }\, J^3  \,\eta_\chi^4} 
\left(\frac{\alpha_J}{\mu }\right)^{\frac{2} {J} }   B_x^2 
 \,\ell^{m,2}_{xx,x}  
+ 
\frac{e^4  \, \tau \, J^5  \, \mu^2  } {64  \, \pi  \,  v_z}
\left(\frac{\alpha_J}{\mu }\right)^{\frac{4} {J} } B_z^2 \, \ell^{m,2}_{xx,z} \,.
\end{align}
Here, $\ell^{bc, 2}_{xx}$, $\ell^{bc,2}_{xx,x}$, and $\ell^{bc,2}_{xx,z}$
designate the BC-only (i.e., $l^2_{1xx}$- and $ t^2_{1xx}$-contributed) parts accompanying $\chi \, B_z $, $B_x^2 $, and $ B_z^2 $, respectively. Similarly, $\ell^{m}_{xx, 23} $, $\ell^{m,2}_{xx,x}$, and $\ell^{m,2}_{xx,z}$ demarcate the OMM [i.e., $l^2_{2xx}$- and $ (t^2_{2xx}+t^2_{3xx} )$-contributed] parts proportional to $\chi \, B_z $, $B_x^2 $, and $ B_z^2 $, respectively.

First, let us spell out the explicit forms of the linear-in-$B$ terms, which turn out to be:
\begin{align}
\label{ellxxset2linear}
& \ell^{bc, 2}_{xx} =  \frac{(2-3 \, J) \, \Gamma \left(3-\frac{1}{J}\right)
\,_2\tilde{F}_1\left(2-\frac{1}{J},\frac{5}{2}-\frac{1}{J};\frac{9}{2}-\frac{1}{J}; \eta_\chi^2\right)
}  {2} \,,\nn
& \ell^{m, 2}_{xx} =  \frac{ 2 \, \Gamma \left(2-\frac{1}{J}\right)
\,\Big [
(3 \,J+2) \;_2F_1\left(2-\frac{1}{J},\frac{5}{2}-\frac{1}{J};\frac{7}{2}-\frac{1}{J}; \eta_\chi^2\right)
-3  \,J \;_3F_2\left(\frac{5}{2},2-\frac{1}{J},\frac{5}{2}-\frac{1}{J};\frac{3}{2},
\frac{9}{2}-\frac{1}{J}; \eta_\chi^2\right)  }
{(5 \,J-2) \,\Gamma \left(\frac{3}{2}-\frac{1}{J}\right)} \,.
\end{align}
Since it is difficult to decipher the $J$-dependent behaviour from the above generic expressions, we illustrate them via the curves in Fig.~\ref{figxxset2type1} (see the top panel) of Sec.~\ref{set2long_type1}.

Next, let us elaborate on the forms of the terms accompanying $B_x^2$ in the following:
\begin{align}
\label{ellxxset2Bx}
& \frac{ \ell^{bc,2}_{xx,x}} { (J-1) \, J^2 \, \Gamma \left(\frac{J-1}{J}\right) }
=
\Big [ J  \,  \eta_\chi^4 
\left \lbrace  \,  J   \, \eta_\chi^2+J  \, (122  \, J-115)+207\right \rbrace
-4 \,  (J-2)  \, (2  \, J-1)  \, (9 \,  J-7)  \,  \eta_\chi^2
\nn & \hspace{ 3.5 cm}
+\frac{(J-2)  \, (2 \,  J-1)  \, (3 \,  J-1) (5  \, J-2)} {J}
 \Big ] \,_2\tilde{F}_1\left(\frac{1}{2}-\frac{1}{J},\frac{J-1}{J};\frac{5}{2}-\frac{1}{J};
\eta_\chi^2\right)
\nn & \hspace{ 3.25 cm}
+ \left  [
 -30  \, J^2 + 8  \, J  \, (2 \,  J+9)  \,  \eta_\chi^4
+ \left \lbrace J  \, (54  \, J-77)+25 \right \rbrace   \eta_\chi^2
+37 \,  J+\frac{2}{J}-15
\right ]
\nn & \hspace{ 3.75 cm} \times
 \left(\eta_\chi^2-1\right) \,(2-J)
\,_2\tilde{F}_1\left(\frac{3}{2}-\frac{1}{J},\frac{J-1}{J};\frac{5}{2}-\frac{1}{J};\eta_\chi^2\right),
\nn &  \frac{\ell^{m,2}_{xx,x}} 
{\frac{ \sqrt{\pi} \,  J^2 \,(7 \,J-2)  \,
\Gamma \left(6-\frac{2}{J}\right)}
{ 2^{ 6 -\frac{2}{J} } \,
(2 \, J-1)  \,\Gamma\left(\frac{9}{2}-\frac{1}{J}\right)}} =
\Big [
\frac{(J-2)^2 \, (2 \, J-1) \, (3 \, J-1) \, (4 \, J-1) \, (5\,  J-2)}
{J^2}
+ \left [ J \left \lbrace J 
\left (2 \,  J\,  (32 \, J+57)-461 \right ) + 59 
\right \rbrace + 30 \right ]  \eta_\chi^4
\nn & \hspace{ 3.75 cm} 
-\frac{4 \, (J-2) \, (2 \, J-1) \,  (3 \, J-2) \left \lbrace J \, (2 \, J-9)+3 \right \rbrace
\, \eta_\chi^2}{J} \Big ]
\,_2\tilde{F}_1\left(\frac{1}{2}-\frac{1}{J},\frac{J-1}{J};\frac{5}{2}-\frac{1}{J};
 \eta_\chi^2\right)
\nn & \hspace{ 3.5 cm} + \Big [
217 -99 \,  \eta_\chi^2 
+ J \left \lbrace
-120 \, J^2+32\,  (\, (J-7) J+2) \,  \eta_\chi^4
+ (157-2 \, J \, (12 \, J+25) ) \,  \eta_\chi^2
+ 418  \,  J-453 \right \rbrace
\nn & \hspace{ 4.2 cm}
+\frac{4}{J^2}+\frac{6 \left(3 \,  \eta_\chi^2-8\right)}{J}
\Big ] \,
(J-2) \left( \eta_\chi^2-1\right) \,
  _2\tilde{F}_1\left(\frac{3}{2}-\frac{1}{J},\frac{J-1}{J};\frac{5}{2}-\frac{1}{J};
    \eta_\chi^2\right).
\end{align}
Finally, on evaluating the terms accompanying $B_z^2$, we obtain
\begin{align}
\label{ellxxset2Bz}
 \ell^{bc ,2}_{xx,z} & = 
\Gamma \left(4-\frac{2}{J}\right) \,
  _3\tilde{F}_2\left(\frac{3}{2},\frac{3}{2}-\frac{2}{J},2-\frac{2}{J};
   \frac{1}{2},\frac{11}{2}-\frac{2}{J};
   \eta_\chi^2\right), \nn
 \ell^{m ,2}_{xx,z} &  =  \begin{cases}
&  - \, \frac{ 8} {5 \,\pi }   \hspace{ 11.75 cm} \text{ for }  J=1 \\ & \\
& \frac{ 3 \, \Gamma \left(2-\frac{2}{J}\right)}
{4  \,\sqrt{\pi }  \,J^2  \,\Gamma \left(\frac{11}{2}-\frac{2}{J}\right)}
\, \Big [
\frac{(7  \,J-4) \, (9 \, J-4)} {3} \left(
3-\frac{2  \,(J+2)}{J^2}\right) \,
  _3F_2\left(\frac{3}{2},\frac{3}{2}
   -\frac{2}{J},2-\frac{2}{J};\frac{1}{2},\frac{7}{2}-\frac{2}{J};\eta_\chi^2\right)
\Big ]
\nn & \hspace{ 2.5  cm }  
+ \, 5 \, (3  \,J-4) \, (5  \,J-2) \,_3F_2\left(\frac{7}{2},\frac{3}{2}-\frac{2}{J},2-\frac{2}{J};
\frac{1}{2},\frac{11}{2}-\frac{2}{J};\eta_\chi^2\right)
\nn & \hspace{ 2.5  cm }
-2 \, (7  \,J-10)  \,(9  \,J-4) \,
  _3F_2\left(\frac{5}{2},\frac{3}{2}-\frac{2}{J},2-\frac{2}{J};\frac{1}{2},
\frac{9}{2}-\frac{2}{J}; \eta_\chi^2\right) 
\Big ]  \hspace{1.1 cm } \text{ for } J>1
\end{cases} \;.
\end{align}
Since all these final expressions are quite complicated, we illustrate the net behaviour by plotting the corresponding curves in Fig.~\ref{figxxset2type1}. The resulting behaviour is discussed in Sec.~\ref{set2long_type1} of the main text.

\subsubsection{Results for the type-II phase for $\mu>0$}
\label{appset2long_type2}

In the type-II phase, both the conduction and valence bands contribute for any given $ \mu $. The contributions are divided up into BC-only and OMM parts as
\begin{align}
 \sigma_{xx}^{ (\chi, bc) } & =
\frac{e^3 \, \tau \, J^2 \,\mu^2  \,\eta_\chi \, \chi  \, B_z  }
{16 \, \pi^{2} \, v_z} \left(\frac{\alpha_J}{\mu }\right)^{\frac{2} {J} }
 \left( \varrho^{bc, 2}_{xx} + \varsigma^{bc, 2}_{xx} \right)
+ \frac{e^4\, \tau \, J^5\,v_z} {3840 \, \pi^2} 
\left(\frac{\alpha_J}{\mu }\right)^{\frac{2} {J} }  B_x^2 
\left( \varrho^{bc,2}_{xx,x}  +  \varsigma^{bc,2}_{xx,x} \right )
 \nn &  \,\quad
+ \frac{e^4 \,\tau\, J^9 \, \mu^2 } {1920 \,\pi^2 \,v_z}
\left(\frac{\alpha_J} {\mu }\right)^{\frac{4} {J} } B_z^2 
 \left( \varrho^{bc,2}_{xx,z}  +  \varsigma^{bc,2}_{xx,z} \right ) ,
\nn  \sigma_{xx}^{ (\chi, m) } & = 
\frac{e^3 \, \tau \, J^2 \,\mu^2  \,\eta_\chi \, \chi  \, B_z  }
{16 \, \pi^{2} \, v_z} \left(\frac{\alpha_J}{\mu }\right)^{\frac{2} {J} }
\left( \varrho^{m, 2}_{xx} + \varsigma^{m, 2}_{xx} \right)
 + \frac{e^4\, \tau \, J^5\,v_z} {3840 \, \pi^2}
\left(\frac{\alpha_J}{\mu }\right)^{\frac{2} {J} }   B_x^2 
\left( \varrho^{m,2}_{xx,x}  +  \varsigma^{m,2}_{xx,x} \right )
 \nn &  \,\quad 
+ \frac{e^4 \,\tau\, J^9 \, \mu^2 } {1920 \,\pi^2 \,v_z}
\left(\frac{\alpha_J} {\mu }\right)^{\frac{4} {J} } B_z^2 
 \left( \varrho^{m,2}_{xx,z}  +  \varsigma^{m,2}_{xx,z} \right ).
\end{align}
The symbols used above indicate the following: (1) $ \varrho^{bc,2}_{xx} $ ($ \varsigma^{bc,2}_{xx} $) represents the BC-only part proportional to $ \chi B_z $, arising from the $s=1$ ($s=2$) band; (2) $ \varrho^{bc,2}_{xx,x} $ ($ \varsigma^{bc,2}_{xx,x} $) represents the BC-only part proportional to $ B_x^2 $, arising from the $s=1$ ($s=2$) band; (3) $ \varrho^{bc,2}_{xx,z} $ ($ \varsigma^{bc,2}_{xx,z} $) represents the BC-only part proportional to $ B_z^2 $, arising from the $s=1$ ($s=2$) band; (4) $\varrho^{m,2}_{xx} $ ($ \varsigma^{m,2}_{xx} $) represents the OMM part proportional to $ \chi \,B_z $, arising from the $s=1$ ($s=2$) band; (5) $\varrho^{m,2}_{xx,x} $ ($ \varsigma^{m,2}_{xx,x} $) represents the OMM part proportional to $ B_x^2 $, arising from the $s=1$ ($s=2$) band;
(6) $\varrho^{m,2}_{xx,z} $ ($ \varsigma^{m,2}_{xx,z} $) represents the OMM part proportional to $ B_z^2 $, arising from the $s=1$ ($s=2$) band. Since the integrals are quite complicated, the final answers for all the cases are extracted by performing them separately for each value of $J$.

Dealing with the $\propto \chi B_z$ parts, we find that they are divergent in $\Lambda/\mu $, with the degree of divergence increasing with $J$. The expressions for the dominant singular terms are described below:
\begin{enumerate}

\item $J=1$:
\begin{align}
\label{eqlong_type2_lin_j1}
& \left \lbrace \varrho^{bc, 2}_{xx} ,\, \varrho^{m, 2}_{xx},\, \varsigma^{bc, 2}_{xx},
\,  \varsigma^{m, 2}_{xx} \right \rbrace
= \left \lbrace \bar \varrho^{bc, 2}_{xx} ,\, \bar \varrho^{m, 2}_{xx},\,
\bar  \varsigma^{bc, 2}_{xx}, \, \bar  \varsigma^{m, 2}_{xx} \right \rbrace
\ln \bigg(\frac{\Lambda} {\mu} \bigg) + \order{ \left( \Lambda/\mu \right)^0 }\,,
\nn & \bar \varrho^{bc, 2}_{xx}  = \frac{ 1 -\eta_\chi^2} {\eta_\chi^5} \,,\quad
\bar \varrho^{m, 2}_{xx}  = \frac{5 \left(\eta_\chi^2-1\right)} {\eta_\chi^5}
\; \Rightarrow \;
\bar \varrho^{bc, 2}_{xx} + \bar \varrho^{m, 2}_{xx} = \frac{4 \left(\eta_\chi^2-1\right)}{\eta_\chi^5}\,,
\nn & \bar  \varsigma^{bc, 2}_{xx}  = \frac{\eta_\chi^2-1}{\eta_\chi^5}\,,\quad
\bar \varsigma^{m, 2}_{xx}  = \frac{3 \left(1-\eta_\chi^2\right)}{\eta_\chi^5}
\; \Rightarrow \;
\bar \varsigma^{bc, 2}_{xx} + \bar \varsigma^{m, 2}_{xx} = 
\frac{2 \left(1-\eta_\chi^2\right)}{\eta_\chi^5} \,.
\end{align}

\item $J=2$:
\begin{align}
\label{eqlong_type2_lin_j2}
& \left \lbrace \varrho^{bc, 2}_{xx} ,\, \varrho^{m, 2}_{xx},\, \varsigma^{bc, 2}_{xx},
\,  \varsigma^{m, 2}_{xx} \right \rbrace
= \left \lbrace \bar \varrho^{bc, 2}_{xx} ,\, \bar \varrho^{m, 2}_{xx},\,
\bar  \varsigma^{bc, 2}_{xx}, \, \bar  \varsigma^{m, 2}_{xx} \right \rbrace
\frac{\Lambda} {\mu} + \order{ \ln \bigg( {\Lambda} / {\mu} \bigg) }\,,
\nn & \bar \varrho^{bc, 2}_{xx}  = \frac{ -\,2 \left(\eta_\chi^2 - 1\right)^{\frac{3} {2} }}
{\eta_\chi^7}\,,\quad
\bar \varrho^{m, 2}_{xx}  = \frac{4 \left(2 \, \eta_\chi^4-5  \, \eta_\chi^2+3 \right)}
{\eta_\chi^7  \, \sqrt{\eta_\chi^2-1}}
\; \Rightarrow \;
\bar \varrho^{bc, 2}_{xx} + \bar \varrho^{m, 2}_{xx} = 
\frac{2 \left(3  \, \eta_\chi^4-8  \, \eta_\chi^2+5 \right)}
{\eta_\chi^7  \, \sqrt{\eta_\chi^2-1}}\,,
\nn & \bar  \varsigma^{bc, 2}_{xx}  = \frac{2 \left(\eta_\chi^2-1\right)^{\frac{3} {2} }}
{\eta_\chi^7}
\,,\quad
\bar \varsigma^{m, 2}_{xx}  = \frac{4 \sqrt{\eta_\chi^2-1}} {\eta_\chi^7}
\; \Rightarrow \;
\bar \varsigma^{bc, 2}_{xx} + \bar \varsigma^{m, 2}_{xx} = 
\frac{2 \left(\eta_\chi^4-1\right)} {\eta_\chi^7 \, \sqrt{\eta_\chi^2-1}} \,.
\end{align}

\item $J=3$:
\begin{align}
\label{eqlong_type2_lin_j3}
& \left \lbrace \varrho^{bc, 2}_{xx} ,\, \varrho^{m, 2}_{xx},\, \varsigma^{bc, 2}_{xx},
\,  \varsigma^{m, 2}_{xx} \right \rbrace
= \left \lbrace \bar \varrho^{bc, 2}_{xx} ,\, \bar \varrho^{m, 2}_{xx},\,
\bar  \varsigma^{bc, 2}_{xx}, \, \bar  \varsigma^{m, 2}_{xx} \right \rbrace
\bigg(\frac{\Lambda} {\mu} \bigg)^{\frac{4}{3}}
+ \order{  \bigg( {\Lambda} / {\mu} \bigg)^{\frac{1}{3}} }\,,
\nn & \bar \varrho^{bc, 2}_{xx}  =
\frac{ -\, 9 \left( \eta_\chi^2 -1 \right)^{ \frac{5} {3} }}
{4 \,\eta_\chi^{ \frac{23} {3} }}
\,,\quad
\bar \varrho^{m, 2}_{xx}  = 
\frac{3 \left(\eta_\chi^2-1\right)^{ \frac{2} {3} }
 \left(11 \,\eta_\chi^2-19\right)} {4 \,\eta_\chi^{ \frac{23} {3} }}
\; \Rightarrow \;
\bar \varrho^{bc, 2}_{xx} + \bar \varrho^{m, 2}_{xx} = 
\frac{3 \left(\eta_\chi^2-1\right)^{ \frac{2} {3} } 
\left [ 11 \, \eta_\chi^2-3 \left(\eta_\chi^2-1\right)-19\right ]}
{4 \, \eta_\chi^{ \frac{23} {3} }}\,,
\nn & \bar  \varsigma^{bc, 2}_{xx}  = 
\frac{9 \left(\eta_\chi^2-1\right)^{ \frac{5} {3} }}
{4 \,\eta_\chi^{ \frac{23} {3} }}\,,\quad
\bar \varsigma^{m, 2}_{xx}  = 
\frac{3 \left(\eta_\chi^2-1\right)^{ \frac{2} {3} }  \left(3 \,\eta_\chi^2+5\right)}
{4 \,\eta_\chi^{ \frac{23} {3} }}
\; \Rightarrow \;
\bar \varsigma^{bc, 2}_{xx} + \bar \varsigma^{m, 2}_{xx} =
\frac{3 \left(\eta_\chi^2-1\right)^{ \frac{2} {3} } 
 \left(3 \,\eta_\chi^2+1\right)}
{2 \, \eta_\chi^{ \frac{23} {3} }}  \,.
\end{align}

\end{enumerate}
For all $J$'s, we find that $\bar  \varrho^{bc, 2}_{xx} +  \bar  \varsigma^{bc, 2}_{xx} = 0$, such that
the net contribution arises solely from $\bar  \varrho^{m, 2}_{xx} +  \bar  \varsigma^{m, 2}_{xx}$.

Next come the $B_x^2$ parts, which are all non-divergent. Their explicit forms are given below:
\begin{enumerate}

\item $J=1$:
\begin{align}
\label{eqxx_type2_Bx_j1}
& \varrho^{bc,2}_{xx,x} = 128 + 60 \, \eta_\chi^3 + 208 \, \eta_\chi^2
+ 255  \, \eta_\chi +\frac{25}{\eta_\chi}-\frac{7}{\eta_\chi^3}+\frac{3}{\eta_\chi^5}\,, \quad
\varrho^{m,2}_{xx,x} = - \, 64 -128 \,  \eta_\chi^2-225  \, \eta_\chi
+\frac{25}{\eta_\chi}
+\frac{17}{\eta_\chi^3}-\frac{9}{\eta_\chi^5} \,,
\nn &\varsigma^{bc, 2}_{xx, x} = 128 -60  \, \eta_\chi^3 + 208  \, \eta_\chi^2
-255  \, \eta_\chi
-\frac{25}{\eta_\chi}+\frac{7}{\eta_\chi^3}-\frac{3}{\eta_\chi^5} \,,\quad
\varsigma^{m,2}_{xx,x} = 224-64  \, \eta_\chi^2+15  \, \eta_\chi-\frac{215}{\eta_\chi}
+\frac{49} {\eta_\chi^3}-\frac{9}{\eta_\chi^5}\,.
\end{align}

\item $J=2$:
\begin{align}
\label{eqxx_type2_Bx_j2}
& \varrho^{bc,2}_{xx,x} =
\frac{\sqrt{\eta_\chi^2-1} \left( 656 \,  \eta_\chi^6 +
223  \, \eta_\chi^4-294 \,  \eta_\chi^2+120\right)} {32  \, \eta_\chi^6}
+
\frac{15 \left(16  \, \eta_\chi^2+31\right) } {16} 
\cot^{-1}\left(\frac{\eta_\chi-1}{\sqrt{\eta_\chi^2-1}}\right) , \nn &
\varrho^{m,2}_{xx,x} = \frac{\sqrt{\eta_\chi^2-1} 
\left( 42 -128  \, \eta_\chi^4+101  \, \eta_\chi^2 \right)}{8  \, \eta_\chi^4}
+ 
\frac{15}{4} \cot^{-1}\left(\frac{\eta_\chi-1}{\sqrt{\eta_\chi^2-1}}\right),
\nn &\varsigma^{bc, 2}_{xx, x} = \frac{\sqrt{\eta_\chi^2-1} 
\left(656 \,  \eta_\chi^6 +223  \, \eta_\chi^4-294  \, \eta_\chi^2+120\right)}
{32 \,  \eta_\chi^6} \left(16 \,  \eta_\chi^2+31\right) 
-
\frac{15  \left(16 \,  \eta_\chi^2+31 \right)  } {16}
\cot^{-1}\left(\frac{\eta_\chi+1} {\sqrt{\eta_\chi^2-1}}\right),\nn &
\varsigma^{m,2}_{xx,x} = \frac{\sqrt{\eta_\chi^2-1} 
\left( 42 -128 \,  \eta_\chi^4 + 101 \,\eta_\chi^2 \right)} {8  \, \eta_\chi^4}
-\frac{15}{4} \cot^{-1}\left(\frac{\eta_\chi+1}{\sqrt{\eta_\chi^2-1}}\right).
\end{align}

\item $J=3$:
\begin{align}
\label{eqxx_type2_Bx_j3}
& \frac{19683 \, \eta_\chi^{\frac{19}{3}}  \varrho^{bc,2}_{xx,x}} {\sqrt{\pi }}  = 
\frac{160  \,\sqrt{3} \, \Gamma \left(\frac{5}{6}\right) 
\sqrt[3]{\eta_\chi \left(\eta_\chi^2-1\right)}
 \left [ 9 \left(27
  \,  \eta_\chi^4+20  \, \eta_\chi^2-26\right) \eta_\chi^2+91\right ]}
   {\Gamma \left(\frac{4}{3}\right)} \nn
& \hspace{ 2.85 cm} +
\frac{180  \times 2^{\frac{2}{3}}  \, \Gamma \left(\frac{5}{3}\right)
 \left(\eta_\chi+1\right)^{\frac{2}{3}} 
\left(3  \,\eta_\chi^2 \left(81 \eta_\chi^6 + 135 \, \eta_\chi^4
-90  \,\eta_\chi^2+91\right)-91\right) \,
   _2F_1\left(\frac{1}{3},\frac{2}{3};\frac{4}{3};\frac{\eta_\chi+1}{1-\eta_\chi}\right)}
 {\Gamma \left(\frac{7}{6}\right) \sqrt[3]{\eta_\chi^2-1}}   , \nn &
\frac{ 19683  \,\sqrt{3}  \, \eta_\chi^{\frac{19}{3}} 
\sqrt[3]{\eta_\chi^2-1 }  \,
\Gamma \left(\frac{1}{6}\right)  \,\Gamma \left(\frac{1}{3}\right) \varrho^{m,2}_{xx,x} } 
{160 \,\pi^{\frac{3} {2} }} \nn & =
 6 \left(\eta_\chi^2-1\right)^{\frac{2}{3}}
 \left( 59-432  \,\eta_\chi^4 + 432  \,\eta_\chi^2\right) \eta_\chi^{\frac{7}{3}}
+ 2002 \sqrt[3]{\eta_\chi \left(\eta_\chi^2-1\right)^2}
\nn & \, \quad
+
2^{\frac{2}{3}} \left(\eta_\chi+1\right)^{\frac{2}{3}} 
\left [ 9  \,\eta_\chi^2 
\left(468 \, \eta_\chi^4-123  \,\eta_\chi^2+28\right)-1001 \right ]
\, _2F_1\left(\frac{1}{3},\frac{2}{3};\frac{4}{3};\frac{\eta_\chi+1}
{1-\eta_\chi}\right ) ,
\nn &
\frac{19683 \, \eta_\chi^{\frac{19}{3}}  \sqrt[3]{\eta_\chi^2-1} \, \varsigma^{bc, 2}_{xx, x}}
{20  \,  \Gamma \left(\frac{5}{3}\right)}
 =
\frac{54 \,\Gamma \left(\frac{5}{6}\right) \sqrt[3]{\eta_\chi} 
\left(\eta_\chi^2-1\right)^{\frac{2}{3}} 
\left[ 9 \left(27\,  \eta_\chi^4+20 \,\eta_\chi^2-26\right) \eta_\chi^2+91\right ]}
   {\sqrt{\pi }} 
\nn & \hspace{ 4.25 cm }
-\frac{18 \,\Gamma \left(\frac{2}{3}\right) 
\left(\eta_\chi-1\right)^{\frac{2}{3}} 
\left[3 \,\eta_\chi^2 \left(81\, \eta_\chi^6+135\, \eta_\chi^4-90\, \eta_\chi^2+91 \right)
-91\right ] \,
   _2F_1\left(\frac{1}{3},\frac{2}{3};\frac{4}{3};\frac{2}{\eta_\chi+1}-1\right)}
{\Gamma \left(\frac{4}{3}\right)}
 ,\nn &
\frac{ - \,19683 \, \sqrt{\pi }\, \eta_\chi^{\frac{19}{3}}  \sqrt[3]{\eta_\chi^2-1}
\, \varsigma^{m,2}_{xx,x}}
{40 \,\Gamma \left(\frac{2}{3}\right)  \, \Gamma\left(\frac{5}{6}\right)}
 =
 2  \, \sqrt[3]{\eta_\chi} \left(\eta_\chi^2-1\right)^{\frac{2}{3}} 
 \left(2592 \,  \eta_\chi^6-378 \,  \eta_\chi^4-2877  \, \eta_\chi^2+637\right) 
\nn &  \hspace{5 cm} +
 2^{\frac{2}{3}} \left(\eta_\chi-1\right)^{\frac{2}{3}}
  \left[ 9 \left(198  \, \eta_\chi^4+ 87  \, \eta_\chi^2-350\right) 
 \eta_\chi^2+637\right ] \, _2F_1\left(\frac{1}{3},\frac{2}{3};\frac{4}{3};\frac{2}{\eta_\chi+1}-1\right).
\end{align}

\end{enumerate}

Lastly, we have the $\propto B_z^2$ parts, which we find to be divergent in $\Lambda/\mu $ for $J\geq 2$, with the degree of divergence increasing with $J$. The expressions for the regular/singular terms are described below:
\begin{enumerate}

\item $J=1$:
\begin{align}
\label{eqxx_type2_Bz_j1}
\nn & \varrho^{bc, 2}_{xx, z}  =  8 + 5 \,\eta_\chi
+\frac{5}{\eta_\chi^3}-\frac{2}{\eta_\chi^5} \,,\;
 \varrho^{m, 2}_{xx, z}  =  3 \left(- 8 -5 \,\eta_\chi-\frac{5}{\eta_\chi^3}
 +\frac{2}{\eta_\chi^5}-8\right)
\; \Rightarrow \;
\varrho^{bc, 2}_{xx, z} + \varrho^{m, 2}_{xx, z} 
= 2 \left( - 8-5 \,\eta_\chi-\frac{5}{\eta_\chi^3}+\frac{2}{\eta_\chi^5} \right) ,
\nn & \varsigma^{bc, 2}_{xx, z}  = 
8 -5 \,\eta_\chi-\frac{5}{\eta_\chi^3}+\frac{2}{\eta_\chi^5}
\,,\;
\varsigma^{m, 2}_{xx, z}  =
3 \left( 8-5 \,\eta_\chi-\frac{5}{\eta_\chi^3}+\frac{2}{\eta_\chi^5} \right)
\; \Rightarrow \;
\varsigma^{bc, 2}_{xx, z} + \varsigma^{m, 2}_{xx, z} = 
4 \left( 8-5 \,\eta_\chi-\frac{5}{\eta_\chi^3}+\frac{2}{\eta_\chi^5} \right).
\end{align}

\item $J=2$:
\begin{align}
\label{eqxx_type2_Bz_j2}
& \left \lbrace \varrho^{bc, 2}_{xx, z} ,\, \varrho^{m, 2}_{xx, z},\, 
\varsigma^{bc, 2}_{xx, z}, \,  \varsigma^{m, 2}_{xx, z} \right \rbrace
= \left \lbrace \bar \varrho^{bc, 2}_{xx, z} ,\, \bar \varrho^{m, 2}_{xx, z},\,
\bar  \varsigma^{bc, 2}_{xx, z}, \, \bar  \varsigma^{m, 2}_{xx, z} \right \rbrace
\ln \bigg( {\Lambda} / {\mu} \bigg) 
+ \order{  \bigg( {\Lambda} / {\mu} \bigg)^0 },
\nn & \bar \varrho^{bc, 2}_{xx, z}  =
 \frac{15 \left(\eta_\chi^2-1\right)^2} {4 \, \eta_\chi^7}\,,\;
\bar \varrho^{m, 2}_{xx, z}  = \frac{15 \left(\eta_\chi^2-2\right)^2}
{4 \,\eta_\chi^7}
\; \Rightarrow \;
\bar \varrho^{bc, 2}_{xx, z} + \bar \varrho^{m, 2}_{xx, z} = 
\frac{15 \left(2 \,\eta_\chi^4-6 \,\eta_\chi^2+5\right)} {4 \,\eta_\chi^7}\,,
\nn & \bar  \varsigma^{bc, 2}_{xx, z}  = \frac{ - \,15 \left(\eta_\chi^2-1\right)^2}
{4 \,\eta_\chi^7}\,,\;
\bar \varsigma^{m, 2}_{xx, z}  = \frac{15 \left(\eta_\chi^4-4\, \eta_\chi^2+2\right)}
{4 \,\eta_\chi^7}
\; \Rightarrow \;
\bar \varsigma^{bc, 2}_{xx, z} + \bar \varsigma^{m, 2}_{xx, z} = 
\frac{15-30 \,\eta_\chi^2} {4 \,\eta_\chi^7} \,.
\end{align}

\item $J=3$:
\begin{align}
\label{eqxx_type2_Bz_j3}
& \left \lbrace \varrho^{bc, 2}_{xx, z} ,\, \varrho^{m, 2}_{xx, z},\, 
\varsigma^{bc, 2}_{xx, z}, \,  \varsigma^{m, 2}_{xx, z} \right \rbrace
= \left \lbrace \bar \varrho^{bc, 2}_{xx, z} ,\, \bar \varrho^{m, 2}_{xx, z},\,
\bar  \varsigma^{bc, 2}_{xx, z}, \, \bar  \varsigma^{m, 2}_{xx, z} \right \rbrace
\bigg(\frac{\Lambda} {\mu} \bigg)^{\frac{2}{3}}
+ \order{  \bigg( {\Lambda} / {\mu} \bigg)^0 },
\nn & \bar \varrho^{bc, 2}_{xx}  = \frac{10 \left(\eta_\chi^2-1\right)^{\frac{7} {3} }}
{9 \,\eta_\chi^{\frac{25} {3} }}\,,\;
\bar \varrho^{m, 2}_{xx, z}  = \frac{10 \sqrt[3]{\eta_\chi^2-1}
 \, \left( 65 + 17  \,\eta_\chi^4-66  \,\eta_\chi^2 \right)}
 {81  \,\eta_\chi^{\frac{25} {3} }}
\; \Rightarrow \;
\bar \varrho^{bc, 2}_{xx} + \bar \varrho^{m, 2}_{xx} = 
\frac{20 \sqrt[3]{\eta_\chi^2-1} \left(13  \,\eta_\chi^4-42 \, \eta_\chi^2+37\right)}
{81  \,\eta_\chi^{\frac{25} {3} }} \,,
\nn & \bar  \varsigma^{bc, 2}_{xx, z}  = \frac{ - \,10 \left(\eta_\chi^2-1\right)^{\frac{7} {3} }}
{9  \,\eta_\chi^{\frac{25} {3} }} \,,\;
\bar \varsigma^{m, 2}_{xx, z}  = \frac{10 \sqrt[3]{\eta_\chi^2-1} 
\left( 25-7  \,\eta_\chi^4-34  \,\eta_\chi^2 \right)}
{81  \,\eta_\chi^{\frac{25} {3} }}
\; \Rightarrow \;
\bar \varsigma^{bc, 2}_{xx, z} + \bar \varsigma^{m, 2}_{xx, z} =
\frac{160 \sqrt[3]{\eta_\chi^2-1} \left( 1-\eta_\chi^4-\eta_\chi^2 \right)}
{81  \,\eta_\chi^{\frac{25} {3} }} \,.
\end{align}

\end{enumerate}
We note that $ \varrho^{bc, 2}_{xx,z} +   \varsigma^{bc, 2}_{xx,z} =16 $ for $J=1$, and $ \bar \varrho^{bc, 2}_{xx,z} +  \bar \varsigma^{bc, 2}_{xx,z} = 0$ for $J= 2 \text{ and } 3 $.

All the above equations are pictorially represented in Figs.~\ref{figset2long_type2_lin} and \ref{figset2long_type2} of Sec.~\ref{set2long_type2}.

\subsection{Set-up II: In-plane transverse components}
\label{appset2_inplane_trs}

The starting expression is captured by
\begin{align}  
\label{eqappset2_zx}
\sigma_{zx}^{\chi, s}  & =  
\frac{e^3 \, \tau  \, J^2\,\alpha_J^2 \,v_z^2\, \chi \, B_x}  {4} 
\int \frac{d \epsilon \, d \gamma } {(2\,\pi)^2 } \,
\frac{ k_\perp^{2 J-2}} {\epsilon^5}
\left(\zeta \, l^2_{1zx}   + \upsilon \,l^2_{2zx}  \right ) \mathcal J
\nn & \quad  
+ \frac{e^4 \, \tau \, J^3 \, \alpha_J^4 \, v_z^2 \,B_x \, B_z }   {8 }
\int \frac{d \epsilon \, d \gamma } {(2\,\pi)^2 } \,
\frac{  k_\perp^{4 J-4 }}  {\epsilon^8}
\left(\zeta^2 \, t^2_{1zx} +  \upsilon^2 \,t^2_{2zx} + \zeta \,\upsilon \,t^2_{3zx} \right)
\mathcal J  ,\nn 
 l^2_{1zx} & = \alpha_J^2 \, k_\perp^{2 J} 
 \left( k_z \, v_z+\epsilon  \, \eta_\chi\right)
 +2  \,  \epsilon  \left [ (-1)^s  \, \epsilon  
 \left ( \eta_\chi^2 + 1\right) k_z \,  v_z
 -\eta_\chi  \left( k_z^2 \,  v_z^2 + \epsilon^2\right)
 \right ] \delta (\epsilon -\mu ) \,, \nn
 l^2_{2zx} & =
 2  \, k_z \,  v_z \left[ (-1)^s  \, \alpha_J^2  \, k_\perp^{2 J}
 -k_z  \, v_z \left\lbrace (-1)^s  \, k_z  \, v_z-\epsilon  \, \eta_\chi\right \rbrace
 \right ] \delta (\epsilon -\mu ) 
 +
 (-1)^s  \, \epsilon  \,  \alpha_J^2  \, k_\perp^{2 J} 
 \left[ (-1)^s \,  k_z  \, v_z-\epsilon \,  \eta_\chi \right]  
 \delta^\prime (\epsilon -\mu ) \,, \nn 
t^2_{1zx} & = J \left [
2  \, \epsilon  \left(\epsilon^2 -k_z^2  \, v_z^2 \right)
   \left \lbrace \epsilon -(-1)^s \, \eta_\chi  \, k_z  \, v_z\right \rbrace
 - \alpha_J^2 \,  k_\perp^{2 J} \left \lbrace k_z \,  v_z
  \left( (-1)^s  \, \epsilon \, \eta_\chi-2  \, k_z  \, v_z\right)
+ \epsilon^2\right \rbrace
   \right ] \delta (\epsilon -\mu ) \,, \nn
t^2_{2zx} & = \left[ 2  \, (J+2)  \, \epsilon^2 \,  k_z^2  \, v_z^2 
-8  \, J  \, k_z^4  \, v_z^4 \right ] \delta (\epsilon -\mu )  \nn
& \quad + \epsilon \left [
J \,  (-1)^s  \, \alpha_J^2  \, k_\perp^{2 J} 
\left( 4 \,  k_z^2  \, v_z^2-\epsilon^2 \right) 
-2  \, k_z  \,  \, v_z \left(\epsilon^2-2 \,  J \,  k_z^2  \, v_z^2\right) 
\left(\epsilon \eta_\chi-(-1)^s  \, k_z  \, v_z\right)
\right ] \delta^\prime (\epsilon -\mu ) \nn
& \quad + J  \, \epsilon^2  \, \alpha_J^2  \, k_z \,  v_z \,  k_\perp^{2 J}
\left[ k_z \,  v_z-(-1)^s  \, \epsilon  \, \eta_\chi \right] 
\delta^{\prime \prime} (\epsilon -\mu ) \,,\nn
t^2_{3zx} & = \left [ \alpha_J^2 \, k_\perp^{2 J} 
\left\lbrace (J+1)  \, \epsilon^2-4  \, J  \, k_z^2  \, v_z^2 \right \rbrace
+2 \left \lbrace 
\epsilon^2 \,  k_z^2  \, v_z^2 -(-1)^s  \, \epsilon \,  \eta_\chi  \, k_z  \, v_z 
\left(3 \,  J  \, k_z^2  \, v_z^2-2 \,  J \,  \epsilon^2+\epsilon^2 \right)
+2  \, J  \,  k_z^4  \, v_z^4 - J  \, \epsilon^4  \right \rbrace
 \right ] \delta (\epsilon -\mu )    \nn 
& \quad -J  \, \epsilon \left [
\alpha_J^2 \,k_\perp^{2 J} \left \lbrace
(-1)^s \left(2  \, k_z^2  \, v_z^2-\epsilon^2 \right )
-\epsilon  \, \eta_\chi  \, k_z \,  v_z\right \rbrace
+ 2 \,  \epsilon  \,  k_z  \, v_z \left \lbrace 
\epsilon^2 \, \eta_\chi -(-1)^s \epsilon  
  \left(\eta_\chi^2 + 1 \right) k_z  \, v_z + \eta_\chi  \, k_z^2  \, v_z^2
\right \rbrace
\right ] \delta^\prime (\epsilon -\mu ) \,.
\end{align} 
Here, we observe that the conductivity comprises terms which are linear-in-$B$ as well quadratic-in-$ B $. The former are caused by a nonzero current proportional to $\chi \left( \mathbf E \cdot \mathbf B \right) {\mathbf{\hat z}} $.

\subsubsection{Results for the type-I phase for $\mu>0$}
\label{appset2_inplane_type1}

For $\mu>0$, only the conduction band contributes in the type-I phase. The contributions are divided up as shown below:
\begin{align}
 & \sigma_{zx}^{ (\chi, bc) } = 
 \frac{e^3 \, \tau \, J\, v_z} {48 \,\pi^2 \, \eta_\chi^4} \, \chi \,B_x\, \ell^{bc, l}_{zx,2}
 + \frac{e^4 \,\tau\, J \, v_z} {32\, \pi^{ \frac{3} {2} } }
\left (\frac {\alpha_J} {\mu} \right)^{\frac {2} {J}} B_x \, B_z\, \ell^{bc, q}_{zx,2}\,,
\nn &
\sigma_{zx}^{ (\chi, m) } = 
 \frac{e^3 \, \tau \, J\, v_z} {48 \,\pi^2 \, \eta_\chi^4} \, \chi \,B_x\, \ell^{m, l}_{zx,2}
 +
\frac{e^4 \,\tau\, J \, v_z} {32\, \pi^{ \frac{3} {2} } }
\left (\frac {\alpha_J} {\mu} \right)^{\frac {2} {J}} B_x \, B_z\,\ell^{m, q}_{zx,2} \, .
\end{align}
Here, $ \ell^{bc, l}_{zx,2}$ (i.e., $l^2_{1zx}$-contributed) and $ \ell^{bc, q}_{zx,2} $ (i.e., $t^2_{1zx}$-contributed) represent the BC-only parts proportional to $\chi  B_x $ and $B_x  B_z $, respectively. Similarly, $ \ell^{m, l}_{zx,2} $ (i.e., $l^2_{2zx}$-contributed) and $ \ell^{m, q}_{zx,2} $ [i.e., $ ( t^2_{2zx} + t^2_{3zx} ) $-contributed] represent the OMM parts proportional to $\chi B_x $ and $B_x  B_z $, respectively.

The coefficients accompanying $J \chi  B_x $ are $J$-independent and take the following forms:
\begin{align}
\label{eqzx_set2_type1_lin}
& \ell^{bc, l}_{zx,2} = 
-2 \,  \eta_\chi \left(6  \, \eta_\chi^4 - 5  \, \eta_\chi^2 + 3\right) 
6 \left(\eta_\chi^2-1\right)^2 \tanh^{-1} \eta_\chi\,,\quad
\ell^{m, l}_{zx,2} = 30  \, \eta_\chi-26 \,  \eta_\chi^3
-6 \left(\eta_\chi^4-6  \, \eta_\chi^2+5\right) \tanh^{-1}\eta_\chi \,.
\end{align}
Clearly, $\ell^{bc, l}_{zx,2} <0$, $\ell^{m, l}_{zx,2} >0$, and $\ell^{bc, l}_{zx,2} + \ell^{m, l}_{zx,2}<0 $.
As for the $B_x B_z$-coefficients, we end up with
\begin{align}
\label{eqzx_set2_type1_q}
 & \ell^{bc, q}_{zx,2} = \frac{J \, \Gamma \left(3-\frac{1}{J}\right)} {15 \, \eta_\chi^2} \,
 \Big [ 
 \left(  \eta_\chi^2 + 6 \,  J  \, \eta_\chi^2 + 6 \,  J + \frac{4}{J}-14 \right)
 \,_2\tilde{F}_1\left(\frac{1}{2}-\frac{1}{J},\frac{J-1}{J};\frac{7}{2}-\frac{1}{J};\eta_\chi^2\right)
 \nn & \hspace{ 3.25 cm }
+ (J-2) \left(\eta_\chi^2-1\right) \left( 6-9  \, \eta_\chi^2-\frac{2}{J} \right) 
\,_2\tilde{F}_1\left(\frac{3}{2}-\frac{1}{J},\frac{J-1}{J};\frac{7}{2}-\frac{1}{J};\eta_\chi^2\right)
\Big ] ,\nn
& \ell^{m, q}_{zx,2}  =  \frac{ \Gamma \left(2-\frac{1}{J}\right)} {2} \, 
\Big [ 
-2  \, J \,(J-1)  \,_2\tilde{F}_1\left(\frac{1}{2}-\frac{1}{J},\frac{J-1}{J};\frac{5}{2}-\frac{1}{J};
\eta_\chi^2\right)
+ (J-2)  \,(J+4) \, \eta_\chi^2 \;_2\tilde{F}_1\left(\frac{3}{2}-\frac{1}{J},\frac{J-1}{J};
\frac{7}{2}-\frac{1}{J};\eta_\chi^2 \right) 
 \nn & \hspace{ 3 cm } +
\sqrt{\pi }  \,J 
\left \lbrace 2  \,(J-2) \, \eta_\chi^2+3  \,J-7\right \rbrace  \;
  _3\tilde{F}_2\left(\frac{3}{2},\frac{1}{2}-\frac{1}{J},\frac{J-1}{J};\frac{1}{2},\frac{7}{2}-\frac{1}{J};
   \eta_\chi^2\right) 
 \nn & \hspace{ 3 cm } 
-3  \,\sqrt{\pi }  \,(J-2)  \,(4 \, J-1) \,
  _3\tilde{F}_2\left(\frac{5}{2},\frac{1}{2}-\frac{1}{J},\frac{J-1}{J};\frac{1}{2},\frac{9}{2}-\frac{1}{J};
 \eta_\chi^2\right)  
\nn  & \hspace{ 3 cm } 
-\frac{3  \,\sqrt{\pi }  \,(J-2)  \, \left \lbrace J (3 J+8)-4 \right \rbrace 
\eta_\chi^2 } {4  \, J} 
\; _3\tilde{F}_2\left(\frac{5}{2},\frac{J-1}{J},\frac{3}{2}-\frac{1}{J};\frac{3}{2},
 \frac{9}{2}-\frac{1}{J}; \eta_\chi^2\right)
\Big ] .
\end{align}
The resulting characteristics are further elucidated in Sec.~\ref{set2_inplane_type1}. We illustrate the net behaviour for the linear-in-$ B$ parts and $J=3$ node's quadratic-in-$B$ dependence via Fig.~\ref{figzxset2type1}.

\subsubsection{Results for the type-II phase for $\mu>0$}
\label{appset2_inplane_type2}

In the type-II phase, both the conduction and valence bands contribute for any given $ \mu $. The contributions are divided up into BC-only and OMM parts as
\begin{align}
 & \sigma_{zx}^{ (\chi, bc) } = 
 \frac{e^3 \, \tau \, J\, v_z} {48 \,\pi^2 \, \eta_\chi^4} \, \chi \,B_x 
 \left(  \varrho^{bc, l}_{zx,2} + \varsigma^{bc, l}_{zx,2} \right)
 +
 \frac {e^4  \, J^2 \,\tau  \, v_z} {32 \, \pi^2} 
\left (\frac {\alpha_J} {\mu} \right)^{\frac {2} {J} }
B_x \, B_z
\left ( \varrho^{bc, q}_{zx,2} + \varsigma^{bc, q}_{zx,2} \right) \text{ and}
\nn & \sigma_{zx}^{ (\chi, m) } = 
\frac{e^3 \, \tau \, J\, v_z} {48 \,\pi^2 \, \eta_\chi^4}\, \chi \,B_x
 \left(  \varrho^{m, l}_{zx,2} + \varsigma^{m, l}_{zx,2} \right)
 +
\frac {e^4  \, J^2 \,\tau  \, v_z} {32 \, \pi^2}
\left (\frac {\alpha_J} {\mu} \right)^{\frac {2} {J}}
B_x \, B_z
\left ( \varrho^{m, q}_{zx,2} + \varsigma^{m, q}_{zx,2} \right) .
\end{align}
The symbols used above indicate the following: (1) $ \varrho^{bc, l}_{zx,2} $ ($ \varsigma^{bc, l}_{zx,2} $) represents the BC-only part proportional to $ \chi  B_x $, arising from the $s=1$ ($s=2$) band; (2) $ \varrho^{m, l}_{zx,2} $ ($ \varsigma^{m, l}_{zx,2} $) represents the OMM part proportional to $ \chi  B_x $, arising from the $s=1$ ($s=2$) band; (3) $\varrho^{bc, q}_{zx,2} $ ($ \varsigma^{bc, q}_{zx,2} $) represents the BC-only part proportional to $ B_x B_z $, arising from the $s=1$ ($s=2$) band; (4) $\varrho^{m, q}_{zx,2} $ ($ \varsigma^{m, q}_{zx,2} $) represents the OMM part proportional to $ B_x B_z $, arising from the $s=1$ ($s=2$) band.

The coefficients accompanying $J \chi  B_x $ are $J$-independent and are logarithmically divergent.
Defining
$$ \left \lbrace \varrho^{bc, l}_{zx,2}, \, \varrho^{m, l}_{zx,2} , \,
\varsigma^{bc, l}_{zx,2}, \, \varsigma^{m, l}_{zx,2} \right \rbrace 
= \left \lbrace 
\bar \varrho^{bc, l}_{zx,2}, \,\bar \varrho^{m, l}_{zx,2} , \,
\bar \varsigma^{bc, l}_{zx,2}, \, \bar \varsigma^{m, l}_{zx,2}
  \right \rbrace  \ln(\Lambda/ \mu) 
+ \order{ \left( {\mu}/ {\Lambda}  \right)^0} \,, $$
we have
\begin{align}
\label{eqzx_set2_type2_lin}
& {\bar \varrho}^{bc, l}_{zx,2} = 3 \left(\eta_\chi^2-1\right)^2\,,\quad
{\bar \varrho}^{m, l}_{zx,2} = 3 \left(5-\eta_\chi^2\right) \left(\eta_\chi^2-1\right) \quad 
\Rightarrow  \quad
{\bar \varrho}^{bc, l}_{zx,2}  +  {\bar \varrho}^{m, l}_{zx,2}  = 2 \left(\eta_\chi^2-1\right),
\nn & {\bar \varsigma}^{bc, l}_{zx,2} = -3 \left(\eta_\chi^2-1\right)^2\,,\quad
{\bar \varsigma}^{m, l}_{zx,2} = -3 \left(\eta_\chi^2-1\right) \left(\eta_\chi^2+3\right)\quad 
\Rightarrow  \quad
{\bar \varsigma}^{bc, l}_{zx,2}  +  {\bar \varsigma}^{m, l}_{zx,2}  = 
6\left (1- \eta_\chi^4 \right) .
\end{align}
Clearly, $\bar \varrho^{bc, l}_{zx,2} > 0$, $\bar \varrho^{m, l}_{zx,2}  $ goes from positive to negative at $\eta_\chi = \sqrt 5 $, and $ \bar \varrho^{bc, l}_{zx,2}  + \bar \varrho^{m, l}_{zx,2}> 0 $. Furthermore, 
$\bar \varsigma^{bc, l}_{zx,2} = - \,\bar \varrho^{bc, l}_{zx,2} $ and $\bar  \varsigma^{m, l}_{zx,2} <0 $. The sum over the two bands shows that the BC-only part is always zero, so that the OMM parts yield the net contribution of $-\,6\left (\eta_\chi^2 - 1 \right)^2$, which is of course negative.

As for the $B_x B_z$-coefficients, since the integrals are quite complicated, the final expressions are extracted by performing them separately for each value of $J$, which turn out to be divergent. The three cases are discussed below, evaluated upto $ \order{ \left({\mu}/ {\Lambda} \right)^0} $:
\begin{enumerate}

\item $J=1$: The $s=1$ band has completely convergent/non-singular contributions. However, the $s=2$ band gives a singularity in the form of a logarithmic divergence, which comes from the BC-only part (i.e., the OMM part is non-singular). The logarithmic part (arising from $ \varsigma^{bc, q}_{zx,2} $), thus, dominates the overall response, which amounts to
$ \ln (\Lambda/\mu) \; {\mu^2 \left(\eta_\chi^2-1\right)^2} / {\eta_\chi^5} $.

\item $J=2$: Although the $s=1$ band has completely convergent/non-singular contributions, the $s=2$ band's contributions contain singularities in the forms of a logarithmic divergence and a linear-in-$\Lambda/\mu$ term, which exclusively originate from the BC-only part (i.e., the OMM part is non-singular). The dominating contribution then comes from $ \varsigma^{bc, q}_{zx,2} $, whose singular parts are captured by
\begin{align}
\frac{4 \,\mu^2 \left(\eta_\chi^2-1\right)^{ \frac{5} {2} }}{\eta_\chi^7} \,\frac{\Lambda}{\mu}
+ \frac{4 \, \mu  \left(\eta_\chi^2-1\right)^{ \frac{3} {2} } 
\left [\mu  \left(\eta_\chi^2-5\right)-6\right ]}
{\eta_\chi^6} \,\ln \bigg( \frac{\Lambda}{\mu} \bigg) \,.
\end{align}

\item $J=3$: Again, the $s=1$ band has completely convergent/non-singular contributions. On the other hand, the $s=2$ band's contributions harbour singularities that go as $\Lambda^{ \frac{4} {3} }$ and $\Lambda^{1/3}$, which originate from the BC-only part (i.e., the OMM part is non-singular). The dominating contribution comes from $ 
\varsigma^{bc, q}_{zx,2}  $, whose singular parts are now captured by
\begin{align}
\frac{27 \,\mu^2 \left(\eta_\chi^2-1\right)^{ \frac{8} {3} }}
{4 \, \eta_\chi^{\frac{23} {3} }  }   \bigg( \frac{\Lambda}{\mu} \bigg)^{ \frac{4} {3} }
+
\frac{9 \, \mu  \left(\eta_\chi^2-1\right)^{ \frac{5} {3} } 
\left [(3 \, \mu -1) \,\eta_\chi^2-2\, (8 \,\mu + 9 )\right] }
{\eta_\chi^{ \frac{20} {3} }}   \bigg( \frac{\Lambda}{\mu} \bigg)^{ \frac{1} {3} } .
\end{align}

 \end{enumerate}
The above details explain the summaries discussed in Sec.~\ref{set2_inplane_type2}.

\subsection{Set-up II: Out-of-plane transverse components}
\label{appset2_outplane}

All the out-of-plane components vanish, i.e., $ \sigma_{yx}^{ \chi, s } =  0\,.$

\section{Set-up III --- $\mathbf E = E \, {\mathbf{\hat z}}$, $\mathbf B = B_x \, {\mathbf{\hat x}} + B_z \, {\mathbf{\hat z}}  $}
\label{appset3}

In set-up III, as shown in Fig.~\ref{figsetup}(b), the tilt-axis is parallel to $\mathbf E $, but not to $\mathbf B $. We choose $ \hat{\mathbf r}_E =  {\mathbf{\hat z}}  $ and $\hat{\mathbf r}_B = \cos \theta \, {\mathbf{\hat x}} 
+ \sin \theta \, {\mathbf{\hat z}}  $, such that $\mathbf E = E \, {\mathbf{\hat z}}$ and $\mathbf B = B_x \, {\mathbf{\hat x}} 
+ B_z \, {\mathbf{\hat z}} \equiv B  \,\hat{\mathbf r}_B  $.
In the following, we will include a prefactor of $\zeta $ ($\upsilon $) for each factor of a component of BC (OMM). This helps us distinguish whether the term originates from BC or OMM or both.

\subsection{Set-up III: Longitudinal components}
\label{appset3long}

The starting expression is captured by
\begin{align}  
\label{eqappset3_long}
  \sigma_{zz}^{\chi, s}  & =  \sigma^{\chi, s}_{{\rm Drude}, zz} +
\frac{e^3 \, \tau \,J^2 \, \alpha_J^2 \,v_z^2 \, \chi \, B_z}  {2} 
\int \frac{d \epsilon \, d \gamma } {(2\,\pi)^2 } \,
\frac{ k_\perp^{2 J-2}} {\epsilon^5}
\left(\zeta \, l^3_{1zz}   + \upsilon \,l^3_{2zz}  \right ) \mathcal J
\nn & \quad  
+\frac{e^4 \, \tau \, J^2 \, \alpha_J^4 \, v_z^2}   { 16 }
\int \frac{d \epsilon \, d \gamma } {(2\,\pi)^2 } \,
\frac{  k_\perp^{4 J-4}}  {\epsilon^8}
\left(\zeta^2 \, t^3_{1zz} +  \upsilon^2 \,t^3_{2zz} + \zeta \,\upsilon \,t^3_{3zz} \right)
\mathcal J ,\nn 
 l^3_{1zz} & = \left[
 (-1)^s  \, \epsilon^2 \left(\eta_\chi^2+2\right) k_z  \, v_z
 + k_z^3  \, v_z^3-2 \,  \epsilon^3  \,  \eta_\chi  
 \right ]  \delta (\epsilon - \mu)  \,,
\nn l^3_{2zz}  & = 
2 \left ( 2 \, k_z^2  \,v_z^2-\epsilon^2\right) 
\left[ (-1)^s \, k_z  \,v_z-\epsilon  \, \eta_\chi \right] \delta (\epsilon - \mu) 
 +
\epsilon  \, k_z  \,v_z \left[
\epsilon^2  \,\eta_\chi^2 - 2  \,(-1)^s  \,\epsilon   \,\eta_\chi  \, k_z  \, v_z
+ k_z^2  \,v_z^2  \right ] \delta^\prime (\epsilon - \mu)   \,,\nn
t^3_{1zz}  & = 4 \, J^2 \,B_z^2 \left(\epsilon^2-k_z^2 \,  v_z^2\right)^2 \delta (\epsilon - \mu)
+
2 \,  B_x^2  \, k_\perp^2  \, v_z^2 
\left[ \epsilon^2  \, \eta_\chi^2 - 2   \, (-1)^s  \, \epsilon  \,  \eta_\chi  \,  k_z  \, v_z
+ k_z^2 \,  v_z^2 \right ] \delta (\epsilon - \mu)\,,  
\nn t^3_{2zz}  & =  \left[ 4 \,  J^2  \, B_z^2 \left(\epsilon^2-2 \,  k_z^2  \, v_z^2\right)^2  
+ 8  \, B_x^2  \, k_\perp^2  \, k_z^2  \, v_z^4 \right ] \delta (\epsilon - \mu) 
\nn & \quad  +
+ 8  \, \epsilon  \,  k_z \,  v_z 
\left[ (-1)^s  \, k_z  \, v_z-\epsilon  \,  \eta_\chi \right ]
\left [ J^2  \, B_z^2 \left(2  \, k_z^2 \,  v_z^2-\epsilon^2\right)
+ B_x^2 \,  k_\perp^2 \,  v_z^2 \right  ] \delta^\prime (\epsilon - \mu) 
\nn & \quad
+ \epsilon^2  \,v_z^2 \left(2  \,J^2  \,B_z^2 \, k_z^2+B_x^2  \,k_\perp^2\right) 
\left [ \epsilon^2  \,\eta_\chi^2 
-2 (-1)^s \epsilon  \, \eta_\chi  \, k_z  \,v_z + k_z^2  \,v_z^2
\right ] \delta^{\prime \prime} (\epsilon - \mu) \,,  
\nn t^3_{3zz}  & =
-\, 4  \, J \,  B_z^2 \left[ 2 \, J \,  \epsilon^4
-2  \, (J-2) \,  (-1)^s \,  \epsilon^3  \, \eta_\chi  \,  k_z  \, v_z
-4 \,  (J+1) \,  \epsilon^2  \, k_z^2  \, v_z^2 + 4  \,  J  \, k_z^4  \, v_z^4 
\right ] \delta (\epsilon - \mu)
\nn & \quad
 \,-8  \, B_x^2 \,  k_\perp^2  \, k_z  \, v_z^3 
\left[ k_z  \, v_z-(-1)^s  \, \epsilon  \,  \eta_\chi \right ] \delta (\epsilon - \mu)
\nn & \quad
\,-4  \, J^2  \, \epsilon  \,  B_z^2  \, k_z \,  v_z \left [
(-1)^s \left \lbrace k_z^3  \, v_z^3-\epsilon^2 \left(\eta_\chi^2+2\right) 
k_z \,  v_z\right \rbrace
+2  \, \epsilon^3  \, \eta_\chi \right ] \delta^\prime (\epsilon - \mu) 
\nn & \quad
\,-2  \, \epsilon \,   B_x^2  \, k_\perp^2  \,  v_z^2 
\left[ (-1)^s \left(k_z^2  \, v_z^2 + \epsilon^2  \, \eta_\chi^2\right)
-2  \, \epsilon  \,  \eta_\chi \, k_z \,  v_z\right ] \delta^\prime (\epsilon - \mu)\,.
\end{align}  
We find that $\sigma_{zz}^{ \chi, s } $ contains linear-in-$B$ as well as quadratic-in-$ B $ terms. 
The former are caused by nonzero currents proportional to $\chi \left( \mathbf E \cdot \mathbf B \right) {\mathbf{\hat z}} $ and/or $\chi \left( \mathbf B \cdot  {\mathbf{\hat z}} \right ) \mathbf E $ (since, of course, $\mathbf E $ is parallel to the tilt-axis).

\subsubsection{Results for the type-I phase for $\mu>0$}
\label{appset3long_type1}

For $\mu>0$, only the conduction band contributes in the type-I phase. The contributions are divided up as shown below:
\begin{align}
 & \sigma_{zz}^{ (\chi, bc) } =  \frac {e^3 \, J\, \tau \, v_z}
{ 4 \, \pi^2} \, \chi  \, B_z \,\ell^{bc, 3}_{zz}
+
 \frac {e^4 \, \tau \, J   \, v_z^3}
{ 128 \, \pi^2\, \mu^2} \, B_x^2 \,\ell^{bc, 3}_{zz, x}
+ \frac {e^4 \, \tau \, J^3\, v_z} {16\, \pi^{\frac{3}{2}} } 
\left (\frac {\alpha_J} {\mu} \right)^{\frac {2}  {J}}
 B_z^2 \,\ell^{bc, 3}_{zz, z}\,,
\nn & \sigma_{zz}^{ (\chi, m) } = \frac {e^3 \, J\, \tau \, v_z}
{ 4 \, \pi^2} \, \chi  \, B_z \,\ell^{m, 3}_{zz}
+
 \frac {e^4 \, \, \tau\, J \, \tau  \, v_z^3}
{ 128 \, \pi^2\, \mu^2} \, B_x^2 \,\ell^{m, 3}_{zz, x}
+ \frac {e^4 \, \tau \, J^3\, v_z} {16\, \pi^{\frac{3}{2}} } 
 \left (\frac {\alpha_J} {\mu} \right)^{\frac {2}  {J}}
B_z^2 \,\ell^{m, 3}_{zz, z} \,.
\end{align}
Here, $ \ell^{bc, 3}_{zz} $, $ \ell^{bc, 3}_{zz, x} $, and $ \ell^{bc, 3}_{zz, z} $
represent the BC-only (i.e., $l^3_{1zz}$- and $t^3_{1zz}$-contributed) parts proportional to $\chi B_z $, $B_x^2 $, $ B_z^2 $, respectively. Similarly, $ \ell^{m, 3}_{zz} $, $ \ell^{m, 3}_{zz, x} $, and $ \ell^{m, 3}_{zz, z} $ represent the OMM [i.e., $l^3_{2zz}$- and $( t^3_{2zz} + t^3_{3zz} )$-contributed] parts proportional to $\chi B_z $, $B_x^2 $, $ B_z^2 $, respectively.

The coefficients accompanying $J \chi  B_z $ are $J$-independent and take the following forms:
\begin{align}
\label{eqzz_set3_type1_lin}
& \ell^{bc, 3}_{zz} = - \, \eta_\chi -\frac{5} {3  \, \eta_\chi }+\frac{1}{\eta_\chi^3}
-\frac{\left(\eta_\chi^2-1\right)^2 \, \tanh^{-1} \eta_\chi  }
{\eta_\chi^4}\,,\quad
\ell^{m, 3}_{zz} = - \, \eta_\chi +\frac{19} {3 \,  \eta_\chi }-\frac{5} {\eta_\chi^3}
+ \left( 3 -\frac{8} {\eta_\chi^2}+\frac{5} {\eta_\chi^4} \right) \tanh^{-1} \eta_\chi  .
\end{align}
Clearly, $ \ell^{bc, 3}_{zz} <0 $, $ \ell^{m, 3}_{zz} > 0$, and $ \ell^{bc, 3}_{zz} + \ell^{m, 3}_{zz} < 0 $.
Likewise, the coefficients accompanying $ J B_x^2 $ are also $J$-independent and can be expressed as:
\begin{align}
\label{eqzx_set2_type1_Bx}
& \ell^{bc, 3}_{zz, x} =  \frac{16 \left( 1 + 7 \, \eta_\chi^2 \right)}{15} \,,\quad
\ell^{m, 3}_{zz, x} =  \frac{ - \, 16 \left( 3 + \eta_\chi^2 \right)}{15} 
\quad
\Rightarrow \quad
\ell^{bc, 3}_{zz, x}  + \ell^{m, 3}_{zz, x} = 
\frac{- \,32\left( 1 - 3 \, \eta_\chi^2 \right ) }  {15}  \,.
\end{align}
The above equations imply that $\ell^{bc, 3}_{zz, x} > 0$ and $  \ell^{m, 3}_{zz, x} < 0$. Their sum goes from negative to positive at $\eta_\chi = 1/\sqrt 3$.
Finally, for the $ B_z^2 $-coefficients, we end up with
\begin{align}
\label{eqzz_set3_type1_q}
  \ell^{bc, 3}_{zz,z} &= 
 \Gamma \left(4-\frac{1}{J}\right) \,
  _2\tilde{F}_1\left(\frac{1}{2}-\frac{1}{J},\frac{J-1}{J};\frac{9}{2}-\frac{1}{J};\eta
   _{\chi }^2\right) ,\nn
 \ell^{m, 3}_{zz,z}  & = \begin{cases}
& -\, \frac{2  \, \left( 4 \,+ \, 7 \,  \eta_\chi^2\right) }   {15 \, \sqrt{\pi }}
  \hspace{ 11.8 cm} \text{ for }  J=1 \\ & \\
&   \Big [
6 \left(28  \, \eta_\chi^6 - 97  \, \eta_\chi^4+130 \,  \eta_\chi^2-60\right)
+\frac{ 948  \, \eta_\chi^2 - 342  \, \eta_\chi^4 -578}{J^4}
+\frac{ 1557 -246  \, \eta_\chi^6+1935  \, \eta_\chi^4-3272 \,\eta_\chi^2 }{J^3}   
\\ &
\; \;   +\frac{783  \, \eta_\chi^6-3834  \, \eta_\chi^4+5208  \, \eta_\chi^2-2195}{J^2}
+\frac{ 1494-666 \,  \eta_\chi^6 + 2910  \, \eta_\chi^4-3638  \, \eta_\chi^2 }{J}
-\frac{8}{J^6}+\frac{108-104   \, \eta_\chi^2}{J^5}
\Big ]\, 
\frac{ - \,\left(1-\eta_\chi^2\right)^{\frac{1}{J}}}
{45   \, \eta_\chi^6   \, \Gamma \left(\frac{1}{2}-\frac{1}{J}\right)} \\
 & + \,  \Big [
(J-2)^2  \, ( 1-2  \, J)  \, (3  \, J-2)  \, (3  \, J-1) \,  (4  \, J-1) \,  (5 \,  J-2)
\\ & \quad \;
+ \, 3  \, J\, (J-2)  \,  (2  \, J-1) \,  (3  \, J-2) 
 \left \lbrace J  \, (J  \, (50  \, J-167)+103) \, -18 \right \rbrace
 \eta_\chi^2
\\ & \quad \;
- \,15  \,J^2 \,  (J-2)  \left \lbrace J  \, 
(2 \,  J  \, (J  \, (27 \,  J-128)+159)- 153)+26 \right \rbrace \eta_\chi^4
\\ & \quad \;
+ \,15 \, J^3 \, (J-1)  \left \lbrace J \,  (2  \, J  \, (9 \,J-55)+149)
-50 \right \rbrace \eta_\chi^6
+ 45  \, J^4  \, (J-2)  \,  (2 \,  J-1)  \, \eta_\chi^8 \Big ]
 \\& \quad \times 
\frac{\Gamma \left(6-\frac{2}{J}\right) \,
   _2\tilde{F}_1\left(\frac{1}{2}-\frac{1}{J},\frac{J-1}{J};\frac{3}{2}-\frac{1}{J}
   ;\eta_\chi^2\right)}
 {2880 \,J^6 \, \eta_\chi^6 
  \, \Gamma \left(\frac{7}{2}-\frac{1}{J}\right)}
\, \frac{4^{\frac{1}{J}} \, \sqrt\pi} {2 \, J-1} 
\hspace{ 7.8 cm } \text{ for } J>1
\end{cases} \,.
\end{align}
The resulting characteristics are further elucidated in Sec.~\ref{set3long_type1}. We illustrate the net behaviour for the linear-in-$ B$ parts and $J=3$ node's quadratic-in-$B$ dependence via Fig.~\ref{figzzset3type1}.

\subsubsection{Results for the type-II phase for $\mu>0$}
\label{appset3long_type2}

In the type-II phase, both the conduction and valence bands contribute for any given $ \mu $. The contributions are divided up into BC-only and OMM parts as
\begin{align}
 \sigma_{zz}^{ (\chi, bc) } & = \frac{e^3 \, \tau \,J \,  v_z}
 {4 \, \pi^2} \, \chi  \, B_z
 \left( \varrho^{bc, 3}_{zz} + \varsigma^{bc, 3}_{zz} \right)
+ \frac{e^4 \, \tau \, J\, v_z^3} {128\, \pi^2 \,\mu^2}\, B_x^2 
\left( \varrho^{bc,3}_{zz,x}  +  \varsigma^{bc,3}_{zz,x} \right )
+ \frac{e^4 \, \tau \,J \, v_z} {128 \, \pi^2}
\left(\frac{\alpha_J} {\mu }\right)^{\frac{2} {J} } B_z^2 
 \left( \varrho^{bc,3}_{zz,z}  +  \varsigma^{bc,3}_{zz,z} \right ) \text{ and}
\nn  \sigma_{zz}^{ (\chi, m) } & = \frac{e^3 \, \tau \,J \,  v_z}
 {4 \, \pi^2} \, \chi  \, B_z
 \left( \varrho^{m, 3}_{zz} + \varsigma^{m, 3}_{zz} \right)
+ \frac{e^4 \, \tau \, J\, v_z^3} {128\, \pi^2 \,\mu^2}\, B_x^2 
\left( \varrho^{m,3}_{zz,x}  +  \varsigma^{m,3}_{zz,x} \right )
+ \frac{e^4 \, \tau \,J \, v_z} {128 \, \pi^2}
\left(\frac{\alpha_J} {\mu }\right)^{\frac{2} {J} } B_z^2 
 \left( \varrho^{m,3}_{zz,z}  +  \varsigma^{m,3}_{zz,z} \right ).
\end{align}
The symbols used above indicate the following: (1) $ \varrho^{bc,3}_{zz} $ ($ \varsigma^{bc,3}_{zz} $) represents the BC-only part proportional to $ \chi B_z $, arising from the $s=1$ ($s=2$) band; (2) $ \varrho^{bc,3}_{zz,x} $ ($ \varsigma^{bc,3}_{zz,x} $) represents the BC-only part proportional to $ B_x^2 $, arising from the $s=1$ ($s=2$) band; (3) $ \varrho^{bc,3}_{zz,z} $ ($ \varsigma^{bc,3}_{zz,z} $) represents the BC-only part proportional to $ B_z^2 $, arising from the $s=1$ ($s=2$) band; (4) $\varrho^{m,3}_{zz} $ ($ \varsigma^{m,3}_{zz} $) represents the OMM part proportional to $ \chi B_z $, arising from the $s=1$ ($s=2$) band; (5) $\varrho^{m,3}_{zz,x} $ ($ \varsigma^{m,3}_{zz,x} $) represents the OMM part proportional to $ B_x^2 $, arising from the $s=1$ ($s=2$) band;
(6) $\varrho^{m,3}_{zz,z} $ ($ \varsigma^{m,3}_{zz,z} $) represents the OMM part proportional to $ B_z^2 $, arising from the $s=1$ ($s=2$) band.

The coefficients accompanying $J \chi  B_z $ are $J$-independent and are logarithmically divergent.
Defining
$$ \left \lbrace \varrho^{bc, 3}_{zz}, \, \varrho^{m, 3}_{zz} , \,
\varsigma^{bc, 3}_{zz}, \, \varsigma^{bc, 3}_{zz} \right \rbrace 
= 
\left \lbrace 
\bar \varrho^{bc, 3}_{zz}, \, \bar \varrho^{m, 3}_{zz} , \,
\bar \varsigma^{bc, 3}_{zz}, \,  \bar \varsigma^{bc, 3}_{zz} 
\right \rbrace  \ln(\Lambda/ \mu) 
+ \order{ \left( {\mu}/ {\Lambda}  \right)^0} \,, $$
we have
\begin{align}
\label{eqzx_set2_type2_lin}
& \bar \varrho^{bc, 3}_{zz} = \frac{ - \, \left(\eta_\chi^2-1\right)^2}
{2 \, \eta_\chi^4} \,,\quad
\bar \varrho^{m, 3}_{zz} = \frac{3}{2}-\frac{4}{\eta_\chi^2}+\frac{5}{2 \, \eta_\chi^4}
\quad \Rightarrow  \quad
\bar \varrho^{bc, 3}_{zz}  +  \bar \varrho^{m, 3}_{zz} =
1-\frac{3}{\eta_\chi^2}+\frac{2}{\eta_\chi^4} \, ,
\nn & \bar \varsigma^{bc, 3}_{zz} = \frac{\left(\eta_\chi^2-1\right){}^2}
{2 \, \eta_\chi^4}\,,\quad
\bar \varsigma^{bc, 3}_{zz} = -\,\frac{1}{2}
+\frac{2}{\eta_\chi^2}-\frac{3}{2 \, \eta_\chi^4}
\quad \Rightarrow  \quad
\bar \varsigma^{bc, 3}_{zz} +  {\bar \varsigma}^{m, l}_{zz,2}  = 
\frac{\eta_\chi^2-1}{\eta_\chi^4}\, .
\end{align}
Clearly, $\bar \varrho^{bc, l}_{zx,2} + \bar \varsigma^{bc, 3}_{zz} =  0$, such that the OMM parts yield the net contribution of $ {\left(\eta_\chi^2-1\right)^2} /{\eta_\chi^4}$, which is positive.
For the $ J B_x^2 $-dependent parts, the net response is again $J$-independent. Their explicit expressions are given below:
\begin{align}
\label{eqzx_set2_type2_Bx}
& \varrho^{bc,3}_{zz,x} = \frac{\left(\eta_\chi+1\right)^5 
\left(15 \, \eta_\chi^3-19 \, \eta_\chi^2+10  \,\eta_\chi-2\right)}
{15 \eta_\chi^5}\,,\quad
\varrho^{m,3}_{zz,x} = \frac{\left(\eta_\chi+1\right)^3 
\left(15  \,\eta_\chi^5-53  \,\eta_\chi^4+39  \,\eta_\chi^3+3  \,\eta_\chi^2
-18   \,\eta_\chi+6\right)}       {15 \, \eta_\chi^5}\,,\nn
& \varsigma^{bc,3}_{zz,x} = \frac{ - \,\left(\eta_\chi-1\right)^5
 \left(15  \,\eta_\chi^3+19  \,\eta_\chi^2+10  \,\eta_\chi+2\right)}
 {15  \,\eta_\chi^5}\,,\quad
\varsigma^{m,3}_{zz,x} = \frac{15  \,\eta_\chi^8-56  \,\eta_\chi^7+45  \,\eta_\chi^6
+24  \,\eta_\chi^5-25  \,\eta_\chi^4-9 \eta_\chi^2+6} {15 \, \eta_\chi^5} \,.
\end{align}
For the $ B_z^2 $-dependent parts, since the integrals are quite complicated, the final expressions are evaluated by performing them separately for each value of $J$, which turn out to be non-divergent. The three cases are listed below:
\begin{enumerate}

\item $J=1$:
\begin{align}
\label{eqzx_type2_Bz1}
& \varrho^{bc,3}_{zz,z} = \frac{4 \left(\eta_\chi+1\right)^4 
\left(5  \, \eta_\chi^2-4  \, \eta_\chi+1\right)}{15  \, \eta_\chi^5}\,,
\quad \varrho^{m,3}_{zz,z} = 
\frac{ - \,2 \left(15 \,  \eta_\chi^8+28  \, \eta_\chi^7+16  \, \eta_\chi^5
+55  \, \eta_\chi^4-32  \, \eta_\chi^2+6\right)}{15 \, \eta_\chi^5} \,,
\nn & \varsigma^{bc,3}_{zz,z}  = -\frac{4 \left(\eta_\chi-1\right)^4 
\left(5  \,\eta_\chi^2 + 4  \,\eta_\chi+1\right)} {15 \,\eta_\chi^5}\,,
\quad \varsigma^{m,3}_{zz,z} = \frac{4 
\left(4 \, \eta_\chi^7-10  \,\eta_\chi^6 + 18  \,\eta_\chi^5
-25  \,\eta_\chi^4 + 21  \,\eta_\chi^2-8\right)}
{15 \,  \eta_\chi^5} \,.
\end{align}

\item $J=2$:
\begin{align}
\label{eqzx_type2_Bz2}
& \varrho^{bc,3}_{zz,z} = \frac{2  \, \sqrt{\eta_\chi^2-1} 
\left(33  \, \eta_\chi^4-26 \, \eta_\chi^2+8\right)} {3  \, \eta_\chi^6}
+20 \cot^{-1}\left(\frac{\eta_\chi-1}{\sqrt{\eta_\chi^2-1}}\right),
\nn & \varrho^{m,3}_{zz,z} = \frac{8 \left(2-3  \, \eta_\chi^2\right) 
\sqrt{\eta_\chi^2-1}}{\eta_\chi^4}
-16 \cot^{-1}\left(\frac{\eta_\chi-1} {\sqrt{\eta_\chi^2-1}}\right),
\nn & \varsigma^{bc,3}_{zz,z}= \frac{2 \,\sqrt{\eta_\chi^2-1} 
\left(33 \, \eta_\chi^4-26 \, \eta_\chi^2+8\right)} {3 \, \eta_\chi^6}
-20 \cot^{-1}\left(\frac{\eta_\chi+1}{\sqrt{\eta_\chi^2-1}}\right),
\nn & \varsigma^{bc,3}_{zz,z} = \frac{8 \, \sqrt{\eta_\chi^2-1} 
\left(9  \, \eta_\chi^4-14  \,\eta_\chi^2+8\right)} {3  \, \eta_\chi^6}
-16 \cot^{-1}\left(\frac{\eta_\chi+1}{\sqrt{\eta_\chi^2-1}}\right).
\end{align}

\item $J=3 $:
\begin{align}
\label{eqzx_type2_Bz31}
&  \frac{729  \, \sqrt{3} \,\Gamma^2\left(\frac{1}{3}\right)
 \, \eta_\chi^{\frac{19} {3} } \sqrt[3]{\eta_\chi^2-1} }    
{ 8  \, \pi  \,  \Gamma \left(\frac{5}{3}\right) 
\left(\eta_\chi+1\right)^{ \frac{2} {3} } } 
\left( \varrho^{bc,3}_{zz,z} + \varrho^{m,3}_{zz,z} \right)
\nn & = 
\sqrt[3]{ 2 \, \eta_\chi \left(\eta_\chi-1\right)^2 } \,
\left( 14560 -3483 \, \eta_\chi^6 + 17091  \, \eta_\chi^4-25134 \, \eta_\chi^2  \right)
\nn & \quad +
\left[  \eta_\chi^2 \left(135  \,\eta_\chi^6 + 438  \, \eta_\chi^4
-971  \,\eta_\chi^2+1162\right)-14560 \right ]
\, _2F_1\left(\frac{1}{3},\frac{2}{3};\frac{4}{3};\frac{\eta_\chi+1}{1-\eta_\chi}\right),
\nn & \frac{2187 \,\sqrt{3} \; \Gamma \left(\frac{7}{6}\right)\, 
\Gamma \left(\frac{7}{3}\right) 
\left(\eta_\chi-1\right)^{ \frac{2} {3} } \,
 \eta_\chi^{ \frac{20} {3} }}{128 \, 2^{ \frac{2} {3} } \, \pi^{ \frac{3} {2} }}
\,\varrho^{bc,3}_{zz,z} 
   \nn &  =
 2 \,\sqrt[3]{\eta_\chi} \left(\eta_\chi^2-1\right)^{ \frac{4} {3} }
 \left(135 \,   \eta_\chi^4-198  \, \eta_\chi^2+91\right) \, 
   _2F_1\left(\frac{1}{3},\frac{2}{3};\frac{4}{3};\frac{\eta_\chi+1}{1-\eta_\chi}\right)  
 \nn & \quad
 +
 \,_2F_1\left(\frac{1}{3},\frac{5}{3};\frac{7}{3};\frac{\eta_\chi+1}{1-\eta_\chi}\right)  
 \sqrt[9]{\eta_\chi} \, \sqrt[3]{\eta_\chi^2-1} 
 \,\sqrt[3]{297 \left(\eta_\chi^2-1\right)^{ \frac{2} {3} } \eta_\chi^5 
 + 297  \, \eta_\chi^4-276 \,  \eta_\chi^3-276 \,  \eta_\chi^2 
 + 91 \,  \eta_\chi + 91 }  \,,
\end{align}
\begin{align}
\label{eqzx_type2_Bz32}
& \frac{59049\, \sqrt{3} \,\Gamma \left(\frac{4}{3}\right) 
\,\Gamma \left(\frac{13}{6}\right) 
\eta_\chi^{ \frac{19} {3} } \sqrt[3] {\eta_\chi+1} }
{28  \, \pi^{ \frac{3} {2} } \sqrt[3]{\eta_\chi-1}}
\left( \varsigma^{bc,3}_{zz,z} +\varsigma^{m,3}_{zz,z} \right)
\nn & =
 2 \;\sqrt[3]{\eta_\chi} \left(\eta_\chi+1\right)^{ \frac{2} {3} } 
 \left( 1456 -7857 \,  \eta_\chi^6 + 5589 \,  \eta_\chi^4 + 246  \, \eta_\chi^2 \right)
\nn & \quad +
 2^{ \frac{2} {3} } \left [27 \left(27  \, \eta_\chi^6
- 222 \,\eta_\chi^4 + 197  \, \eta_\chi^2-14\right) \eta_\chi^2
 + 1456 \right ]
 \, _2F_1\left(\frac{1}{3},\frac{2}{3};\frac{4}{3};\frac{2}{\eta_\chi+1}-1\right),
\nn & \frac{6561 \sqrt{3} \,\Gamma \left(\frac{4}{3}\right) 
\,\Gamma \left(\frac{13}{6}\right) 
\eta_\chi^{ \frac{19} {3} } \sqrt[3]{\eta_\chi^2-1}}
{224\, 2^{ \frac{2} {3} } \,\pi ^{ \frac{3} {2} } \left(\eta_\chi-1\right)^{ \frac{2} {3} }}
\,\varsigma^{bc,3}_{zz,z} 
   \nn &  = 
\sqrt[3]{ 2\, \eta_\chi \left(\eta_\chi+1\right)^2}
\, \left(-297 \,  \eta_\chi^4 + 276  \, \eta_\chi^2-91\right)
+ \left [ 45  \, \eta_\chi^2 \left(9 \,  \eta_\chi^4-9  \, \eta_\chi^2+7\right)-91\right ]\,
   _2F_1\left(\frac{1}{3},\frac{2}{3};\frac{4}{3};\frac{2}{\eta_\chi+1}-1\right) .
\end{align}

\end{enumerate}
Since all these terms are non-divergent, while measuring the $B_z$-dependence, they will be masked by the log-dependent terms in the $ \propto \chi B_z $-parts. All the above expressions furnish the contents of Sec.~\ref{set3long_type2} and Fig.~\ref{figzzset3type2} therein.

\subsection{Set-up III: In-plane transverse components}
\label{appset3_inplane_trs}

The starting expression is captured by
\begin{align}  
\label{eqappset3_xz}
\sigma_{xz}^{\chi, s}  & =  
\frac{e^3 \, \tau  \, J^2\,\alpha_J^2 \,v_z^2\, \chi \, B_x}  {4} 
\int \frac{d \epsilon \, d \gamma } {(2\,\pi)^2 } \,
\frac{ k_\perp^{2 J-2}} {\epsilon^5}
\left(\zeta \, l^3_{1xz}   + \upsilon \,l^3_{2xz}  \right ) \mathcal J
\nn & \quad  
+ \frac{e^4 \, \tau \, J^3 \, \alpha_J^4 \, v_z^2 \,B_x \, B_z }   {8 }
\int \frac{d \epsilon \, d \gamma } {(2\,\pi)^2 } \,
\frac{  k_\perp^{4 J-4 }}  {\epsilon^8}
\left(\zeta^2 \, t^3_{1xz} +  \upsilon^2 \,t^3_{2xz} + \zeta \,\upsilon \,t^3_{3xz} \right)
\mathcal J  ,\nn 
 l^3_{1xz} & =  l^2_{1zx}\,, \quad  l^3_{2xz}  = l^2_{2zx}\,, \quad
t^3_{1xz} = t^2_{1zx} \,, \quad t^3_{2xz}  =  t^2_{2zx}\,, \quad t^3_{3xz}  = t^2_{3zx} \,.
\end{align} 
The final expressions show that that all the terms are the same as those for the $zx$-components obtained for set-up II [cf. Eq.~\eqref{eqappset2_zx}]. Hence, the same behaviour that is outlined in Appendix~\ref{appset2_inplane_type2} applies here. In particular, the conductivity contains terms which are linear-in-$B$ as well those which are quadratic-in-$ B $. The former are caused by nonzero currents $ \propto \chi \left( \mathbf E \cdot  {\mathbf{\hat z}} \right ) B_x \, {\mathbf{\hat x}} $.

\subsection{Set-up III: Out-of-plane transverse components}
\label{appset3_outplane}

All the out-of-plane components vanish, i.e., $ \sigma_{yz}^{ \chi, s } =  0\,.$

 
\bibliography{ref_dir}

\begin{thebibliography}{80}%
\makeatletter
\providecommand \@ifxundefined [1]{%
 \@ifx{#1\undefined}
}%
\providecommand \@ifnum [1]{%
 \ifnum #1\expandafter \@firstoftwo
 \else \expandafter \@secondoftwo
 \fi
}%
\providecommand \@ifx [1]{%
 \ifx #1\expandafter \@firstoftwo
 \else \expandafter \@secondoftwo
 \fi
}%
\providecommand \natexlab [1]{#1}%
\providecommand \enquote  [1]{``#1''}%
\providecommand \bibnamefont  [1]{#1}%
\providecommand \bibfnamefont [1]{#1}%
\providecommand \citenamefont [1]{#1}%
\providecommand \href@noop [0]{\@secondoftwo}%
\providecommand \href [0]{\begingroup \@sanitize@url \@href}%
\providecommand \@href[1]{\@@startlink{#1}\@@href}%
\providecommand \@@href[1]{\endgroup#1\@@endlink}%
\providecommand \@sanitize@url [0]{\catcode `\\12\catcode `\$12\catcode
  `\&12\catcode `\#12\catcode `\^12\catcode `\_12\catcode `\%12\relax}%
\providecommand \@@startlink[1]{}%
\providecommand \@@endlink[0]{}%
\providecommand \url  [0]{\begingroup\@sanitize@url \@url }%
\providecommand \@url [1]{\endgroup\@href {#1}{\urlprefix }}%
\providecommand \urlprefix  [0]{URL }%
\providecommand \Eprint [0]{\href }%
\providecommand \doibase [0]{https://doi.org/}%
\providecommand \selectlanguage [0]{\@gobble}%
\providecommand \bibinfo  [0]{\@secondoftwo}%
\providecommand \bibfield  [0]{\@secondoftwo}%
\providecommand \translation [1]{[#1]}%
\providecommand \BibitemOpen [0]{}%
\providecommand \bibitemStop [0]{}%
\providecommand \bibitemNoStop [0]{.\EOS\space}%
\providecommand \EOS [0]{\spacefactor3000\relax}%
\providecommand \BibitemShut  [1]{\csname bibitem#1\endcsname}%
\let\auto@bib@innerbib\@empty
\bibitem [{\citenamefont {Burkov}\ and\ \citenamefont
  {Balents}(2011)}]{burkov11_weyl}%
  \BibitemOpen
  \bibfield  {author} {\bibinfo {author} {\bibfnamefont {A.~A.}\ \bibnamefont
  {Burkov}}\ and\ \bibinfo {author} {\bibfnamefont {L.}~\bibnamefont
  {Balents}},\ }\bibfield  {title} {\bibinfo {title} {Weyl semimetal in a
  topological insulator multilayer},\ }\href
  {https://doi.org/10.1103/PhysRevLett.107.127205} {\bibfield  {journal}
  {\bibinfo  {journal} {Phys. Rev. Lett.}\ }\textbf {\bibinfo {volume} {107}},\
  \bibinfo {pages} {127205} (\bibinfo {year} {2011})}\BibitemShut {NoStop}%
\bibitem [{\citenamefont {Yan}\ and\ \citenamefont
  {Felser}(2017)}]{yan17_topological}%
  \BibitemOpen
  \bibfield  {author} {\bibinfo {author} {\bibfnamefont {B.}~\bibnamefont
  {Yan}}\ and\ \bibinfo {author} {\bibfnamefont {C.}~\bibnamefont {Felser}},\
  }\bibfield  {title} {\bibinfo {title} {Topological materials: {W}eyl
  semimetals},\ }\href
  {https://doi.org/10.1146/annurev-conmatphys-031016-025458} {\bibfield
  {journal} {\bibinfo  {journal} {Annual Review of Condensed Matter Physics}\
  }\textbf {\bibinfo {volume} {8}},\ \bibinfo {pages} {337} (\bibinfo {year}
  {2017})}\BibitemShut {NoStop}%
\bibitem [{\citenamefont {Bradlyn}\ \emph {et~al.}(2016)\citenamefont
  {Bradlyn}, \citenamefont {Cano}, \citenamefont {Wang}, \citenamefont
  {Vergniory}, \citenamefont {Felser}, \citenamefont {Cava},\ and\
  \citenamefont {Bernevig}}]{bernevig}%
  \BibitemOpen
  \bibfield  {author} {\bibinfo {author} {\bibfnamefont {B.}~\bibnamefont
  {Bradlyn}}, \bibinfo {author} {\bibfnamefont {J.}~\bibnamefont {Cano}},
  \bibinfo {author} {\bibfnamefont {Z.}~\bibnamefont {Wang}}, \bibinfo {author}
  {\bibfnamefont {M.~G.}\ \bibnamefont {Vergniory}}, \bibinfo {author}
  {\bibfnamefont {C.}~\bibnamefont {Felser}}, \bibinfo {author} {\bibfnamefont
  {R.~J.}\ \bibnamefont {Cava}},\ and\ \bibinfo {author} {\bibfnamefont
  {B.~A.}\ \bibnamefont {Bernevig}},\ }\bibfield  {title} {\bibinfo {title}
  {{Beyond Dirac and Weyl fermions: Unconventional quasiparticles in
  conventional crystals}},\ }\href
  {https://science.sciencemag.org/content/353/6299/aaf5037} {\bibfield
  {journal} {\bibinfo  {journal} {Science}\ }\textbf {\bibinfo {volume} {353}}
  (\bibinfo {year} {2016})}\BibitemShut {NoStop}%
\bibitem [{\citenamefont {Fang}\ \emph {et~al.}(2012)\citenamefont {Fang},
  \citenamefont {Gilbert}, \citenamefont {Dai},\ and\ \citenamefont
  {Bernevig}}]{bernevig2}%
  \BibitemOpen
  \bibfield  {author} {\bibinfo {author} {\bibfnamefont {C.}~\bibnamefont
  {Fang}}, \bibinfo {author} {\bibfnamefont {M.~J.}\ \bibnamefont {Gilbert}},
  \bibinfo {author} {\bibfnamefont {X.}~\bibnamefont {Dai}},\ and\ \bibinfo
  {author} {\bibfnamefont {B.~A.}\ \bibnamefont {Bernevig}},\ }\bibfield
  {title} {\bibinfo {title} {Multi-{W}eyl topological semimetals stabilized by
  point group symmetry},\ }\href
  {https://doi.org/10.1103/PhysRevLett.108.266802} {\bibfield  {journal}
  {\bibinfo  {journal} {Phys. Rev. Lett.}\ }\textbf {\bibinfo {volume} {108}},\
  \bibinfo {pages} {266802} (\bibinfo {year} {2012})}\BibitemShut {NoStop}%
\bibitem [{\citenamefont {Dantas}\ \emph {et~al.}(2018)\citenamefont {Dantas},
  \citenamefont {Pena-Benitez}, \citenamefont {Roy},\ and\ \citenamefont
  {Sur{\'o}wka}}]{dantas18_magnetotransport}%
  \BibitemOpen
  \bibfield  {author} {\bibinfo {author} {\bibfnamefont {R.}~\bibnamefont
  {Dantas}}, \bibinfo {author} {\bibfnamefont {F.}~\bibnamefont
  {Pena-Benitez}}, \bibinfo {author} {\bibfnamefont {B.}~\bibnamefont {Roy}},\
  and\ \bibinfo {author} {\bibfnamefont {P.}~\bibnamefont {Sur{\'o}wka}},\
  }\bibfield  {title} {\bibinfo {title} {Magnetotransport in multi-{W}eyl
  semimetals: A kinetic theory approach},\ }\href
  {https://doi.org/10.1007/JHEP12(2018)069} {\bibfield  {journal} {\bibinfo
  {journal} {Journal of High Energy Physics}\ }\textbf {\bibinfo {volume}
  {2018}},\ \bibinfo {pages} {1} (\bibinfo {year} {2018})}\BibitemShut
  {NoStop}%
\bibitem [{\citenamefont {Nielsen}\ and\ \citenamefont
  {Ninomiya}(1981)}]{nielsen}%
  \BibitemOpen
  \bibfield  {author} {\bibinfo {author} {\bibfnamefont {H.}~\bibnamefont
  {Nielsen}}\ and\ \bibinfo {author} {\bibfnamefont {M.}~\bibnamefont
  {Ninomiya}},\ }\bibfield  {title} {\bibinfo {title} {A no-go theorem for
  regularizing chiral fermions},\ }\href
  {https://doi.org/https://doi.org/10.1016/0370-2693(81)91026-1} {\bibfield
  {journal} {\bibinfo  {journal} {Physics Letters B}\ }\textbf {\bibinfo
  {volume} {105}},\ \bibinfo {pages} {219} (\bibinfo {year}
  {1981})}\BibitemShut {NoStop}%
\bibitem [{\citenamefont {Huang}\ \emph {et~al.}(2015)\citenamefont {Huang},
  \citenamefont {Zhao}, \citenamefont {Long}, \citenamefont {Wang},
  \citenamefont {Chen}, \citenamefont {Yang}, \citenamefont {Liang},
  \citenamefont {Xue}, \citenamefont {Weng}, \citenamefont {Fang},
  \citenamefont {Dai},\ and\ \citenamefont {Chen}}]{huang15_observation}%
  \BibitemOpen
  \bibfield  {author} {\bibinfo {author} {\bibfnamefont {X.}~\bibnamefont
  {Huang}}, \bibinfo {author} {\bibfnamefont {L.}~\bibnamefont {Zhao}},
  \bibinfo {author} {\bibfnamefont {Y.}~\bibnamefont {Long}}, \bibinfo {author}
  {\bibfnamefont {P.}~\bibnamefont {Wang}}, \bibinfo {author} {\bibfnamefont
  {D.}~\bibnamefont {Chen}}, \bibinfo {author} {\bibfnamefont {Z.}~\bibnamefont
  {Yang}}, \bibinfo {author} {\bibfnamefont {H.}~\bibnamefont {Liang}},
  \bibinfo {author} {\bibfnamefont {M.}~\bibnamefont {Xue}}, \bibinfo {author}
  {\bibfnamefont {H.}~\bibnamefont {Weng}}, \bibinfo {author} {\bibfnamefont
  {Z.}~\bibnamefont {Fang}}, \bibinfo {author} {\bibfnamefont {X.}~\bibnamefont
  {Dai}},\ and\ \bibinfo {author} {\bibfnamefont {G.}~\bibnamefont {Chen}},\
  }\bibfield  {title} {\bibinfo {title} {Observation of the
  chiral-anomaly-induced negative magnetoresistance in 3d {W}eyl semimetal
  {T}a{A}s},\ }\href {https://doi.org/10.1103/PhysRevX.5.031023} {\bibfield
  {journal} {\bibinfo  {journal} {Phys. Rev. X}\ }\textbf {\bibinfo {volume}
  {5}},\ \bibinfo {pages} {031023} (\bibinfo {year} {2015})}\BibitemShut
  {NoStop}%
\bibitem [{\citenamefont {Lv}\ \emph {et~al.}(2015)\citenamefont {Lv},
  \citenamefont {Weng}, \citenamefont {Fu}, \citenamefont {Wang}, \citenamefont
  {Miao}, \citenamefont {Ma}, \citenamefont {Richard}, \citenamefont {Huang},
  \citenamefont {Zhao}, \citenamefont {Chen}, \citenamefont {Fang},
  \citenamefont {Dai}, \citenamefont {Qian},\ and\ \citenamefont
  {Ding}}]{lv_Weyl}%
  \BibitemOpen
  \bibfield  {author} {\bibinfo {author} {\bibfnamefont {B.~Q.}\ \bibnamefont
  {Lv}}, \bibinfo {author} {\bibfnamefont {H.~M.}\ \bibnamefont {Weng}},
  \bibinfo {author} {\bibfnamefont {B.~B.}\ \bibnamefont {Fu}}, \bibinfo
  {author} {\bibfnamefont {X.~P.}\ \bibnamefont {Wang}}, \bibinfo {author}
  {\bibfnamefont {H.}~\bibnamefont {Miao}}, \bibinfo {author} {\bibfnamefont
  {J.}~\bibnamefont {Ma}}, \bibinfo {author} {\bibfnamefont {P.}~\bibnamefont
  {Richard}}, \bibinfo {author} {\bibfnamefont {X.~C.}\ \bibnamefont {Huang}},
  \bibinfo {author} {\bibfnamefont {L.~X.}\ \bibnamefont {Zhao}}, \bibinfo
  {author} {\bibfnamefont {G.~F.}\ \bibnamefont {Chen}}, \bibinfo {author}
  {\bibfnamefont {Z.}~\bibnamefont {Fang}}, \bibinfo {author} {\bibfnamefont
  {X.}~\bibnamefont {Dai}}, \bibinfo {author} {\bibfnamefont {T.}~\bibnamefont
  {Qian}},\ and\ \bibinfo {author} {\bibfnamefont {H.}~\bibnamefont {Ding}},\
  }\bibfield  {title} {\bibinfo {title} {{Experimental Discovery of Weyl
  Semimetal TaAs}},\ }\href {https://doi.org/10.1103/PhysRevX.5.031013}
  {\bibfield  {journal} {\bibinfo  {journal} {Phys. Rev. X}\ }\textbf {\bibinfo
  {volume} {5}},\ \bibinfo {pages} {031013} (\bibinfo {year}
  {2015})}\BibitemShut {NoStop}%
\bibitem [{\citenamefont {Xu}\ \emph {et~al.}(2015)\citenamefont {Xu},
  \citenamefont {Belopolski}, \citenamefont {Alidoust}, \citenamefont
  {Neupane}, \citenamefont {Bian}, \citenamefont {Zhang}, \citenamefont
  {Sankar}, \citenamefont {Chang}, \citenamefont {Yuan}, \citenamefont {Lee},
  \citenamefont {Huang}, \citenamefont {Zheng}, \citenamefont {Ma},
  \citenamefont {Sanchez}, \citenamefont {Wang}, \citenamefont {Bansil},
  \citenamefont {Chou}, \citenamefont {Shibayev}, \citenamefont {Lin},
  \citenamefont {Jia},\ and\ \citenamefont {Hasan}}]{yang_Weyl}%
  \BibitemOpen
  \bibfield  {author} {\bibinfo {author} {\bibfnamefont {S.-Y.}\ \bibnamefont
  {Xu}}, \bibinfo {author} {\bibfnamefont {I.}~\bibnamefont {Belopolski}},
  \bibinfo {author} {\bibfnamefont {N.}~\bibnamefont {Alidoust}}, \bibinfo
  {author} {\bibfnamefont {M.}~\bibnamefont {Neupane}}, \bibinfo {author}
  {\bibfnamefont {G.}~\bibnamefont {Bian}}, \bibinfo {author} {\bibfnamefont
  {C.}~\bibnamefont {Zhang}}, \bibinfo {author} {\bibfnamefont
  {R.}~\bibnamefont {Sankar}}, \bibinfo {author} {\bibfnamefont
  {G.}~\bibnamefont {Chang}}, \bibinfo {author} {\bibfnamefont
  {Z.}~\bibnamefont {Yuan}}, \bibinfo {author} {\bibfnamefont {C.-C.}\
  \bibnamefont {Lee}}, \bibinfo {author} {\bibfnamefont {S.-M.}\ \bibnamefont
  {Huang}}, \bibinfo {author} {\bibfnamefont {H.}~\bibnamefont {Zheng}},
  \bibinfo {author} {\bibfnamefont {J.}~\bibnamefont {Ma}}, \bibinfo {author}
  {\bibfnamefont {D.~S.}\ \bibnamefont {Sanchez}}, \bibinfo {author}
  {\bibfnamefont {B.}~\bibnamefont {Wang}}, \bibinfo {author} {\bibfnamefont
  {A.}~\bibnamefont {Bansil}}, \bibinfo {author} {\bibfnamefont
  {F.}~\bibnamefont {Chou}}, \bibinfo {author} {\bibfnamefont {P.~P.}\
  \bibnamefont {Shibayev}}, \bibinfo {author} {\bibfnamefont {H.}~\bibnamefont
  {Lin}}, \bibinfo {author} {\bibfnamefont {S.}~\bibnamefont {Jia}},\ and\
  \bibinfo {author} {\bibfnamefont {M.~Z.}\ \bibnamefont {Hasan}},\ }\bibfield
  {title} {\bibinfo {title} {{Discovery of a Weyl fermion semimetal and
  topological Fermi arcs}},\ }\href {https://doi.org/10.1126/science.aaa9297}
  {\bibfield  {journal} {\bibinfo  {journal} {Science}\ }\textbf {\bibinfo
  {volume} {349}},\ \bibinfo {pages} {613} (\bibinfo {year}
  {2015})}\BibitemShut {NoStop}%
\bibitem [{\citenamefont {Ruan}\ \emph {et~al.}(2016)\citenamefont {Ruan},
  \citenamefont {Jian}, \citenamefont {Yao}, \citenamefont {Zhang},
  \citenamefont {Zhang},\ and\ \citenamefont {Xing}}]{ruan_Weyl}%
  \BibitemOpen
  \bibfield  {author} {\bibinfo {author} {\bibfnamefont {J.}~\bibnamefont
  {Ruan}}, \bibinfo {author} {\bibfnamefont {S.-K.}\ \bibnamefont {Jian}},
  \bibinfo {author} {\bibfnamefont {H.}~\bibnamefont {Yao}}, \bibinfo {author}
  {\bibfnamefont {H.}~\bibnamefont {Zhang}}, \bibinfo {author} {\bibfnamefont
  {S.-C.}\ \bibnamefont {Zhang}},\ and\ \bibinfo {author} {\bibfnamefont
  {D.}~\bibnamefont {Xing}},\ }\bibfield  {title} {\bibinfo {title}
  {{Symmetry-protected ideal Weyl semimetal in HgTe-class materials}},\ }\href
  {https://doi.org/10.1038/ncomms11136} {\bibfield  {journal} {\bibinfo
  {journal} {Nature Communications}\ }\textbf {\bibinfo {volume} {7}},\
  \bibinfo {pages} {11136} (\bibinfo {year} {2016})}\BibitemShut {NoStop}%
\bibitem [{\citenamefont {Xu}\ \emph {et~al.}(2011)\citenamefont {Xu},
  \citenamefont {Weng}, \citenamefont {Wang}, \citenamefont {Dai},\ and\
  \citenamefont {Fang}}]{Gang2011}%
  \BibitemOpen
  \bibfield  {author} {\bibinfo {author} {\bibfnamefont {G.}~\bibnamefont
  {Xu}}, \bibinfo {author} {\bibfnamefont {H.}~\bibnamefont {Weng}}, \bibinfo
  {author} {\bibfnamefont {Z.}~\bibnamefont {Wang}}, \bibinfo {author}
  {\bibfnamefont {X.}~\bibnamefont {Dai}},\ and\ \bibinfo {author}
  {\bibfnamefont {Z.}~\bibnamefont {Fang}},\ }\bibfield  {title} {\bibinfo
  {title} {Chern semimetal and the quantized anomalous {H}all effect in
  {H}g{C}r$_{2}${S}e$_{4}$},\ }\href
  {https://doi.org/10.1103/PhysRevLett.107.186806} {\bibfield  {journal}
  {\bibinfo  {journal} {Phys. Rev. Lett.}\ }\textbf {\bibinfo {volume} {107}},\
  \bibinfo {pages} {186806} (\bibinfo {year} {2011})}\BibitemShut {NoStop}%
\bibitem [{\citenamefont {Huang}\ \emph {et~al.}(2016)\citenamefont {Huang},
  \citenamefont {Xu}, \citenamefont {Belopolski}, \citenamefont {Lee},
  \citenamefont {Chang}, \citenamefont {Chang}, \citenamefont {Wang},
  \citenamefont {Alidoust}, \citenamefont {Bian}, \citenamefont {Neupane},
  \citenamefont {Sanchez}, \citenamefont {Zheng}, \citenamefont {Jeng},
  \citenamefont {Bansil}, \citenamefont {Neupert}, \citenamefont {Lin},\ and\
  \citenamefont {Hasan}}]{hasan_mweyl16}%
  \BibitemOpen
  \bibfield  {author} {\bibinfo {author} {\bibfnamefont {S.-M.}\ \bibnamefont
  {Huang}}, \bibinfo {author} {\bibfnamefont {S.-Y.}\ \bibnamefont {Xu}},
  \bibinfo {author} {\bibfnamefont {I.}~\bibnamefont {Belopolski}}, \bibinfo
  {author} {\bibfnamefont {C.-C.}\ \bibnamefont {Lee}}, \bibinfo {author}
  {\bibfnamefont {G.}~\bibnamefont {Chang}}, \bibinfo {author} {\bibfnamefont
  {T.-R.}\ \bibnamefont {Chang}}, \bibinfo {author} {\bibfnamefont
  {B.}~\bibnamefont {Wang}}, \bibinfo {author} {\bibfnamefont {N.}~\bibnamefont
  {Alidoust}}, \bibinfo {author} {\bibfnamefont {G.}~\bibnamefont {Bian}},
  \bibinfo {author} {\bibfnamefont {M.}~\bibnamefont {Neupane}}, \bibinfo
  {author} {\bibfnamefont {D.}~\bibnamefont {Sanchez}}, \bibinfo {author}
  {\bibfnamefont {H.}~\bibnamefont {Zheng}}, \bibinfo {author} {\bibfnamefont
  {H.-T.}\ \bibnamefont {Jeng}}, \bibinfo {author} {\bibfnamefont
  {A.}~\bibnamefont {Bansil}}, \bibinfo {author} {\bibfnamefont
  {T.}~\bibnamefont {Neupert}}, \bibinfo {author} {\bibfnamefont
  {H.}~\bibnamefont {Lin}},\ and\ \bibinfo {author} {\bibfnamefont {M.~Z.}\
  \bibnamefont {Hasan}},\ }\bibfield  {title} {\bibinfo {title} {New type of
  {W}eyl semimetal with quadratic double weyl fermions},\ }\href
  {https://doi.org/10.1073/pnas.1514581113} {\bibfield  {journal} {\bibinfo
  {journal} {Proceedings of the National Academy of Sciences}\ }\textbf
  {\bibinfo {volume} {113}},\ \bibinfo {pages} {1180} (\bibinfo {year}
  {2016})}\BibitemShut {NoStop}%
\bibitem [{\citenamefont {{Singh}}\ \emph {et~al.}(2018)\citenamefont
  {{Singh}}, \citenamefont {{Chang}}, \citenamefont {{Chang}}, \citenamefont
  {{Huang}}, \citenamefont {{Su}}, \citenamefont {{Lin}}, \citenamefont
  {{Lin}},\ and\ \citenamefont {{Bansil}}}]{singh18_tunable}%
  \BibitemOpen
  \bibfield  {author} {\bibinfo {author} {\bibfnamefont {B.}~\bibnamefont
  {{Singh}}}, \bibinfo {author} {\bibfnamefont {G.}~\bibnamefont {{Chang}}},
  \bibinfo {author} {\bibfnamefont {T.-R.}\ \bibnamefont {{Chang}}}, \bibinfo
  {author} {\bibfnamefont {S.-M.}\ \bibnamefont {{Huang}}}, \bibinfo {author}
  {\bibfnamefont {C.}~\bibnamefont {{Su}}}, \bibinfo {author} {\bibfnamefont
  {M.-C.}\ \bibnamefont {{Lin}}}, \bibinfo {author} {\bibfnamefont
  {H.}~\bibnamefont {{Lin}}},\ and\ \bibinfo {author} {\bibfnamefont
  {A.}~\bibnamefont {{Bansil}}},\ }\bibfield  {title} {\bibinfo {title}
  {Tunable double-{W}eyl fermion semimetal state in the {S}r{S}i$_{2}$
  materials class},\ }\href {https://doi.org/10.1038/s41598-018-28644-y}
  {\bibfield  {journal} {\bibinfo  {journal} {Scientific Reports}\ }\textbf
  {\bibinfo {volume} {8}},\ \bibinfo {eid} {10540} (\bibinfo {year}
  {2018})}\BibitemShut {NoStop}%
\bibitem [{\citenamefont {Liu}\ and\ \citenamefont
  {Zunger}(2017)}]{liu2017predicted}%
  \BibitemOpen
  \bibfield  {author} {\bibinfo {author} {\bibfnamefont {Q.}~\bibnamefont
  {Liu}}\ and\ \bibinfo {author} {\bibfnamefont {A.}~\bibnamefont {Zunger}},\
  }\bibfield  {title} {\bibinfo {title} {{Predicted realization of cubic
  {Dirac} fermion in quasi-one-dimensional transition-metal
  monochalcogenides}},\ }\href {https://doi.org/10.1103/PhysRevX.7.021019}
  {\bibfield  {journal} {\bibinfo  {journal} {Phys. Rev. X}\ }\textbf {\bibinfo
  {volume} {7}},\ \bibinfo {pages} {021019} (\bibinfo {year}
  {2017})}\BibitemShut {NoStop}%
\bibitem [{\citenamefont {Son}\ and\ \citenamefont
  {Spivak}(2013)}]{son13_chiral}%
  \BibitemOpen
  \bibfield  {author} {\bibinfo {author} {\bibfnamefont {D.~T.}\ \bibnamefont
  {Son}}\ and\ \bibinfo {author} {\bibfnamefont {B.~Z.}\ \bibnamefont
  {Spivak}},\ }\bibfield  {title} {\bibinfo {title} {{Chiral anomaly and
  classical negative magnetoresistance of Weyl metals}},\ }\href
  {https://doi.org/10.1103/PhysRevB.88.104412} {\bibfield  {journal} {\bibinfo
  {journal} {Phys. Rev. B}\ }\textbf {\bibinfo {volume} {88}},\ \bibinfo
  {pages} {104412} (\bibinfo {year} {2013})}\BibitemShut {NoStop}%
\bibitem [{\citenamefont {Burkov}(2017)}]{burkov17_giant}%
  \BibitemOpen
  \bibfield  {author} {\bibinfo {author} {\bibfnamefont {A.~A.}\ \bibnamefont
  {Burkov}},\ }\bibfield  {title} {\bibinfo {title} {Giant planar {H}all effect
  in topological metals},\ }\href {https://doi.org/10.1103/PhysRevB.96.041110}
  {\bibfield  {journal} {\bibinfo  {journal} {Phys. Rev. B}\ }\textbf {\bibinfo
  {volume} {96}},\ \bibinfo {pages} {041110} (\bibinfo {year}
  {2017})}\BibitemShut {NoStop}%
\bibitem [{\citenamefont {Li}\ \emph {et~al.}(2017)\citenamefont {Li},
  \citenamefont {Wang}, \citenamefont {Li}, \citenamefont {Yang}, \citenamefont
  {Shen}, \citenamefont {Sheng}, \citenamefont {Li}, \citenamefont {Lu},
  \citenamefont {Zheng},\ and\ \citenamefont {Xu}}]{li_nmr17}%
  \BibitemOpen
  \bibfield  {author} {\bibinfo {author} {\bibfnamefont {Y.}~\bibnamefont
  {Li}}, \bibinfo {author} {\bibfnamefont {Z.}~\bibnamefont {Wang}}, \bibinfo
  {author} {\bibfnamefont {P.}~\bibnamefont {Li}}, \bibinfo {author}
  {\bibfnamefont {X.}~\bibnamefont {Yang}}, \bibinfo {author} {\bibfnamefont
  {Z.}~\bibnamefont {Shen}}, \bibinfo {author} {\bibfnamefont {F.}~\bibnamefont
  {Sheng}}, \bibinfo {author} {\bibfnamefont {X.}~\bibnamefont {Li}}, \bibinfo
  {author} {\bibfnamefont {Y.}~\bibnamefont {Lu}}, \bibinfo {author}
  {\bibfnamefont {Y.}~\bibnamefont {Zheng}},\ and\ \bibinfo {author}
  {\bibfnamefont {Z.-A.}\ \bibnamefont {Xu}},\ }\bibfield  {title} {\bibinfo
  {title} {{Negative magnetoresistance in {W}eyl semimetals NbAs and NbP:
  Intrinsic chiral anomaly and extrinsic effects}},\ }\href
  {https://doi.org/10.1007/s11467-016-0636-8} {\bibfield  {journal} {\bibinfo
  {journal} {Frontiers of Physics}\ }\textbf {\bibinfo {volume} {12}},\
  \bibinfo {pages} {127205} (\bibinfo {year} {2017})}\BibitemShut {NoStop}%
\bibitem [{\citenamefont {Nandy}\ \emph {et~al.}(2017)\citenamefont {Nandy},
  \citenamefont {Sharma}, \citenamefont {Taraphder},\ and\ \citenamefont
  {Tewari}}]{nandy_2017_chiral}%
  \BibitemOpen
  \bibfield  {author} {\bibinfo {author} {\bibfnamefont {S.}~\bibnamefont
  {Nandy}}, \bibinfo {author} {\bibfnamefont {G.}~\bibnamefont {Sharma}},
  \bibinfo {author} {\bibfnamefont {A.}~\bibnamefont {Taraphder}},\ and\
  \bibinfo {author} {\bibfnamefont {S.}~\bibnamefont {Tewari}},\ }\bibfield
  {title} {\bibinfo {title} {Chiral anomaly as the origin of the planar {H}all
  effect in {W}eyl semimetals},\ }\href
  {https://doi.org/10.1103/PhysRevLett.119.176804} {\bibfield  {journal}
  {\bibinfo  {journal} {Phys. Rev. Lett.}\ }\textbf {\bibinfo {volume} {119}},\
  \bibinfo {pages} {176804} (\bibinfo {year} {2017})}\BibitemShut {NoStop}%
\bibitem [{\citenamefont {Nandy}\ \emph {et~al.}(2018)\citenamefont {Nandy},
  \citenamefont {Taraphder},\ and\ \citenamefont {Tewari}}]{nandy18_Berry}%
  \BibitemOpen
  \bibfield  {author} {\bibinfo {author} {\bibfnamefont {S.}~\bibnamefont
  {Nandy}}, \bibinfo {author} {\bibfnamefont {A.}~\bibnamefont {Taraphder}},\
  and\ \bibinfo {author} {\bibfnamefont {S.}~\bibnamefont {Tewari}},\
  }\bibfield  {title} {\bibinfo {title} {Berry phase theory of planar {H}all
  effect in topological insulators},\ }\href
  {https://doi.org/10.1038/s41598-018-33258-5} {\bibfield  {journal} {\bibinfo
  {journal} {Scientific Reports}\ }\textbf {\bibinfo {volume} {8}},\ \bibinfo
  {pages} {14983} (\bibinfo {year} {2018})}\BibitemShut {NoStop}%
\bibitem [{\citenamefont {Nag}\ and\ \citenamefont {Nandy}(2020)}]{Nag_2020}%
  \BibitemOpen
  \bibfield  {author} {\bibinfo {author} {\bibfnamefont {T.}~\bibnamefont
  {Nag}}\ and\ \bibinfo {author} {\bibfnamefont {S.}~\bibnamefont {Nandy}},\
  }\bibfield  {title} {\bibinfo {title} {Magneto-transport phenomena of
  type-{I} multi-{W}eyl semimetals in co-planar setups},\ }\href
  {https://doi.org/10.1088/1361-648x/abc310} {\bibfield  {journal} {\bibinfo
  {journal} {Journal of Physics: Condensed Matter}\ }\textbf {\bibinfo {volume}
  {33}},\ \bibinfo {pages} {075504} (\bibinfo {year} {2020})}\BibitemShut
  {NoStop}%
\bibitem [{\citenamefont {{Mandal}}(2025)}]{ips-mwsm-floquet}%
  \BibitemOpen
  \bibfield  {author} {\bibinfo {author} {\bibfnamefont {I.}~\bibnamefont
  {{Mandal}}},\ }\bibfield  {title} {\bibinfo {title} {{Effects of
  time-periodic drive in the linear response for planar-Hall set-ups with Weyl
  and multi-Weyl semimetals}},\ }\href {https://arxiv.org/abs/2503.14406}
  {\bibfield  {journal} {\bibinfo  {journal} {arXiv e-prints}\ } (\bibinfo
  {year} {2025})},\ \Eprint {https://arxiv.org/abs/2503.14406}
  {arXiv:2503.14406 [cond-mat.mes-hall]} \BibitemShut {NoStop}%
\bibitem [{\citenamefont {Nielsen}\ and\ \citenamefont
  {Ninomiya}(1983)}]{chiral_ABJ}%
  \BibitemOpen
  \bibfield  {author} {\bibinfo {author} {\bibfnamefont {H.}~\bibnamefont
  {Nielsen}}\ and\ \bibinfo {author} {\bibfnamefont {M.}~\bibnamefont
  {Ninomiya}},\ }\bibfield  {title} {\bibinfo {title} {{The Adler-Bell-Jackiw
  anomaly and Weyl fermions in a crystal}},\ }\href
  {https://doi.org/https://doi.org/10.1016/0370-2693(83)91529-0} {\bibfield
  {journal} {\bibinfo  {journal} {Physics Letters B}\ }\textbf {\bibinfo
  {volume} {130}},\ \bibinfo {pages} {389} (\bibinfo {year}
  {1983})}\BibitemShut {NoStop}%
\bibitem [{\citenamefont {Huang}\ \emph {et~al.}(2017)\citenamefont {Huang},
  \citenamefont {Zhou},\ and\ \citenamefont {Shen}}]{chiral_ano_mWSM}%
  \BibitemOpen
  \bibfield  {author} {\bibinfo {author} {\bibfnamefont {Z.-M.}\ \bibnamefont
  {Huang}}, \bibinfo {author} {\bibfnamefont {J.}~\bibnamefont {Zhou}},\ and\
  \bibinfo {author} {\bibfnamefont {S.-Q.}\ \bibnamefont {Shen}},\ }\bibfield
  {title} {\bibinfo {title} {{Topological responses from chiral anomaly in
  multi-Weyl semimetals}},\ }\href {https://doi.org/10.1103/PhysRevB.96.085201}
  {\bibfield  {journal} {\bibinfo  {journal} {Phys. Rev. B}\ }\textbf {\bibinfo
  {volume} {96}},\ \bibinfo {pages} {085201} (\bibinfo {year}
  {2017})}\BibitemShut {NoStop}%
\bibitem [{\citenamefont {Das}\ and\ \citenamefont
  {Agarwal}(2019{\natexlab{a}})}]{amit-magneto}%
  \BibitemOpen
  \bibfield  {author} {\bibinfo {author} {\bibfnamefont {K.}~\bibnamefont
  {Das}}\ and\ \bibinfo {author} {\bibfnamefont {A.}~\bibnamefont {Agarwal}},\
  }\bibfield  {title} {\bibinfo {title} {{Linear magnetochiral transport in
  tilted type-I and type-II Weyl semimetals}},\ }\href
  {https://doi.org/10.1103/PhysRevB.99.085405} {\bibfield  {journal} {\bibinfo
  {journal} {Phys. Rev. B}\ }\textbf {\bibinfo {volume} {99}},\ \bibinfo
  {pages} {085405} (\bibinfo {year} {2019}{\natexlab{a}})}\BibitemShut
  {NoStop}%
\bibitem [{\citenamefont {Ghosh}\ and\ \citenamefont
  {Mandal}(2024{\natexlab{a}})}]{ips-rahul-ph}%
  \BibitemOpen
  \bibfield  {author} {\bibinfo {author} {\bibfnamefont {R.}~\bibnamefont
  {Ghosh}}\ and\ \bibinfo {author} {\bibfnamefont {I.}~\bibnamefont {Mandal}},\
  }\bibfield  {title} {\bibinfo {title} {{Electric and thermoelectric response
  for Weyl and multi-Weyl semimetals in planar Hall configurations including
  the effects of strain}},\ }\href
  {https://doi.org/https://doi.org/10.1016/j.physe.2024.115914} {\bibfield
  {journal} {\bibinfo  {journal} {Physica E: Low-dimensional Systems and
  Nanostructures}\ }\textbf {\bibinfo {volume} {159}},\ \bibinfo {pages}
  {115914} (\bibinfo {year} {2024}{\natexlab{a}})}\BibitemShut {NoStop}%
\bibitem [{\citenamefont {Medel~Onofre}\ and\ \citenamefont
  {Mart\'{\i}n-Ruiz}(2023)}]{onofre}%
  \BibitemOpen
  \bibfield  {author} {\bibinfo {author} {\bibfnamefont {L.}~\bibnamefont
  {Medel~Onofre}}\ and\ \bibinfo {author} {\bibfnamefont {A.}~\bibnamefont
  {Mart\'{\i}n-Ruiz}},\ }\bibfield  {title} {\bibinfo {title} {Planar hall
  effect in {W}eyl semimetals induced by pseudoelectromagnetic fields},\ }\href
  {https://doi.org/10.1103/PhysRevB.108.155132} {\bibfield  {journal} {\bibinfo
   {journal} {Phys. Rev. B}\ }\textbf {\bibinfo {volume} {108}},\ \bibinfo
  {pages} {155132} (\bibinfo {year} {2023})}\BibitemShut {NoStop}%
\bibitem [{\citenamefont {Ghosh}\ and\ \citenamefont
  {Mandal}(2024{\natexlab{b}})}]{ips-rahul-tilt}%
  \BibitemOpen
  \bibfield  {author} {\bibinfo {author} {\bibfnamefont {R.}~\bibnamefont
  {Ghosh}}\ and\ \bibinfo {author} {\bibfnamefont {I.}~\bibnamefont {Mandal}},\
  }\bibfield  {title} {\bibinfo {title} {{Direction-dependent conductivity in
  planar Hall set-ups with tilted Weyl/multi-Weyl semimetals}},\ }\href
  {https://doi.org/10.1088/1361-648X/ad38fa} {\bibfield  {journal} {\bibinfo
  {journal} {Journal of Physics Condensed Matter}\ }\textbf {\bibinfo {volume}
  {36}},\ \bibinfo {pages} {275501} (\bibinfo {year}
  {2024}{\natexlab{b}})}\BibitemShut {NoStop}%
\bibitem [{\citenamefont {{Medel}}\ \emph {et~al.}(2024)\citenamefont
  {{Medel}}, \citenamefont {{Ghosh}}, \citenamefont {{Mart{\'\i}n-Ruiz}},\ and\
  \citenamefont {{Mandal}}}]{ips-medel}%
  \BibitemOpen
  \bibfield  {author} {\bibinfo {author} {\bibfnamefont {L.}~\bibnamefont
  {{Medel}}}, \bibinfo {author} {\bibfnamefont {R.}~\bibnamefont {{Ghosh}}},
  \bibinfo {author} {\bibfnamefont {A.}~\bibnamefont {{Mart{\'\i}n-Ruiz}}},\
  and\ \bibinfo {author} {\bibfnamefont {I.}~\bibnamefont {{Mandal}}},\
  }\bibfield  {title} {\bibinfo {title} {{Electric, thermal, and thermoelectric
  magnetoconductivity for Weyl/multi-Weyl semimetals in planar Hall set-ups
  induced by the combined effects of topology and strain}},\ }\href
  {https://doi.org/10.1038/s41598-024-68615-0} {\bibfield  {journal} {\bibinfo
  {journal} {Scientific Reports}\ }\textbf {\bibinfo {volume} {14}},\ \bibinfo
  {eid} {21390} (\bibinfo {year} {2024})}\BibitemShut {NoStop}%
\bibitem [{\citenamefont {Das}\ and\ \citenamefont
  {Agarwal}(2020)}]{das20_thermal}%
  \BibitemOpen
  \bibfield  {author} {\bibinfo {author} {\bibfnamefont {K.}~\bibnamefont
  {Das}}\ and\ \bibinfo {author} {\bibfnamefont {A.}~\bibnamefont {Agarwal}},\
  }\bibfield  {title} {\bibinfo {title} {Thermal and gravitational chiral
  anomaly induced magneto-transport in {W}eyl semimetals},\ }\href
  {https://doi.org/10.1103/PhysRevResearch.2.013088} {\bibfield  {journal}
  {\bibinfo  {journal} {Phys. Rev. Res.}\ }\textbf {\bibinfo {volume} {2}},\
  \bibinfo {pages} {013088} (\bibinfo {year} {2020})}\BibitemShut {NoStop}%
\bibitem [{\citenamefont {Ghosh}\ \emph {et~al.}(2024)\citenamefont {Ghosh},
  \citenamefont {Haidar},\ and\ \citenamefont {Mandal}}]{ips-rsw-ph}%
  \BibitemOpen
  \bibfield  {author} {\bibinfo {author} {\bibfnamefont {R.}~\bibnamefont
  {Ghosh}}, \bibinfo {author} {\bibfnamefont {F.}~\bibnamefont {Haidar}},\ and\
  \bibinfo {author} {\bibfnamefont {I.}~\bibnamefont {Mandal}},\ }\bibfield
  {title} {\bibinfo {title} {{Linear response in planar Hall and thermal Hall
  setups for Rarita-Schwinger-Weyl semimetals}},\ }\href
  {https://doi.org/10.1103/PhysRevB.110.245113} {\bibfield  {journal} {\bibinfo
   {journal} {Phys. Rev. B}\ }\textbf {\bibinfo {volume} {110}},\ \bibinfo
  {pages} {245113} (\bibinfo {year} {2024})}\BibitemShut {NoStop}%
\bibitem [{\citenamefont {Mandal}\ \emph {et~al.}(2025)\citenamefont {Mandal},
  \citenamefont {Saha},\ and\ \citenamefont {Ghosh}}]{ips-shreya}%
  \BibitemOpen
  \bibfield  {author} {\bibinfo {author} {\bibfnamefont {I.}~\bibnamefont
  {Mandal}}, \bibinfo {author} {\bibfnamefont {S.}~\bibnamefont {Saha}},\ and\
  \bibinfo {author} {\bibfnamefont {R.}~\bibnamefont {Ghosh}},\ }\bibfield
  {title} {\bibinfo {title} {{Signatures of topology in generic transport
  measurements for Rarita-Schwinger-Weyl semimetals}},\ }\href
  {https://doi.org/https://doi.org/10.1016/j.ssc.2024.115799} {\bibfield
  {journal} {\bibinfo  {journal} {Solid State Communications}\ }\textbf
  {\bibinfo {volume} {397}},\ \bibinfo {pages} {115799} (\bibinfo {year}
  {2025})}\BibitemShut {NoStop}%
\bibitem [{\citenamefont {Trescher}\ \emph {et~al.}(2015)\citenamefont
  {Trescher}, \citenamefont {Sbierski}, \citenamefont {Brouwer},\ and\
  \citenamefont {Bergholtz}}]{emil_tilted}%
  \BibitemOpen
  \bibfield  {author} {\bibinfo {author} {\bibfnamefont {M.}~\bibnamefont
  {Trescher}}, \bibinfo {author} {\bibfnamefont {B.}~\bibnamefont {Sbierski}},
  \bibinfo {author} {\bibfnamefont {P.~W.}\ \bibnamefont {Brouwer}},\ and\
  \bibinfo {author} {\bibfnamefont {E.~J.}\ \bibnamefont {Bergholtz}},\
  }\bibfield  {title} {\bibinfo {title} {{Quantum transport in Dirac materials:
  Signatures of tilted and anisotropic Dirac and Weyl cones}},\ }\href
  {https://doi.org/10.1103/PhysRevB.91.115135} {\bibfield  {journal} {\bibinfo
  {journal} {Phys. Rev. B}\ }\textbf {\bibinfo {volume} {91}},\ \bibinfo
  {pages} {115135} (\bibinfo {year} {2015})}\BibitemShut {NoStop}%
\bibitem [{\citenamefont {Trescher}\ \emph {et~al.}(2017)\citenamefont
  {Trescher}, \citenamefont {Sbierski}, \citenamefont {Brouwer},\ and\
  \citenamefont {Bergholtz}}]{trescher17_tilted}%
  \BibitemOpen
  \bibfield  {author} {\bibinfo {author} {\bibfnamefont {M.}~\bibnamefont
  {Trescher}}, \bibinfo {author} {\bibfnamefont {B.}~\bibnamefont {Sbierski}},
  \bibinfo {author} {\bibfnamefont {P.~W.}\ \bibnamefont {Brouwer}},\ and\
  \bibinfo {author} {\bibfnamefont {E.~J.}\ \bibnamefont {Bergholtz}},\
  }\bibfield  {title} {\bibinfo {title} {{Tilted disordered Weyl semimetals}},\
  }\href {https://doi.org/10.1103/PhysRevB.95.045139} {\bibfield  {journal}
  {\bibinfo  {journal} {Phys. Rev. B}\ }\textbf {\bibinfo {volume} {95}},\
  \bibinfo {pages} {045139} (\bibinfo {year} {2017})}\BibitemShut {NoStop}%
\bibitem [{\citenamefont {Herring}(1937)}]{herring}%
  \BibitemOpen
  \bibfield  {author} {\bibinfo {author} {\bibfnamefont {C.}~\bibnamefont
  {Herring}},\ }\bibfield  {title} {\bibinfo {title} {Accidental degeneracy in
  the energy bands of crystals},\ }\href
  {https://doi.org/10.1103/PhysRev.52.365} {\bibfield  {journal} {\bibinfo
  {journal} {Phys. Rev.}\ }\textbf {\bibinfo {volume} {52}},\ \bibinfo {pages}
  {365} (\bibinfo {year} {1937})}\BibitemShut {NoStop}%
\bibitem [{\citenamefont {{Soluyanov}}\ \emph {et~al.}(2015)\citenamefont
  {{Soluyanov}}, \citenamefont {{Gresch}}, \citenamefont {{Wang}},
  \citenamefont {{Wu}}, \citenamefont {{Troyer}}, \citenamefont {{Dai}},\ and\
  \citenamefont {{Bernevig}}}]{soluyanov2015type}%
  \BibitemOpen
  \bibfield  {author} {\bibinfo {author} {\bibfnamefont {A.~A.}\ \bibnamefont
  {{Soluyanov}}}, \bibinfo {author} {\bibfnamefont {D.}~\bibnamefont
  {{Gresch}}}, \bibinfo {author} {\bibfnamefont {Z.}~\bibnamefont {{Wang}}},
  \bibinfo {author} {\bibfnamefont {Q.}~\bibnamefont {{Wu}}}, \bibinfo {author}
  {\bibfnamefont {M.}~\bibnamefont {{Troyer}}}, \bibinfo {author}
  {\bibfnamefont {X.}~\bibnamefont {{Dai}}},\ and\ \bibinfo {author}
  {\bibfnamefont {B.~A.}\ \bibnamefont {{Bernevig}}},\ }\bibfield  {title}
  {\bibinfo {title} {{Type-II Weyl semimetals}},\ }\href
  {https://doi.org/10.1038/nature15768} {\bibfield  {journal} {\bibinfo
  {journal} {\nat}\ }\textbf {\bibinfo {volume} {527}},\ \bibinfo {pages} {495}
  (\bibinfo {year} {2015})}\BibitemShut {NoStop}%
\bibitem [{\citenamefont {Ma}\ \emph {et~al.}(2019)\citenamefont {Ma},
  \citenamefont {Jiang}, \citenamefont {Liu},\ and\ \citenamefont
  {Xie}}]{ma19_planar}%
  \BibitemOpen
  \bibfield  {author} {\bibinfo {author} {\bibfnamefont {D.}~\bibnamefont
  {Ma}}, \bibinfo {author} {\bibfnamefont {H.}~\bibnamefont {Jiang}}, \bibinfo
  {author} {\bibfnamefont {H.}~\bibnamefont {Liu}},\ and\ \bibinfo {author}
  {\bibfnamefont {X.~C.}\ \bibnamefont {Xie}},\ }\bibfield  {title} {\bibinfo
  {title} {{Planar Hall effect in tilted Weyl semimetals}},\ }\href
  {https://doi.org/10.1103/PhysRevB.99.115121} {\bibfield  {journal} {\bibinfo
  {journal} {Phys. Rev. B}\ }\textbf {\bibinfo {volume} {99}},\ \bibinfo
  {pages} {115121} (\bibinfo {year} {2019})}\BibitemShut {NoStop}%
\bibitem [{\citenamefont {Kundu}\ \emph {et~al.}(2020)\citenamefont {Kundu},
  \citenamefont {Yang},\ and\ \citenamefont {Jalil}}]{kundu20}%
  \BibitemOpen
  \bibfield  {author} {\bibinfo {author} {\bibfnamefont {A.}~\bibnamefont
  {Kundu}}, \bibinfo {author} {\bibfnamefont {H.}~\bibnamefont {Yang}},\ and\
  \bibinfo {author} {\bibfnamefont {M.~B.~A.}\ \bibnamefont {Jalil}},\
  }\bibfield  {title} {\bibinfo {title} {{Magnetotransport of Weyl semimetals
  with tilted Dirac cones}},\ }\href {https://doi.org/10.1088/1367-2630/aba98d}
  {\bibfield  {journal} {\bibinfo  {journal} {New Journal of Physics}\ }\textbf
  {\bibinfo {volume} {22}},\ \bibinfo {pages} {083081} (\bibinfo {year}
  {2020})}\BibitemShut {NoStop}%
\bibitem [{\citenamefont {K\"onye}\ and\ \citenamefont
  {Ogata}(2021)}]{konye21_microscopic}%
  \BibitemOpen
  \bibfield  {author} {\bibinfo {author} {\bibfnamefont {V.}~\bibnamefont
  {K\"onye}}\ and\ \bibinfo {author} {\bibfnamefont {M.}~\bibnamefont
  {Ogata}},\ }\bibfield  {title} {\bibinfo {title} {{Microscopic theory of
  magnetoconductivity at low magnetic fields in terms of Berry curvature and
  orbital magnetic moment}},\ }\href
  {https://doi.org/10.1103/PhysRevResearch.3.033076} {\bibfield  {journal}
  {\bibinfo  {journal} {Phys. Rev. Res.}\ }\textbf {\bibinfo {volume} {3}},\
  \bibinfo {pages} {033076} (\bibinfo {year} {2021})}\BibitemShut {NoStop}%
\bibitem [{\citenamefont {Shao}\ and\ \citenamefont
  {Yan}(2022)}]{shao22_plane}%
  \BibitemOpen
  \bibfield  {author} {\bibinfo {author} {\bibfnamefont {J.}~\bibnamefont
  {Shao}}\ and\ \bibinfo {author} {\bibfnamefont {L.}~\bibnamefont {Yan}},\
  }\bibfield  {title} {\bibinfo {title} {{In-plane magnetotransport phenomena
  in tilted Weyl semimetals}},\ }\href
  {https://doi.org/10.1088/1361-648X/ac9e35} {\bibfield  {journal} {\bibinfo
  {journal} {Journal of Phys.: Condensed Matter}\ }\textbf {\bibinfo {volume}
  {51}},\ \bibinfo {pages} {025401} (\bibinfo {year} {2022})}\BibitemShut
  {NoStop}%
\bibitem [{\citenamefont {{Xiong}}\ \emph {et~al.}(2016)\citenamefont
  {{Xiong}}, \citenamefont {{Kushwaha}}, \citenamefont {{Krizan}},
  \citenamefont {{Liang}}, \citenamefont {{Cava}},\ and\ \citenamefont
  {{Ong}}}]{params2}%
  \BibitemOpen
  \bibfield  {author} {\bibinfo {author} {\bibfnamefont {J.}~\bibnamefont
  {{Xiong}}}, \bibinfo {author} {\bibfnamefont {S.}~\bibnamefont {{Kushwaha}}},
  \bibinfo {author} {\bibfnamefont {J.}~\bibnamefont {{Krizan}}}, \bibinfo
  {author} {\bibfnamefont {T.}~\bibnamefont {{Liang}}}, \bibinfo {author}
  {\bibfnamefont {R.~J.}\ \bibnamefont {{Cava}}},\ and\ \bibinfo {author}
  {\bibfnamefont {N.~P.}\ \bibnamefont {{Ong}}},\ }\bibfield  {title} {\bibinfo
  {title} {{Anomalous conductivity tensor in the Dirac semimetal Na$_3$Bi}},\
  }\href {https://doi.org/10.1209/0295-5075/114/27002} {\bibfield  {journal}
  {\bibinfo  {journal} {EPL (Europhysics Letters)}\ }\textbf {\bibinfo {volume}
  {114}},\ \bibinfo {pages} {27002} (\bibinfo {year} {2016})}\BibitemShut
  {NoStop}%
\bibitem [{\citenamefont {Xiao}\ \emph {et~al.}(2010)\citenamefont {Xiao},
  \citenamefont {Chang},\ and\ \citenamefont {Niu}}]{xiao_review}%
  \BibitemOpen
  \bibfield  {author} {\bibinfo {author} {\bibfnamefont {D.}~\bibnamefont
  {Xiao}}, \bibinfo {author} {\bibfnamefont {M.-C.}\ \bibnamefont {Chang}},\
  and\ \bibinfo {author} {\bibfnamefont {Q.}~\bibnamefont {Niu}},\ }\bibfield
  {title} {\bibinfo {title} {Berry phase effects on electronic properties},\
  }\href {https://doi.org/10.1103/RevModPhys.82.1959} {\bibfield  {journal}
  {\bibinfo  {journal} {Rev. Mod. Phys.}\ }\textbf {\bibinfo {volume} {82}},\
  \bibinfo {pages} {1959} (\bibinfo {year} {2010})}\BibitemShut {NoStop}%
\bibitem [{\citenamefont {Sundaram}\ and\ \citenamefont
  {Niu}(1999)}]{sundaram99_wavepacket}%
  \BibitemOpen
  \bibfield  {author} {\bibinfo {author} {\bibfnamefont {G.}~\bibnamefont
  {Sundaram}}\ and\ \bibinfo {author} {\bibfnamefont {Q.}~\bibnamefont {Niu}},\
  }\bibfield  {title} {\bibinfo {title} {{Wave-packet dynamics in slowly
  perturbed crystals: Gradient corrections and Berry-phase effects}},\ }\href
  {https://doi.org/10.1103/PhysRevB.59.14915} {\bibfield  {journal} {\bibinfo
  {journal} {Phys. Rev. B}\ }\textbf {\bibinfo {volume} {59}},\ \bibinfo
  {pages} {14915} (\bibinfo {year} {1999})}\BibitemShut {NoStop}%
\bibitem [{\citenamefont {Knoll}\ \emph {et~al.}(2020)\citenamefont {Knoll},
  \citenamefont {Timm},\ and\ \citenamefont {Meng}}]{timm}%
  \BibitemOpen
  \bibfield  {author} {\bibinfo {author} {\bibfnamefont {A.}~\bibnamefont
  {Knoll}}, \bibinfo {author} {\bibfnamefont {C.}~\bibnamefont {Timm}},\ and\
  \bibinfo {author} {\bibfnamefont {T.}~\bibnamefont {Meng}},\ }\bibfield
  {title} {\bibinfo {title} {{Negative longitudinal magnetoconductance at weak
  fields in Weyl semimetals}},\ }\href
  {https://doi.org/10.1103/PhysRevB.101.201402} {\bibfield  {journal} {\bibinfo
   {journal} {Phys. Rev. B}\ }\textbf {\bibinfo {volume} {101}},\ \bibinfo
  {pages} {201402} (\bibinfo {year} {2020})}\BibitemShut {NoStop}%
\bibitem [{\citenamefont {Haldane}(2004)}]{haldane}%
  \BibitemOpen
  \bibfield  {author} {\bibinfo {author} {\bibfnamefont {F.~D.~M.}\
  \bibnamefont {Haldane}},\ }\bibfield  {title} {\bibinfo {title} {Berry
  curvature on the {F}ermi surface: {A}nomalous {H}all effect as a topological
  {F}ermi-liquid property},\ }\href
  {https://doi.org/10.1103/PhysRevLett.93.206602} {\bibfield  {journal}
  {\bibinfo  {journal} {Phys. Rev. Lett.}\ }\textbf {\bibinfo {volume} {93}},\
  \bibinfo {pages} {206602} (\bibinfo {year} {2004})}\BibitemShut {NoStop}%
\bibitem [{\citenamefont {Goswami}\ and\ \citenamefont
  {Tewari}(2013)}]{pallab_axionic}%
  \BibitemOpen
  \bibfield  {author} {\bibinfo {author} {\bibfnamefont {P.}~\bibnamefont
  {Goswami}}\ and\ \bibinfo {author} {\bibfnamefont {S.}~\bibnamefont
  {Tewari}},\ }\bibfield  {title} {\bibinfo {title} {{Axionic field theory of
  $(3+1)$-dimensional Weyl semimetals}},\ }\href
  {https://doi.org/10.1103/PhysRevB.88.245107} {\bibfield  {journal} {\bibinfo
  {journal} {Phys. Rev. B}\ }\textbf {\bibinfo {volume} {88}},\ \bibinfo
  {pages} {245107} (\bibinfo {year} {2013})}\BibitemShut {NoStop}%
\bibitem [{\citenamefont {Burkov}(2014)}]{burkov_intrinsic_hall}%
  \BibitemOpen
  \bibfield  {author} {\bibinfo {author} {\bibfnamefont {A.~A.}\ \bibnamefont
  {Burkov}},\ }\bibfield  {title} {\bibinfo {title} {Anomalous {H}all effect in
  {W}eyl metals},\ }\href {https://doi.org/10.1103/PhysRevLett.113.187202}
  {\bibfield  {journal} {\bibinfo  {journal} {Phys. Rev. Lett.}\ }\textbf
  {\bibinfo {volume} {113}},\ \bibinfo {pages} {187202} (\bibinfo {year}
  {2014})}\BibitemShut {NoStop}%
\bibitem [{\citenamefont {Gusynin}\ \emph {et~al.}(2006)\citenamefont
  {Gusynin}, \citenamefont {Sharapov},\ and\ \citenamefont
  {Carbotte}}]{gusynin06_magneto}%
  \BibitemOpen
  \bibfield  {author} {\bibinfo {author} {\bibfnamefont {V.}~\bibnamefont
  {Gusynin}}, \bibinfo {author} {\bibfnamefont {S.}~\bibnamefont {Sharapov}},\
  and\ \bibinfo {author} {\bibfnamefont {J.}~\bibnamefont {Carbotte}},\
  }\bibfield  {title} {\bibinfo {title} {Magneto-optical conductivity in
  graphene},\ }\href {https://doi.org/10.1088/0953-8984/19/2/026222} {\bibfield
   {journal} {\bibinfo  {journal} {Journal of Physics: Condensed Matter}\
  }\textbf {\bibinfo {volume} {19}},\ \bibinfo {pages} {026222} (\bibinfo
  {year} {2006})}\BibitemShut {NoStop}%
\bibitem [{\citenamefont {St\aa{}lhammar}\ \emph {et~al.}(2020)\citenamefont
  {St\aa{}lhammar}, \citenamefont {Larana-Aragon}, \citenamefont {Knolle},\
  and\ \citenamefont {Bergholtz}}]{marcus-emil}%
  \BibitemOpen
  \bibfield  {author} {\bibinfo {author} {\bibfnamefont {M.}~\bibnamefont
  {St\aa{}lhammar}}, \bibinfo {author} {\bibfnamefont {J.}~\bibnamefont
  {Larana-Aragon}}, \bibinfo {author} {\bibfnamefont {J.}~\bibnamefont
  {Knolle}},\ and\ \bibinfo {author} {\bibfnamefont {E.~J.}\ \bibnamefont
  {Bergholtz}},\ }\bibfield  {title} {\bibinfo {title} {Magneto-optical
  conductivity in generic {W}eyl semimetals},\ }\href
  {https://doi.org/10.1103/PhysRevB.102.235134} {\bibfield  {journal} {\bibinfo
   {journal} {Phys. Rev. B}\ }\textbf {\bibinfo {volume} {102}},\ \bibinfo
  {pages} {235134} (\bibinfo {year} {2020})}\BibitemShut {NoStop}%
\bibitem [{\citenamefont {Yadav}\ \emph {et~al.}(2023)\citenamefont {Yadav},
  \citenamefont {Sekh},\ and\ \citenamefont {Mandal}}]{ips_optical_cond}%
  \BibitemOpen
  \bibfield  {author} {\bibinfo {author} {\bibfnamefont {S.}~\bibnamefont
  {Yadav}}, \bibinfo {author} {\bibfnamefont {S.}~\bibnamefont {Sekh}},\ and\
  \bibinfo {author} {\bibfnamefont {I.}~\bibnamefont {Mandal}},\ }\bibfield
  {title} {\bibinfo {title} {Magneto-optical conductivity in the type-{I} and
  type-{II} phases of {W}eyl/multi-{W}eyl semimetals},\ }\href
  {https://doi.org/10.1016/j.physb.2023.414765} {\bibfield  {journal} {\bibinfo
   {journal} {Physica B: Condensed Matter}\ }\textbf {\bibinfo {volume}
  {656}},\ \bibinfo {pages} {414765} (\bibinfo {year} {2023})}\BibitemShut
  {NoStop}%
\bibitem [{\citenamefont {Papaj}\ and\ \citenamefont
  {Fu}(2019)}]{papaj_magnus}%
  \BibitemOpen
  \bibfield  {author} {\bibinfo {author} {\bibfnamefont {M.}~\bibnamefont
  {Papaj}}\ and\ \bibinfo {author} {\bibfnamefont {L.}~\bibnamefont {Fu}},\
  }\bibfield  {title} {\bibinfo {title} {Magnus {H}all effect},\ }\href
  {https://doi.org/10.1103/PhysRevLett.123.216802} {\bibfield  {journal}
  {\bibinfo  {journal} {Phys. Rev. Lett.}\ }\textbf {\bibinfo {volume} {123}},\
  \bibinfo {pages} {216802} (\bibinfo {year} {2019})}\BibitemShut {NoStop}%
\bibitem [{\citenamefont {Mandal}\ \emph {et~al.}(2020)\citenamefont {Mandal},
  \citenamefont {Das},\ and\ \citenamefont {Agarwal}}]{amit-magnus}%
  \BibitemOpen
  \bibfield  {author} {\bibinfo {author} {\bibfnamefont {D.}~\bibnamefont
  {Mandal}}, \bibinfo {author} {\bibfnamefont {K.}~\bibnamefont {Das}},\ and\
  \bibinfo {author} {\bibfnamefont {A.}~\bibnamefont {Agarwal}},\ }\bibfield
  {title} {\bibinfo {title} {Magnus {N}ernst and thermal {H}all effect},\
  }\href {https://doi.org/10.1103/PhysRevB.102.205414} {\bibfield  {journal}
  {\bibinfo  {journal} {Phys. Rev. B}\ }\textbf {\bibinfo {volume} {102}},\
  \bibinfo {pages} {205414} (\bibinfo {year} {2020})}\BibitemShut {NoStop}%
\bibitem [{\citenamefont {{Sekh, Sajid}}\ and\ \citenamefont {{Mandal,
  Ipsita}}(2022)}]{ips-magnus}%
  \BibitemOpen
  \bibfield  {author} {\bibinfo {author} {\bibnamefont {{Sekh, Sajid}}}\ and\
  \bibinfo {author} {\bibnamefont {{Mandal, Ipsita}}},\ }\bibfield  {title}
  {\bibinfo {title} {Magnus {H}all effect in three-dimensional topological
  semimetals},\ }\href {https://doi.org/10.1140/epjp/s13360-022-02840-2}
  {\bibfield  {journal} {\bibinfo  {journal} {Eur. Phys. J. Plus}\ }\textbf
  {\bibinfo {volume} {137}},\ \bibinfo {pages} {736} (\bibinfo {year}
  {2022})}\BibitemShut {NoStop}%
\bibitem [{\citenamefont {Sekh}\ and\ \citenamefont {Mandal}(2022)}]{ips-cd1}%
  \BibitemOpen
  \bibfield  {author} {\bibinfo {author} {\bibfnamefont {S.}~\bibnamefont
  {Sekh}}\ and\ \bibinfo {author} {\bibfnamefont {I.}~\bibnamefont {Mandal}},\
  }\bibfield  {title} {\bibinfo {title} {{Circular dichroism as a probe for
  topology in three-dimensional semimetals}},\ }\href
  {https://doi.org/10.1103/PhysRevB.105.235403} {\bibfield  {journal} {\bibinfo
   {journal} {Phys. Rev. B}\ }\textbf {\bibinfo {volume} {105}},\ \bibinfo
  {pages} {235403} (\bibinfo {year} {2022})}\BibitemShut {NoStop}%
\bibitem [{\citenamefont {Mandal}(2024)}]{ips-cd}%
  \BibitemOpen
  \bibfield  {author} {\bibinfo {author} {\bibfnamefont {I.}~\bibnamefont
  {Mandal}},\ }\bibfield  {title} {\bibinfo {title} {Signatures of two- and
  three-dimensional semimetals from circular dichroism},\ }\href
  {https://doi.org/10.1142/S0217979224502163} {\bibfield  {journal} {\bibinfo
  {journal} {International Journal of Modern Physics B}\ }\textbf {\bibinfo
  {volume} {38}},\ \bibinfo {pages} {2450216} (\bibinfo {year}
  {2024})}\BibitemShut {NoStop}%
\bibitem [{\citenamefont {Moore}(2018)}]{moore18_optical}%
  \BibitemOpen
  \bibfield  {author} {\bibinfo {author} {\bibfnamefont {J.~E.}\ \bibnamefont
  {Moore}},\ }\bibfield  {title} {\bibinfo {title} {{Optical properties of Weyl
  semimetals}},\ }\href {https://doi.org/10.1093/nsr/nwy164} {\bibfield
  {journal} {\bibinfo  {journal} {National Science Review}\ }\textbf {\bibinfo
  {volume} {6}},\ \bibinfo {pages} {206} (\bibinfo {year} {2018})}\BibitemShut
  {NoStop}%
\bibitem [{\citenamefont {Guo}\ \emph {et~al.}(2023)\citenamefont {Guo},
  \citenamefont {Asadchy}, \citenamefont {Zhao},\ and\ \citenamefont
  {Fan}}]{guo23_light}%
  \BibitemOpen
  \bibfield  {author} {\bibinfo {author} {\bibfnamefont {C.}~\bibnamefont
  {Guo}}, \bibinfo {author} {\bibfnamefont {V.~S.}\ \bibnamefont {Asadchy}},
  \bibinfo {author} {\bibfnamefont {B.}~\bibnamefont {Zhao}},\ and\ \bibinfo
  {author} {\bibfnamefont {S.}~\bibnamefont {Fan}},\ }\bibfield  {title}
  {\bibinfo {title} {{Light control with Weyl semimetals}},\ }\href
  {https://doi.org/10.1186/s43593-022-00036-w} {\bibfield  {journal} {\bibinfo
  {journal} {eLight}\ }\textbf {\bibinfo {volume} {3}},\ \bibinfo {pages} {2}
  (\bibinfo {year} {2023})}\BibitemShut {NoStop}%
\bibitem [{\citenamefont {Avdoshkin}\ \emph {et~al.}(2020)\citenamefont
  {Avdoshkin}, \citenamefont {Kozii},\ and\ \citenamefont {Moore}}]{kozii}%
  \BibitemOpen
  \bibfield  {author} {\bibinfo {author} {\bibfnamefont {A.}~\bibnamefont
  {Avdoshkin}}, \bibinfo {author} {\bibfnamefont {V.}~\bibnamefont {Kozii}},\
  and\ \bibinfo {author} {\bibfnamefont {J.~E.}\ \bibnamefont {Moore}},\
  }\bibfield  {title} {\bibinfo {title} {Interactions remove the quantization
  of the chiral photocurrent at weyl points},\ }\href
  {https://doi.org/10.1103/PhysRevLett.124.196603} {\bibfield  {journal}
  {\bibinfo  {journal} {Phys. Rev. Lett.}\ }\textbf {\bibinfo {volume} {124}},\
  \bibinfo {pages} {196603} (\bibinfo {year} {2020})}\BibitemShut {NoStop}%
\bibitem [{\citenamefont {Mandal}(2020)}]{ips_cpge}%
  \BibitemOpen
  \bibfield  {author} {\bibinfo {author} {\bibfnamefont {I.}~\bibnamefont
  {Mandal}},\ }\bibfield  {title} {\bibinfo {title} {Effect of interactions on
  the quantization of the chiral photocurrent for double-{W}eyl semimetals},\
  }\href {https://www.mdpi.com/2073-8994/12/6/919} {\bibfield  {journal}
  {\bibinfo  {journal} {Symmetry}\ }\textbf {\bibinfo {volume} {12}} (\bibinfo
  {year} {2020})}\BibitemShut {NoStop}%
\bibitem [{\citenamefont {{Mandal}}\ and\ \citenamefont
  {{Sen}}(2021)}]{ips-aritra}%
  \BibitemOpen
  \bibfield  {author} {\bibinfo {author} {\bibfnamefont {I.}~\bibnamefont
  {{Mandal}}}\ and\ \bibinfo {author} {\bibfnamefont {A.}~\bibnamefont
  {{Sen}}},\ }\bibfield  {title} {\bibinfo {title} {{Tunneling of multi-Weyl
  semimetals through a potential barrier under the influence of magnetic
  fields}},\ }\href {https://doi.org/10.1016/j.physleta.2021.127293} {\bibfield
   {journal} {\bibinfo  {journal} {Phys. Lett. A}\ }\textbf {\bibinfo {volume}
  {399}},\ \bibinfo {eid} {127293} (\bibinfo {year} {2021})}\BibitemShut
  {NoStop}%
\bibitem [{\citenamefont {{Bera}}\ and\ \citenamefont
  {{Mandal}}(2021)}]{ips-sandip}%
  \BibitemOpen
  \bibfield  {author} {\bibinfo {author} {\bibfnamefont {S.}~\bibnamefont
  {{Bera}}}\ and\ \bibinfo {author} {\bibfnamefont {I.}~\bibnamefont
  {{Mandal}}},\ }\bibfield  {title} {\bibinfo {title} {{Floquet scattering of
  quadratic band-touching semimetals through a time-periodic potential well}},\
  }\href {https://doi.org/10.1088/1361-648X/ac020a} {\bibfield  {journal}
  {\bibinfo  {journal} {Journal of Physics Condensed Matter}\ }\textbf
  {\bibinfo {volume} {33}},\ \bibinfo {eid} {295502} (\bibinfo {year}
  {2021})}\BibitemShut {NoStop}%
\bibitem [{\citenamefont {{Bera}}\ \emph {et~al.}(2023)\citenamefont {{Bera}},
  \citenamefont {{Sekh}},\ and\ \citenamefont {{Mandal}}}]{ips-sandip-sajid}%
  \BibitemOpen
  \bibfield  {author} {\bibinfo {author} {\bibfnamefont {S.}~\bibnamefont
  {{Bera}}}, \bibinfo {author} {\bibfnamefont {S.}~\bibnamefont {{Sekh}}},\
  and\ \bibinfo {author} {\bibfnamefont {I.}~\bibnamefont {{Mandal}}},\
  }\bibfield  {title} {\bibinfo {title} {{Floquet transmission in
  Weyl/multi-Weyl and nodal-line semimetals through a time-periodic potential
  well}},\ }\href {https://doi.org/10.1002/andp.202200460} {\bibfield
  {journal} {\bibinfo  {journal} {Ann. Phys. (Berlin)}\ }\textbf {\bibinfo
  {volume} {535}},\ \bibinfo {pages} {2200460} (\bibinfo {year}
  {2023})}\BibitemShut {NoStop}%
\bibitem [{\citenamefont {{Mandal}}(2023)}]{ips-jns}%
  \BibitemOpen
  \bibfield  {author} {\bibinfo {author} {\bibfnamefont {I.}~\bibnamefont
  {{Mandal}}},\ }\bibfield  {title} {\bibinfo {title} {{Transmission and
  conductance across junctions of isotropic and anisotropic three-dimensional
  semimetals}},\ }\href {https://doi.org/10.1140/epjp/s13360-023-04652-4}
  {\bibfield  {journal} {\bibinfo  {journal} {Eur. Phys. J. Plus}\ }\textbf
  {\bibinfo {volume} {138}},\ \bibinfo {eid} {1039} (\bibinfo {year}
  {2023})}\BibitemShut {NoStop}%
\bibitem [{\citenamefont {Xiao}\ \emph {et~al.}(2007)\citenamefont {Xiao},
  \citenamefont {Yao},\ and\ \citenamefont {Niu}}]{xiao07_valley}%
  \BibitemOpen
  \bibfield  {author} {\bibinfo {author} {\bibfnamefont {D.}~\bibnamefont
  {Xiao}}, \bibinfo {author} {\bibfnamefont {W.}~\bibnamefont {Yao}},\ and\
  \bibinfo {author} {\bibfnamefont {Q.}~\bibnamefont {Niu}},\ }\bibfield
  {title} {\bibinfo {title} {Valley-contrasting physics in graphene: Magnetic
  moment and topological transport},\ }\href
  {https://doi.org/10.1103/PhysRevLett.99.236809} {\bibfield  {journal}
  {\bibinfo  {journal} {Phys. Rev. Lett.}\ }\textbf {\bibinfo {volume} {99}},\
  \bibinfo {pages} {236809} (\bibinfo {year} {2007})}\BibitemShut {NoStop}%
\bibitem [{\citenamefont {Ashcroft}\ and\ \citenamefont
  {Mermin}(2011)}]{mermin}%
  \BibitemOpen
  \bibfield  {author} {\bibinfo {author} {\bibfnamefont {N.}~\bibnamefont
  {Ashcroft}}\ and\ \bibinfo {author} {\bibfnamefont {N.}~\bibnamefont
  {Mermin}},\ }\href {https://books.google.de/books?id=x\_s\_YAAACAAJ} {\emph
  {\bibinfo {title} {Solid State Physics}}}\ (\bibinfo  {publisher} {Cengage
  Learning},\ \bibinfo {year} {2011})\BibitemShut {NoStop}%
\bibitem [{\citenamefont {{Mandal}}\ and\ \citenamefont
  {{Saha}}(2024)}]{ips-kush-review}%
  \BibitemOpen
  \bibfield  {author} {\bibinfo {author} {\bibfnamefont {I.}~\bibnamefont
  {{Mandal}}}\ and\ \bibinfo {author} {\bibfnamefont {K.}~\bibnamefont
  {{Saha}}},\ }\bibfield  {title} {\bibinfo {title} {{Thermoelectric response
  in nodal-point semimetals}},\ }\href {https://doi.org/10.1002/andp.202400016}
  {\bibfield  {journal} {\bibinfo  {journal} {Ann. Phys. (Berlin)}\ }\textbf
  {\bibinfo {volume} {536}},\ \bibinfo {pages} {2400016} (\bibinfo {year}
  {2024})}\BibitemShut {NoStop}%
\bibitem [{\citenamefont {Mandal}(2025)}]{ips-internode}%
  \BibitemOpen
  \bibfield  {author} {\bibinfo {author} {\bibfnamefont {I.}~\bibnamefont
  {Mandal}},\ }\bibfield  {title} {\bibinfo {title} {Chiral anomaly and
  internode scatterings in multifold semimetals},\ }\href
  {https://doi.org/10.1103/PhysRevB.111.165116} {\bibfield  {journal} {\bibinfo
   {journal} {Phys. Rev. B}\ }\textbf {\bibinfo {volume} {111}},\ \bibinfo
  {pages} {165116} (\bibinfo {year} {2025})}\BibitemShut {NoStop}%
\bibitem [{\citenamefont {Haidar}\ and\ \citenamefont
  {Mandal}(2025)}]{ips-spin1-ph}%
  \BibitemOpen
  \bibfield  {author} {\bibinfo {author} {\bibfnamefont {F.}~\bibnamefont
  {Haidar}}\ and\ \bibinfo {author} {\bibfnamefont {I.}~\bibnamefont
  {Mandal}},\ }\bibfield  {title} {\bibinfo {title} {{Reflections of
  topological properties in the planar-Hall response for semimetals carrying
  pseudospin-1 quantum numbers}},\ }\href
  {https://doi.org/https://doi.org/10.1016/j.aop.2025.170010} {\bibfield
  {journal} {\bibinfo  {journal} {Annals of Physics}\ }\textbf {\bibinfo
  {volume} {478}},\ \bibinfo {pages} {170010} (\bibinfo {year}
  {2025})}\BibitemShut {NoStop}%
\bibitem [{\citenamefont {Das}\ and\ \citenamefont
  {Agarwal}(2019{\natexlab{b}})}]{das19_linear2}%
  \BibitemOpen
  \bibfield  {author} {\bibinfo {author} {\bibfnamefont {K.}~\bibnamefont
  {Das}}\ and\ \bibinfo {author} {\bibfnamefont {A.}~\bibnamefont {Agarwal}},\
  }\bibfield  {title} {\bibinfo {title} {{Berry curvature induced thermopower
  in type-I and type-II Weyl semimetals}},\ }\href
  {https://doi.org/10.1103/PhysRevB.100.085406} {\bibfield  {journal} {\bibinfo
   {journal} {Phys. Rev. B}\ }\textbf {\bibinfo {volume} {100}},\ \bibinfo
  {pages} {085406} (\bibinfo {year} {2019}{\natexlab{b}})}\BibitemShut
  {NoStop}%
\bibitem [{\citenamefont {Tchoumakov}\ \emph {et~al.}(2016)\citenamefont
  {Tchoumakov}, \citenamefont {Civelli},\ and\ \citenamefont
  {Goerbig}}]{goerbig-LL-weyl}%
  \BibitemOpen
  \bibfield  {author} {\bibinfo {author} {\bibfnamefont {S.}~\bibnamefont
  {Tchoumakov}}, \bibinfo {author} {\bibfnamefont {M.}~\bibnamefont
  {Civelli}},\ and\ \bibinfo {author} {\bibfnamefont {M.~O.}\ \bibnamefont
  {Goerbig}},\ }\bibfield  {title} {\bibinfo {title} {Magnetic-field-induced
  relativistic properties in type-{I} and type-{II} {W}eyl semimetals},\ }\href
  {https://doi.org/10.1103/PhysRevLett.117.086402} {\bibfield  {journal}
  {\bibinfo  {journal} {Phys. Rev. Lett.}\ }\textbf {\bibinfo {volume} {117}},\
  \bibinfo {pages} {086402} (\bibinfo {year} {2016})}\BibitemShut {NoStop}%
\bibitem [{\citenamefont {Mandal}\ and\ \citenamefont {Saha}(2020)}]{ips-kush}%
  \BibitemOpen
  \bibfield  {author} {\bibinfo {author} {\bibfnamefont {I.}~\bibnamefont
  {Mandal}}\ and\ \bibinfo {author} {\bibfnamefont {K.}~\bibnamefont {Saha}},\
  }\bibfield  {title} {\bibinfo {title} {Thermopower in an anisotropic
  two-dimensional {W}eyl semimetal},\ }\href
  {https://doi.org/10.1103/PhysRevB.101.045101} {\bibfield  {journal} {\bibinfo
   {journal} {Phys. Rev. B}\ }\textbf {\bibinfo {volume} {101}},\ \bibinfo
  {pages} {045101} (\bibinfo {year} {2020})}\BibitemShut {NoStop}%
\bibitem [{\citenamefont {Fu}\ and\ \citenamefont
  {Wang}(2022)}]{fu22_thermoelectric}%
  \BibitemOpen
  \bibfield  {author} {\bibinfo {author} {\bibfnamefont {L.~X.}\ \bibnamefont
  {Fu}}\ and\ \bibinfo {author} {\bibfnamefont {C.~M.}\ \bibnamefont {Wang}},\
  }\bibfield  {title} {\bibinfo {title} {{Thermoelectric transport of
  multi-Weyl semimetals in the quantum limit}},\ }\href
  {https://doi.org/10.1103/PhysRevB.105.035201} {\bibfield  {journal} {\bibinfo
   {journal} {Phys. Rev. B}\ }\textbf {\bibinfo {volume} {105}},\ \bibinfo
  {pages} {035201} (\bibinfo {year} {2022})}\BibitemShut {NoStop}%
\bibitem [{\citenamefont {{Mandal}}\ and\ \citenamefont
  {{Gemsheim}}(2019)}]{ips-seb}%
  \BibitemOpen
  \bibfield  {author} {\bibinfo {author} {\bibfnamefont {I.}~\bibnamefont
  {{Mandal}}}\ and\ \bibinfo {author} {\bibfnamefont {S.}~\bibnamefont
  {{Gemsheim}}},\ }\bibfield  {title} {\bibinfo {title} {{Emergence of
  topological {M}ott insulators in proximity of quadratic band touching
  points}},\ }\href {https://doi.org/10.5488/CMP.22.13701} {\bibfield
  {journal} {\bibinfo  {journal} {Condensed Matter Physics}\ }\textbf {\bibinfo
  {volume} {22}},\ \bibinfo {pages} {13701} (\bibinfo {year}
  {2019})}\BibitemShut {NoStop}%
\bibitem [{\citenamefont {{Mandal}}(2021)}]{ips-biref}%
  \BibitemOpen
  \bibfield  {author} {\bibinfo {author} {\bibfnamefont {I.}~\bibnamefont
  {{Mandal}}},\ }\bibfield  {title} {\bibinfo {title} {{Robust marginal {F}ermi
  liquid in birefringent semimetals}},\ }\href
  {https://doi.org/10.1016/j.physleta.2021.127707} {\bibfield  {journal}
  {\bibinfo  {journal} {Phys. Lett. A}\ }\textbf {\bibinfo {volume} {418}},\
  \bibinfo {eid} {127707} (\bibinfo {year} {2021})}\BibitemShut {NoStop}%
\bibitem [{\citenamefont {{Mandal}}\ and\ \citenamefont
  {{Ziegler}}(2021)}]{ips-klaus}%
  \BibitemOpen
  \bibfield  {author} {\bibinfo {author} {\bibfnamefont {I.}~\bibnamefont
  {{Mandal}}}\ and\ \bibinfo {author} {\bibfnamefont {K.}~\bibnamefont
  {{Ziegler}}},\ }\bibfield  {title} {\bibinfo {title} {{Robust quantum
  transport at particle-hole symmetry}},\ }\href
  {https://doi.org/10.1209/0295-5075/ac1a25} {\bibfield  {journal} {\bibinfo
  {journal} {EPL (Europhysics Letters)}\ }\textbf {\bibinfo {volume} {135}},\
  \bibinfo {eid} {17001} (\bibinfo {year} {2021})}\BibitemShut {NoStop}%
\bibitem [{\citenamefont {Nandkishore}\ and\ \citenamefont
  {Parameswaran}(2017)}]{rahul-sid}%
  \BibitemOpen
  \bibfield  {author} {\bibinfo {author} {\bibfnamefont {R.~M.}\ \bibnamefont
  {Nandkishore}}\ and\ \bibinfo {author} {\bibfnamefont {S.~A.}\ \bibnamefont
  {Parameswaran}},\ }\bibfield  {title} {\bibinfo {title} {Disorder-driven
  destruction of a non-{F}ermi liquid semimetal studied by renormalization
  group analysis},\ }\href {https://doi.org/10.1103/PhysRevB.95.205106}
  {\bibfield  {journal} {\bibinfo  {journal} {Phys. Rev. B}\ }\textbf {\bibinfo
  {volume} {95}},\ \bibinfo {pages} {205106} (\bibinfo {year}
  {2017})}\BibitemShut {NoStop}%
\bibitem [{\citenamefont {Mandal}\ and\ \citenamefont
  {Nandkishore}(2018)}]{ips-rahul-qbt}%
  \BibitemOpen
  \bibfield  {author} {\bibinfo {author} {\bibfnamefont {I.}~\bibnamefont
  {Mandal}}\ and\ \bibinfo {author} {\bibfnamefont {R.~M.}\ \bibnamefont
  {Nandkishore}},\ }\bibfield  {title} {\bibinfo {title} {{Interplay of Coulomb
  interactions and disorder in three-dimensional quadratic band crossings
  without time-reversal symmetry and with unequal masses for conduction and
  valence bands}},\ }\href {https://doi.org/10.1103/PhysRevB.97.125121}
  {\bibfield  {journal} {\bibinfo  {journal} {Phys. Rev. B}\ }\textbf {\bibinfo
  {volume} {97}},\ \bibinfo {pages} {125121} (\bibinfo {year}
  {2018})}\BibitemShut {NoStop}%
\bibitem [{\citenamefont {Mandal}(2018)}]{ips-qbt-sc}%
  \BibitemOpen
  \bibfield  {author} {\bibinfo {author} {\bibfnamefont {I.}~\bibnamefont
  {Mandal}},\ }\bibfield  {title} {\bibinfo {title} {Fate of superconductivity
  in three-dimensional disordered {L}uttinger semimetals},\ }\href
  {https://doi.org/https://doi.org/10.1016/j.aop.2018.03.004} {\bibfield
  {journal} {\bibinfo  {journal} {Annals of Physics}\ }\textbf {\bibinfo
  {volume} {392}},\ \bibinfo {pages} {179 } (\bibinfo {year}
  {2018})}\BibitemShut {NoStop}%
\bibitem [{\citenamefont {Mandal}(2019)}]{ips-plasmons}%
  \BibitemOpen
  \bibfield  {author} {\bibinfo {author} {\bibfnamefont {I.}~\bibnamefont
  {Mandal}},\ }\bibfield  {title} {\bibinfo {title} {Search for plasmons in
  isotropic {L}uttinger semimetals},\ }\href
  {https://doi.org/https://doi.org/10.1016/j.aop.2019.04.002} {\bibfield
  {journal} {\bibinfo  {journal} {Annals of Physics}\ }\textbf {\bibinfo
  {volume} {406}},\ \bibinfo {pages} {173} (\bibinfo {year}
  {2019})}\BibitemShut {NoStop}%
\bibitem [{\citenamefont {Wang}\ and\ \citenamefont
  {Mandal}(2023)}]{ips-jing-plasmons}%
  \BibitemOpen
  \bibfield  {author} {\bibinfo {author} {\bibfnamefont {J.}~\bibnamefont
  {Wang}}\ and\ \bibinfo {author} {\bibfnamefont {I.}~\bibnamefont {Mandal}},\
  }\bibfield  {title} {\bibinfo {title} {{Anatomy of plasmons in generic
  Luttinger semimetals}},\ }\href
  {https://doi.org/10.1140/epjb/s10051-023-00596-x} {\bibfield  {journal}
  {\bibinfo  {journal} {The European Physical Journal B}\ }\textbf {\bibinfo
  {volume} {96}},\ \bibinfo {pages} {132} (\bibinfo {year} {2023})}\BibitemShut
  {NoStop}%
\bibitem [{\citenamefont {Mandal}\ and\ \citenamefont
  {Freire}(2024)}]{ips-hermann-review}%
  \BibitemOpen
  \bibfield  {author} {\bibinfo {author} {\bibfnamefont {I.}~\bibnamefont
  {Mandal}}\ and\ \bibinfo {author} {\bibfnamefont {H.}~\bibnamefont
  {Freire}},\ }\bibfield  {title} {\bibinfo {title} {{Transport properties in
  non-Fermi liquid phases of nodal-point semimetals}},\ }\href
  {https://doi.org/10.1088/1361-648X/ad665e} {\bibfield  {journal} {\bibinfo
  {journal} {Journal of Physics: Condensed Matter}\ }\textbf {\bibinfo {volume}
  {36}},\ \bibinfo {pages} {443002} (\bibinfo {year} {2024})}\BibitemShut
  {NoStop}%
\end{thebibliography}%


\end{document}